\documentclass[lineno]{jfm}

\makeatletter
\newcommand*{\rom}[1]{\expandafter\@slowromancap\romannumeral #1@}
\makeatother

\newcommand*\dif{\mathop{}\!\mathrm{d}}
\newcommand*\Dif{\mathop{}\!\mathrm{D}}
\usepackage{siunitx}
\usepackage{subfig}
\DeclareMathAlphabet\mathbfcal{OMS}{cmsy}{b}{n}

\usepackage{graphicx}
\usepackage{newtxtext}
\usepackage{newtxmath}
\usepackage{soul,xcolor}
\usepackage[sort&compress]{natbib}
\usepackage{hyperref}
\hypersetup{
    colorlinks = true,
    urlcolor   = blue,
    citecolor  = black,
}

\newcommand{\RomanNumeralCaps}[1]
\linenumbers

\title{A comparison between the FENE-P and sPTT constitutive models in Large Amplitude Oscillatory Shear (LAOS)}

\author{T.P. John\aff{1}
  \corresp{\email{thomas.john@manchester.ac.uk}},
  R.J. Poole\aff{2},
  A.J. Kowalski\aff{3}
 \and C.P. Fonte\aff{1}}

\affiliation{\aff{1}Department of Chemical Engineering, The University of Manchester, Manchester, M13 9PL, UK
\aff{2}School of Engineering, The University of Liverpool, Brownlow Street, Liverpool, L69 3GH, UK
\aff{3}Unilever R{\&}D, Port Sunlight Laboratory, Quarry Road East, Bebington, Wirral, CH63 3JW, UK}

\begin{document}
\maketitle
\setlength{\parskip}{0.5pt}

\setstcolor{red}

\begin{abstract}
The FENE-P and sPTT viscoelastic models are widely used for modelling of complex fluids. Although they are derived from distinct micro-structural theories, these models can become mathematically identical in steady and homogeneous flows with a particular choice of the value of the model parameters. However, even with this choice of parameter values, the model responses are known to differ from each other in transient flows. In this work, we investigate the responses of the FENE-P and sPTT constitutive models in Large Amplitude Oscillatory Shear (LAOS). In steady-shear, the shear stress scales with the non-dimensional group $Wi/(aL) \ (Wi\sqrt{\epsilon})$ for the FENE-P (sPTT) model, where $Wi$ is the Weissenberg number, $L^2$ is the limit of extensibility in the FENE-P model ($a$ being $L^2/(L^2-3)$) and $\epsilon$ is the extensibility parameter in the sPTT model. Our numerical and analytical results show that, in LAOS, the FENE-P model only shows this universality for large values of $L^2$ whereas the sPTT model shows this universality for all values of$\epsilon$. In the strongly non-linear region, there is a drastic difference between the responses of the two models, with the FENE-P model exhibiting strong shear stress overshoots which manifest as self-intersecting secondary loops in the viscous Lissajous curves. We quantify the non-linearity exhibited by each constitutive model using the Sequence of Physical Processes framework. Despite the high degree of non-linearity exhibited by the FENE-P model, we also show using fully non-linear 1D simulations that it does not shear band in LAOS within the range of conditions studied.
\end{abstract}


{\bf MSC Codes }  {\it(Optional)} Please enter your MSC Codes here

\clearpage

\section{Introduction}

Viscoelastic fluids are ubiquitous in many industrial sectors including fast moving consumer goods, food, and health-care, among others, and it is of significant importance that we are able to correctly model these flows in a wide range of geometries. The \mbox{Oldroyd-B} model \citep{Oldroyd1950} is given as

\begin{equation}
\boldsymbol{\tau}_p + \lambda \overset{\kern0em\triangledown}{\boldsymbol{\tau}}_p =  2\eta_p\mathsfbi{D} \label{oldroydB}
\end{equation}

\noindent where $\boldsymbol{\tau}_p$ is the polymeric stress, $\lambda$ is the viscoelastic relaxation time, $\eta_p$ is the polymeric viscosity and  $\mathsfbi{D}$ is the rate-of-strain tensor given by $ \mathsfbi{D} = 1/2(\nabla \boldsymbol{u}+ \nabla \boldsymbol{u}^{\mathrm{T}})$. $\overset{\kern0em\triangledown}{\boldsymbol{\tau}}_p$ denotes the upper-convected time derivative of the polymeric stress tensor, which is given as ${\overset{\kern0em\triangledown}{\boldsymbol{\tau}}_p = \Dif \boldsymbol{\tau}_p / \Dif t - \boldsymbol{\tau}_p \bcdot \nabla \boldsymbol{u} - \nabla \boldsymbol{u}^{\text{T}} \bcdot \boldsymbol{\tau}_p}$. For many viscoelastic models, including the \mbox{Oldroyd-B} model, the total stress $\boldsymbol{\sigma}$ appearing in the momentum equation is related to the polymeric stress by $\boldsymbol{\sigma} = \boldsymbol{\tau} - p\mathsfbi{I} = \boldsymbol{\tau}_p + 2\eta_s \mathsfbi{D} - p\mathsfbi{I}$, where $\boldsymbol{\tau}$ is the extra-stress tensor, $p$ is the pressure and $\mathsfbi{I}$ is the identity tensor. The solvent contribution to the stress is expressed as a Newtonian fluid with a viscosity $\eta_s$.

Whilst the simplicity of the Oldroyd-B model makes it particularly useful for solving problems analytically \citep{Rajagopal1995, Ghosh2021, Boyko2022, Zhao2013, Qi2007, Norouzi2018} and testing and validating computational codes \citep{Mompean1997, Duarte2008, Habla2014}, it has a number of well-known short-comings. Likely the most well-known short-coming is that the elasticity has no limit of extensibility. During steady and homogeneous extensional flow, this causes an unphysical singularity in the extensional viscosity as the strain rate is increased \citep{Bird1987}. In transient and homogeneous extensional flow, however, the extensional viscosity grows exponentially in time and the singularity is not present. A vast number of viscoelastic models have since been developed to overcome the problems associated with the \mbox{Oldroyd-B} model. Many of these models are derived from micro-structural theories in order to better capture the underlying physics observed during deformation.  Two such models are the FENE-P (Finitely Extensible Non-linear Elastic with Peterlin closure) \citep{Bird1980} and the sPTT (simplified Phan-Thien-Tanner) \citep{PhanThien1977} models. 

The original FENE model \citep{Warner1972} is derived using kinetic theory for bead-spring dumbbells in which each polymer molecule is assumed to take the form of two beads connected together by a finitely extensible spring.  Therefore, the FENE-P model is most often employed for the modelling of dilute polymer solutions where there is no significant interaction between each polymer molecule. The FENE-P model uses a self-consistent pre-averaging approximation, known as the Peterlin approximation, to close the original FENE model \citep{Bird1980, Keunings1997}. The springs are finitely extensible since the elastic stress increases non-linearly during deformation as the stretching of the spring approaches its prescribed limit.  The sPTT model is derived from a Lodge-Yamamoto type of network theory, where the springs are inter-connected via junction points. It is therefore most applicable for concentrated polymer solutions and melts where there are strong interactions between polymer molecules. Under large deformation, the junctions in the sPTT model can be simultaneously created and destroyed, limiting the build-up of elastic stresses and providing finite extensibility.

Whilst the Oldroyd-B model has a constant shear viscosity in steady and homogeneous shear flow, both the FENE-P and sPTT models are shear-thinning. The first normal stress difference $N_1 = \sigma_{11} - \sigma_{22}$ grows quadratically with shear rate in the Oldroyd-B model for steady and homogeneous shear flow, but in the FENE-P and sPTT models it grows quadratically with shear rate only at low shear rates before exhibiting shear-thinning.  In steady and homogeneous extensional flow, the FENE-P and sPTT models exhibit strain thickening for low strain rates, however a plateau is reached in the extensional viscosity for higher strain rates due to the finite extensibility. The value of the extensional viscosity at the plateau is proportional to $L^2$ ($1/\epsilon$) for the FENE-P (sPTT) model, where $L^2$ and $\epsilon$ represent the respective extensibility parameters in the FENE-P and sPTT models.  

The  FENE-P constitutive model is given in stress tensor form as

\begin{equation}
\boldsymbol{\tau}_p + \lambda \overset{\kern0em\triangledown}{\bigg( \frac{\boldsymbol{\tau}_p}{F(\tau_p)}\bigg)} = 2a\eta_p\mathsfbi{D}\bigg(\frac{1}{F(\tau_p)}\bigg) - a\eta_p \mathsfbi{I} \frac{\Dif}{\Dif t} \bigg(\frac{1}{F(\tau_p)}\bigg) \label{fenep0}
\end{equation}

\noindent where $\tau_p \equiv \mathrm{tr}(\boldsymbol{\tau}_p)$, or equivalently 

\begin{equation}
\begin{split}
\frac{F(\tau_p)}{a} \, \boldsymbol{\tau}_p + \frac{\lambda_1}{a} \  \overset{\kern0em\triangledown}{\boldsymbol{\tau}}_p = 2\eta_p\mathsfbi{D} - F(\tau_p) \bigg[\frac{\lambda}{a} \boldsymbol{\tau}_p + \eta_p\mathsfbi{I}\bigg] \frac{\Dif}{\Dif t} \bigg(\frac{1}{F(\tau_p)}\bigg) \\ \text{where} \quad F(\tau_p) \equiv a + \frac{\displaystyle \lambda}{\displaystyle L^2 \eta_p} \ \mathrm{tr}(\boldsymbol{\tau}_p) \quad \text{and} \quad a \equiv \frac{L^2}{L^2 - 3}
\end{split} 
 \label{fenep1}
\end{equation}

\noindent For steady and homogeneous flows, the substantial derivative term in Equations \eqref{fenep0} and \eqref{fenep1} is equal to zero, and the FENE-P model can be re-written as

\begin{equation}
\frac{F(\tau_p)}{a} \, \boldsymbol{\tau}_p + \frac{\lambda}{a} \  \overset{\kern0em\triangledown}{\boldsymbol{\tau}}_p = 2\eta_p\mathsfbi{D} \quad \text{where} \quad \frac{F(\tau_p)}{a} =  1 + \frac{\lambda}{aL^2 \eta_p} \ \mathrm{tr}(\boldsymbol{\tau}_p)  \label{fenep3}
\end{equation} 

\noindent The sPTT model is given as follows

\begin{equation}
F(\tau_p) \, \boldsymbol{\tau}_p + \lambda \  \overset{\kern0em\triangledown}{\boldsymbol{\tau}}_p = 2\eta_p\mathsfbi{D} \quad \mathrm{where} \quad F(\tau_p) \equiv 1 + \frac{\epsilon \ \lambda}{\eta_p} \ \mathrm{tr}(\boldsymbol{\tau}_p) \label{ptt1}
\end{equation}

\noindent The scalar function $F(\tau_p)$ is then defined on a per-model basis as

\begin{equation}
  F(\tau_p) \equiv \begin{cases}
      a + \frac{\displaystyle \lambda}{\displaystyle L^2\eta_p}\mathrm{tr}(\boldsymbol{\tau}_p) & \text{FENE-P} \\[10pt]
      1 + \frac{\displaystyle\epsilon\lambda}{ \displaystyle\eta_p}\mathrm{tr}(\boldsymbol{\tau}_p) & \text{sPTT}
  \end{cases}
\end{equation}

\noindent The original PTT model employs the Gordon-Schowalter derivative of the polymeric stress $\overset{\kern0em \Box}{\boldsymbol{\tau}}_p = \overset{\kern0em \triangledown}{\boldsymbol{\tau}}_p + \zeta(\boldsymbol{\tau}_p \bcdot \mathsfbi{D} + \mathsfbi{D} \bcdot \boldsymbol{\tau}_p)$ which allows for non-affine transformations between the junction points and the solvent fluid through the slip parameter $\zeta$. The sPTT model refers to the case for the PTT model where $\zeta = 0$ and so $\overset{\kern0em \Box}{\boldsymbol{\tau}}_p = \overset{\kern0em \triangledown}{\boldsymbol{\tau}}_p$.  It should also be noted that the sPTT model (Equation \eqref{ptt1}) uses a linear term for the destruction of the junctions, as does the original PTT model, however there have since been modifications to this where the linear term is replaced by exponential \citep{PhanThien1978} or even generalised \citep{Ferras2019} terms, which are believed to help the model perform better under strong deformations. In this study, we will only use the sPTT model with the linear function (Equation \eqref{ptt1}), we shall always refer to this as the sPTT model. For clarity, we often use the subscripts FP and sPTT to denote the FENE-P and sPTT models respectively.

Upon comparison of Equations \eqref{fenep3} and \eqref{ptt1}, it is observed that with the parameter substitutions $\epsilon = 1/L^2$ and $\lambda_{\mathrm{sPTT}} = \lambda_{\mathrm{FP}} / a$, the FENE-P and sPTT-Linear models become mathematically identical for steady and homogeneous flows. The equivalence of these two models was first noted in the study of \citet{Cruz2005}, who derived analytical solutions for fully-developed pipe and channel flows with the FENE-P and sPTT models.  \cite{Latreche2021} also established analytical solutions for steady, fully-developed, flows of the FENE-P and sPTT models in flat and circular ducts using the aforementioned substitution of parameter values. \cite{Davoodi2022} then investigated the FENE-P and sPTT models for a range of steady and homogeneous flows, as well as unsteady and in-homogeneous flows. Due to the presence of the Lagrangian derivative term in the stress tensor form of the FENE-P model, significant differences were observed between the FENE-P and sPTT responses for the transient flows. Notably, the FENE-P model produced pronounced shear stress overshoots in start-up shear flow, and, during start-up extensional flow, the extensional viscosity grew much more sharply in time for the FENE-P model response than for the sPTT model response. One of the geometries studied by \cite{Davoodi2022} was the cross-slot. For viscoelastic flows in the cross-slot, the elastic stresses causes a symmetry-breaking instability to occur at a critical $Wi$, which has previously been well studied and characterised \citep{Afonso2010, Cruz2014, Davoodi2019, Davoodi2021, Haward2012, Poole2007,Rocha2009, Xi2009}. \cite{Davoodi2022} observed that that the critical value of $Wi$ for the onset of the asymmetry is lower for the FENE-P model than for the sPTT model when relatively low (high) values of $L^2 (\epsilon)$ are used for the FENE-P (sPTT) model, again highlighting that the complex nature of the flow causes a discrepancy between the model responses even though the flow is Eulerian steady. Many industrial processes and flows involve complex geometries which might induce Lagrangian unsteadiness, even for an Eulerian steady flow. \citet{Varchanis2022} recently highlighted that even the Oldroyd-B model exhibits complex rheological behaviour in Lagrangian transient flows, which has significant consequences on, for example, the understanding of pressure drop measurements across a contraction. It is therefore of significant importance to compare and understand how these non-linear models behave in transient flows. 

An ideal way of probing the transient non-linear response of viscoelastic materials and models, and in particular classifying complex fluids \citep{Hyun2002}, is with Large Amplitude Oscillatory Shear (LAOS), which has become a widely used technique for characterising non-linear viscoelasticity experimentally \citep{Hyun2011, Szopinski2016,  Leblanc2008, Sun2011}, theoretically \citep{Kate2012, Bae2017, Khair2016, Kammer2020}, and numerically \citep{Ewoldt2010, Cordasco2016, DAvino2013}. In Small Amplitude Oscillatory Shear (SAOS) the shear stress response of a material or constitutive model is approximately linear and given by $\tau_{p,12} = \gamma_0[G^{'} \mathrm{sin}(\omega t) + G^{''} \mathrm{cos}(\omega t)]$ where $\gamma_0$ and $\omega$ are the amplitude and angular frequency of the oscillation respectively. $G^{'}$ and $G^{''}$ represent the storage and loss moduli respectively. Due to the linearity of the shear stress response, SAOS is one of the most popular techniques for extracting information regarding linear viscoelasticity. For example, $\lambda$ is very often estimated as the inverse of the frequency at which $G^{'}$ and $G^{''}$ cross over in a frequency sweep. However, as $\gamma_0$ increases, flow-induced micro-structural changes take place during the oscillation \citep{Gilbert2016}, and the periodic response of the material (or constitutive model) deviates from linearity. This behaviour can then be interpreted in terms of higher-order harmonics in the shear-stress waveform.  Therefore,  in LAOS,  the stress response cannot be accurately reconstructed using a single mode of $G^{'}$ and $G^{''}$.  Multiple frameworks have been developed for quantitative analysis of the non-linear stress response obtained from LAOS, namely these are Fourier Tranform Rheology \citep{Wilhelm1998}, Stress Decomposition \citep{Cho2005} with Chebyshev analysis \citep{Ewoldt2008}, and a Sequence of Physical Processes \citep{Rogers2011}. LAOS is considered to be especially useful for the purpose of fitting constitutive models to experimental data \citep{Bae2015}. \citet{Calin2010} used LAOS to fit the spectrum of the tensorial mobility parameter of the Giesekus model \citep{Giesekus1982} to experimental data using an iterative numerical solution. \citet{Kate2012} then derived an asymptotic solution for the Giesekus model in oscillatory shear, which they use to easily fit the tensorial mobility parameter to experimental data obtained in the Medium Amplitude Oscillatory Shear (MAOS) regime, where the asymptotic solution is valid. Asymptotic solutions in oscillatory shear have also been derived for the Pom-Pom model \citep{Hoyle2014}, the co-rotational Maxwell Model \citep{Giacomin2015}, and the White-Metzner model \citep{Merger2016}, among others. \cite{Hyun2007} compared the responses of the exponential PTT model, the Giesekus model, and the Pom-Pom model in MAOS, as well as the experimental MAOS response of linear and branched polymers. For perspective of LAOS tests, the reader is referred to the comprehensive reviews by \cite{Hyun2011} and \cite{Kamkar2022}.

For a purely oscillatory shear flow where the strain rate is uniform in space, the strain $\gamma(t)$ and strain rate $\dot{\gamma}(t)$ are given by $\gamma(t) = \gamma_0 \ \mathrm{sin}(\omega t)$ and $\dot{\gamma}(t) = \gamma_0 \omega \ \mathrm{cos}(\omega t)$, respectively. We define here the non-dimensional polymeric stress $\boldsymbol{\tau}_p^* = \boldsymbol{\tau}_p / (\gamma_0 \omega[\eta_p +\eta_s])$, the non-dimensional velocity gradient $(\nabla \boldsymbol{u})^* = \nabla \boldsymbol{u} / (\gamma_0\omega)$,  and the non-dimensional time ${t^* = t\omega}$. We also define the Weissenberg number as $Wi = \lambda \gamma_0 \omega$ and the Deborah number as $De = \lambda \omega$. As usual, we define the dimensionless parameter $\beta$ as the ratio of the solvent viscosity to the total viscosity (polymeric viscosity plus solvent viscosity), such that $\beta = \eta_s / (\eta_s + \eta_p)$.  Using these definitions to non-dimensionalise the FENE-P (Equation \eqref{fenep1}) and sPTT (Equation \eqref{ptt1}) models,  and dropping the asterisks upon non-dimensionalisation, we have respectively

\begin{multline}
\frac{F(\tau_p)}{a} \boldsymbol{\tau}_p + \frac{De}{a} \ \frac{\partial}{\partial t}\boldsymbol{\tau}_p - \frac{Wi}{a}(\boldsymbol{\tau}_p \bcdot \nabla \boldsymbol{u} + \nabla \boldsymbol{u}^{\text{T}} \bcdot \boldsymbol{\tau}_p - \boldsymbol{u} \bcdot \nabla \boldsymbol{\tau}_p) = 2(1-\beta)\mathsfbi{D} \\ - F(\tau_p) \bigg[\frac{Wi}{a}\boldsymbol{\tau}_p + (1-\beta)\mathsfbi{I} \bigg]\bigg(\frac{De}{Wi} \frac{\partial}{\partial t} \bigg(\frac{1}{F(\tau_p)}\bigg) + \boldsymbol{u}\bcdot \nabla \bigg(\frac{1}{F(\tau_p)}\bigg) \bigg)
\label{fenePNonDim}
\end{multline}

\begin{equation}
F(\tau_p) \, \boldsymbol{\tau}_p + De \ \frac{\partial}{\partial t}\boldsymbol{\tau}_p - Wi(\boldsymbol{\tau}_p \bcdot \nabla \boldsymbol{ u} + \nabla \boldsymbol{u}^{\text{T}} \bcdot \boldsymbol{\tau}_p- \boldsymbol{u} \bcdot \nabla \boldsymbol{\tau}_p) = 2(1-\beta)\mathsfbi{D} \label{pttNonDim}
\end{equation}

\noindent where 

\begin{equation}
  F(\tau_p) \equiv \begin{cases}
      a + \frac{\displaystyle Wi}{\displaystyle L^2(1-\beta)}\mathrm{tr}(\boldsymbol{\tau}_p) & \text{FENE-P} \\[10pt]
      1 + \frac{\displaystyle\epsilon Wi}{ \displaystyle(1-\beta)}\mathrm{tr}(\boldsymbol{\tau}_p) & \text{sPTT}
  \end{cases}
\end{equation}

\noindent For all of the models discussed in this study, including the Oldroyd-B model, the extra-stress tensor is given as $\boldsymbol{\tau} = \boldsymbol{\tau}_p + 2\beta\mathsfbi{D}$. In dimensionless form, it is upon substitution of $\epsilon = 1/L^2$ and $Wi_{\mathrm{sPTT}} = Wi_{\mathrm{FP}} / a$ that the FENE-P and sPTT models become mathematically identical for steady ($De = 0$) and homogeneous ($ \boldsymbol{u} \bcdot \nabla \boldsymbol{\tau}_p = 0 $) flows.

With regards to the definitions of $De$ and $Wi$ for LAOS, \citet{Kamani2023} recently highlighted that using a time-independent value of $De$ might seem unphysical in some cases since the true ratio of the flow time-scale (the inverse of the oscillation frequency) and the material time-scale may not necessarily be constant during the oscillation in certain conditions. This requires that $De$, according to its physical interpretation, be a time-dependent value rather than constant value. Whilst the FENE-P and sPTT models have constant relaxation times, and constant values of $De$ and $Wi$ naturally appear from the non-dimensionalisation of the equations for LAOS, the White-Metzner model,  on the contrary, contains a strain-rate dependent relaxation time. In this case, transient values of $De$ and $Wi$ would naturally appear from the equations. In the FENE-P and sPTT models, one might also think of an "effective" relaxation time based on $\lambda$ and $F(\tau_p)$. Therefore, we note that, depending on the model or material in question, one may start to question the correct choice of definition for $De$ and $Wi$ in LAOS, and whether they should be indeed constant or not during an oscillation. However, this is outside of the scope of the current study, and we only use the time-independent values for $De$ and $Wi$ defined previously.

In the limit that $De \ (De / a) \rightarrow 0$ and $Wi \ (Wi / a) \rightarrow 0$, the sPTT (FENE-P) models, as well as the Oldroyd-B model, reduce to that of a Newtonian fluid. Note that $\lim_{Wi \rightarrow 0} (1/F(\tau_p)_{\mathrm{FP}}) = 1/a$ and $\Dif (1/a) /\Dif t = 0$. In the limit that $De \rightarrow 0$,  the response of each model reduces to its respective steady state response.  In the case that $F(\tau_p)_{\mathrm{sPTT}} \rightarrow 1$ for the sPTT model, or $F(\tau_p)_{\mathrm{FP}}/a \rightarrow 1$ for the FENE-P model, the Oldroyd-B model is obtained. Note that for the FENE-P model $F(\tau_p)_{\mathrm{FP}}/a \rightarrow 1$ is equivalent to $F(\tau_p)_{\mathrm{FP}} \rightarrow a$ and therefore $\partial /\partial t \, (1/F(\tau_p)_{\mathrm{FP}}) \rightarrow 0$ and $\boldsymbol{u} \bcdot \nabla (1/F(\tau_p)_{\mathrm{FP}}) \rightarrow 0$ and so the last term on the right hand side of Equation \eqref{fenePNonDim} vanishes in this limit.

Viscoelastic constitutive models can also be written for the conformation tensor, $\mathsfbi{A}$. For dumbbell models such as the FENE-P model, $\mathsfbi{A}$ can be given as $\mathsfbi{A} = \langle \boldsymbol{Q}\boldsymbol{Q} \rangle / Q_{eq}^2$ where $\boldsymbol{Q}$ is the end-to-end vector of an individual dumbbell (the angled brackets represents the ensemble average) and $Q_{eq}^2$ is the square of the magnitude at equilibrium given as $Q_{eq}^2 = \langle \boldsymbol{Q} \bcdot \boldsymbol{Q} \rangle_{eq} / 3$ \citep{Alves2021}.  In general, for other viscoelastic models, $\boldsymbol{Q}$ might represent the end-to-end vector of polymer chains or sub-chains \citep{Hoyle2016}, rather than the dumbbell vector specifically. The dimensionless FENE-P model is given in conformation tensor form as follows

\begin{multline}
De \frac{\partial}{\partial t} \mathsfbi{A} - Wi (\mathsfbi{A} \bcdot \nabla \boldsymbol{u} + \nabla \boldsymbol{u}^{\mathrm{T}} \bcdot \mathsfbi{A} - \boldsymbol{u} \bcdot \nabla \mathsfbi{A}) = - (F(A) \mathsfbi{A} - a \mathsfbi{I}) \\ \text{where} \quad F(A) = \frac{L^2}{L^2 - \mathrm{tr}(\mathsfbi{A})} \label{feneConf1}
\end{multline}

\noindent which can also be re-written as

\begin{multline}
\frac{De}{a} \frac{\partial}{\partial t} \mathsfbi{A} - \frac{Wi}{a} (\mathsfbi{A} \cdot \nabla \boldsymbol{u} + \nabla \boldsymbol{u}^{\mathrm{T}} \cdot \mathsfbi{A} - \boldsymbol{u} \bcdot \nabla \mathsfbi{A}) = - \bigg( \frac{F(A)}{a} \mathsfbi{A} -  \mathsfbi{I} \bigg)  \\ \text{where} \quad \frac{F(A)}{a} = \frac{L^2 - 3}{L^2 - \mathrm{tr}(\mathsfbi{A})} \label{feneConf2}
\end{multline}

\noindent The sPTT model is given in conformation tensor form as

\begin{multline}
De \frac{\partial}{\partial t} \mathsfbi{A} - Wi (\mathsfbi{A} \bcdot \nabla \boldsymbol{u} + \nabla \boldsymbol{u}^{\mathrm{T}} \bcdot \mathsfbi{A} - \boldsymbol{u} \bcdot \nabla \mathsfbi{A}) = - F(A)(\mathsfbi{A} - \mathsfbi{I}) \\ \text{where} \quad F(A) = [1+\epsilon(\mathrm{tr}(\mathsfbi{A})-3)] \label{pttConf1}
\end{multline}

\noindent Note that $A \equiv \mathrm{tr}(\mathsfbi{A})$. $\boldsymbol{\tau}_p$ is then recovered from the solutions of Equations \eqref{feneConf1} to \eqref{pttConf1} as follows

\begin{equation}
\boldsymbol{\tau}_p = 
\begin{cases}
\frac{\displaystyle a(1-\beta)}{\displaystyle Wi} \bigg( \frac{\displaystyle F(A)}{\displaystyle a}\mathsfbi{A} - \mathsfbi{I} \bigg)& \mathrm{FENE-P} \\[10pt]
\frac{\displaystyle (1-\beta)}{\displaystyle Wi}(\mathsfbi{A} - \mathsfbi{I}) & \mathrm{sPTT}
\end{cases}
\label{cramersForm}
\end{equation}

\noindent $F(A)$ is therefore also be defined on a per-model basis as 

\begin{equation}
    F(A) = \begin{cases}
        \frac{\displaystyle L^2}{\displaystyle L^2 - \mathrm{tr}(\mathsfbi{A})} & \text{FENE-P} \\[10pt]
        1+\epsilon(\mathrm{tr}(\mathsfbi{A}-3)) & \text{sPTT} \\
    \end{cases}
\label{extensibilityFunctionConf}
\end{equation}
  
\noindent As highlighted by \citet{Davoodi2022}, the evolution equation for $\mathsfbi{A}$ in network theory models follows a general form given by

\begin{equation}
De \frac{\partial}{\partial t} \mathsfbi{A} - Wi (\mathsfbi{A} \bcdot \nabla \boldsymbol{u} + \nabla \boldsymbol{u}^{\mathrm{T}} \bcdot \mathsfbi{A} - \boldsymbol{u} \bcdot \nabla \mathsfbi{A}) = - (D(A)\mathsfbi{A}-C(A)\mathsfbi{I}) \label{pttrates}
\end{equation}

\noindent with 

\begin{equation}
\boldsymbol{\tau}_p = \frac{(1-\beta)}{Wi}(\mathsfbi{A}-\mathsfbi{I})
\end{equation}

\noindent where $D(A)$ and $C(A)$ represent, respectively, the rates of destruction and creation of micro-structures.  For the sPTT model,  $D(A) = C(A) = F(A)_{\mathrm{sPTT}}$. It is therefore observed, given Equation \eqref{feneConf2}, that the FENE-P model might be considered as a type of network model in which, under large deformations, the rate of destruction of micro-structures is faster than the rate of creation of micro-structures. Network models with faster destruction rates than creation rates are expected, and have been observed, to exhibit large amounts of elastic recoil \citep{Davoodi2022}.

Generally, the LAOS response of a viscoelastic material or model can be classified as one of four archetypes: \rom{1} - strain thinning, \rom{2} - strain hardening (or strain thickening), \rom{3} - weak strain overshoot, and \rom{4} - strong strain overshoot \citep{Hyun2002}. Physically, each classification is believed to correspond to a particular type of underlying micro-structural interaction. \cite{Sim2003} investigated numerically the LAOS response of a general network model and found that the classification of the LAOS response varied depending on the choice of the parameters defining the rates of creation and destruction of junctions. \cite{Townsend2018} simulated the LAOS response of a Newtonian solvent with suspended dumbbells, where the dumbbells are implemented in Stokesian Dynamics, thus, forming a viscoelastic medium.  They compare the simulation results for FENE dumbbells with the LAOS response of the FENE-P constitutive model, which they obtain numerically. For $De = 0.56$, they observe that the FENE-P constitutive model shows purely strain thinning behaviour whereas the FENE dumbbell simulations show some weak strain overshoot for the elastic or storage modulus $G^{'}$. With increased oscillation frequency, the FENE-P response changed to a type \rom{3} response where $G^{''}$ exhibited a strain overshoot, whereas the FENE dumbbell simulations showed a type \rom{1} response. Recently, some authors have also used a micro-macro approach for modelling of standard FENE dumbbells and FENE-type networks in LAOS using a technique known as the Brownian Configuration Field Method (BCFM) \citep{Lopez2019, Vargas2023}. In the FENE-type network model response, self-intersecting secondary loops were observed in the viscous Lissajous curves when the rate of destruction of micro-structures was faster than the rate of creation of micro-structures.  As already mentioned, in the context of the PTT model framework (Equation \eqref{pttrates}), this causes the model to appear more similar in form to the FENE-P model and likely leads to more elastic recoil in the transient model response. Self intersecting secondary loops, which will be discussed in more detail later, are known to be related specifically to large amounts of elastic recoil \citep{Ewoldt2010}. \cite{Ng2011} performed LAOS experiments with a Gluten dough, which they then modelled with a transient network model. The rate of destruction of the junctions was modelled by a term which is essentially a blend between $F(A)_{\mathrm{sPTT}}$ at low stretching and $F(A)_{\mathrm{FP}}$ at high stretching. They also include $F(A)_{\mathrm{FP}}$ in the $\boldsymbol{\tau}_p$-$\mathsfbi{A}$ relationship and so the constitutive model represents a FENE-type network model. The model was able to at least qualitatively predict the experimental Lissajous curves, however, the authors note that the stress overshoots were grossly over-predicted. They attribute this to the functional form of the spring function (essentially $F(A)_{\mathrm{FP}}$) and they introduce a modified function, which diverges to infinity before the FENE limit is approached to empirically temper the magnitude of the stress overshoots. \cite{Keunings1997} shows that these transient stress overshoots in the FENE-P model arise from the pre-averaging Peterlin approximation used to close the original FENE model. The response of the sPTT model in LAOS was obtained and studied recently by \citet{Ofei2020}, who showed that with increasing $De$ and $Wi$ the sPTT response clearly deviates away from the linear UCM/Oldroyd-B response. However, in this study, no quantitative analysis of the generated waveforms was conducted.

Despite the fact that there has been significant recent interest in the similarities and differences between the FENE-P and sPTT model responses in steady and unsteady (or complex flows), and that the LAOS responses of these models have been studied independently, there has yet to be an explicit comparison made between the responses of the models in LAOS when the parameters are chosen such that models provide the same steady and homogeneous response. The aim of this study is to compare the responses of the FENE-P and sPTT constitutive models specifically in LAOS and to understand and highlight any differences observed in the responses.

\section{Numerical Methodology}

\subsection{0D modelling}

The majority of the results in this study are obtained by solving constitutive equations assuming that $\mathsfbi{A}$, $\boldsymbol{\tau}_p$, and $\dot{\gamma}$ are uniform in space. We denote this approach to solving the equations as the 0D method. This methodology will now be detailed.

For an ideal oscillatory shear flow, the dimensionless constitutive model can be solved using 

\begin{equation}
\nabla \boldsymbol{u}(t) = 
\begin{bmatrix}
0 & 0 & 0\\
\dot{\gamma}(t) & 0 & 0 \\
0 & 0 & 0
\end{bmatrix} = \begin{bmatrix}
0 & 0 & 0\\
\mathrm{cos}(t) & 0 & 0 \\
0 & 0 & 0
\end{bmatrix}
\end{equation}

\noindent Since the FENE-P model in stress tensor form (Equation \eqref{fenep1}) cannot be easily expressed as a set of ordinary differential equations (ODEs) for an oscillatory shear flow, we solve the models in conformation tensor form.  From this point on, we only work with dimensionless variables and we re-confirm that the asterisks denoting the dimensionless variables have been dropped for brevity. The following system of ODEs are obtained for the time evolution of $\mathsfbi{A}$ according to the FENE-P model

\begin{subequations}
\begin{align}
\begin{split}
\frac{\dif {A}_{11}}{\dif t} & = 2\bigg(\frac{Wi}{De}\bigg){A}_{12} \ \mathrm{cos}(t) \\
 & - \frac{1}{De} \bigg(\bigg[\frac{L^2}{L^2 - ( {A}_{11}+ {A}_{22}+ {A}_{33})}\bigg]{A}_{11} - a  \bigg)
\end{split}
\\[12pt]
\frac{\dif {A}_{12}}{\dif t} & =\bigg(\frac{Wi}{De}\bigg) {A}_{22} \ \mathrm{cos}(t) - \frac{1}{De} \bigg(\bigg[\frac{L^2}{L^2 - ( {A}_{11}+ {A}_{22}+ {A}_{33})}\bigg]A_{12}\bigg)
\\[12pt]
\frac{\dif {A}_{22}}{\dif t} & = - \frac{1}{De} \bigg(\bigg[\frac{L^2}{L^2 - ( {A}_{11}+ {A}_{22}+ {A}_{33})}\bigg]{A}_{22} - a  \bigg)
\\[12pt]
\frac{\dif {A}_{33}}{\dif t} & = - \frac{1}{De} \bigg(\bigg[\frac{L^2}{L^2 - ( {A}_{11}+ {A}_{22}+ {A}_{33})}\bigg]{A}_{33} - a  \bigg)
\end{align}
\label{feneposcillatoryshear}
\end{subequations}

\noindent and according to the sPTT model

\begin{subequations}
\begin{align}
\begin{split}
\frac{\dif {A}_{11}}{\dif t} &= 2 \bigg(\frac{Wi}{De}\bigg) {A}_{12}  \ \mathrm{cos}(t) \\ 
&- \frac{1}{De} (1+\epsilon({A}_{11}+ {A}_{22}+ {A}_{33} -3))({A}_{11} - 1)
\end{split}
\\[12pt]
\begin{split}
\frac{\dif {A}_{12}}{\dif t} &= \bigg(\frac{Wi}{De}\bigg) {A}_{22} \ \mathrm{cos}(t) \\ 
&- \frac{1}{De} (1+\epsilon({A}_{11}+ {A}_{22}+ {A}_{33} -3))({A}_{12})
\end{split}
\\[12pt]
\frac{\dif {A}_{22}}{\dif t} &= - \frac{1}{De} (1+\epsilon({A}_{11}+ {A}_{22}+ {A}_{33} -3))({A}_{22}-1)
\\[12pt]
\frac{\dif {A}_{33}}{\dif t} &= - \frac{1}{De} (1+\epsilon({A}_{11}+ {A}_{22}+ {A}_{33} -3))({A}_{33}-1)
\end{align}
\label{pttoscillatoryshear}
\end{subequations}

\noindent where $\boldsymbol{\tau}_p$ is recovered from $\mathsfbi{A}$ with Equation \eqref{cramersForm}. With an initial condition of $\mathsfbi{A} = \mathsfbi{I}$ (ie. $\boldsymbol{\tau}_p = 0$), $\dif A_{22}/\dif t = \dif A_{33}/\dif t = 0$ at all times for the sPTT model, and so $A_{22}$ and $A_{33}$ remain fixed at unity. However, for the FENE-P model, under large oscillatory deformations $A_{22}$ and $A_{33}$ will become time-dependent and lower than unity, although it is still the case that $A_{22}=A_{33}$. We note here that, for the FENE-P model, $A_{22} = A_{33} = a/F(A)_{\mathrm{FP}}$ in steady-state conditions, and so $\tau_{p,22} = \tau_{p,33} = 0$ according to Equation \eqref{cramersForm}. $A_{22}$ and $A_{33}$ therefore deviate from unity in the FENE-P response in steady shear, even though the corresponding stresses are still zero. In LAOS, however, the unsteadiness of the flow implies that $\tau_{p,22}$ and $\tau_{p,33}$ are also nonzero.

For all 0D simulations, we omit the solvent contribution to the stress by setting $\beta = 0$ so that we only study the response of the viscoelastic constitutive model itself. Therefore, the Oldroyd-B model is here-on-in denoted as the Upper-Convected Maxwell (UCM) model.  For the FENE-P and sPTT models, we performed simulations for five values of $L^2~ =~1/\epsilon~=~[3.1, 5,  10, 100, 1000]$. Equations \eqref{feneposcillatoryshear} and \eqref{pttoscillatoryshear} were solved in MATLAB using the \textit{ode15s} solver, which uses built-in adaptive time-stepping. The simulations were run until a steady-periodic state was reached. For the results in Section \ref{results}, data is only plotted for the final oscillation when the system is steady-periodic (i.e. the limit cycle).

\subsection{1D modelling}

We also use a 1D modelling approach by solving both the momentum equation and the constitutive model in a 1D gap of fluid.  This is more representative of an actual shear rheometry experiment in which the velocity gradient can become non-uniform in the gap due to phenomena such as shear banding. To solve the equations in the 1D approach, we use the Method Of Lines (MOL) technique, in which spacial derivatives of flow variables are discretised (in this case using finite difference approximations).

The top and bottom walls of the gap are parallel to the $x$-direction. The first and second order spacial derivatives of a scalar variable $\phi$ are discretised using a fourth-order finite difference scheme, respectively, as

\begin{equation}
\bigg(\frac{\partial \phi}{\partial y} \bigg)_i= 
\begin{cases}
\frac{\displaystyle (-3\phi_{i-1}-10\phi_i+18\phi_{i+1}-6\phi_{i+2}+\phi_{i+3})}{\displaystyle 12\Delta} & i = 2 \\[5pt] \frac{\displaystyle (\phi_{i-2}-8\phi_{i-1}+8\phi_{i+1}-\phi_{i+2})}{\displaystyle 12\Delta} & 2 < i < (N_y-1) \\[5pt] \frac{\displaystyle (-\phi_{i-3}+6\phi_{i-2}-18\phi_{i-1}+10\phi_i+3\phi_{i+1})}{\displaystyle 12\Delta} & i = N_y - 1
\end{cases}
\label{firstderivative}
\end{equation}

\begin{equation}
\bigg( \frac{\partial^2 \phi}{\partial y^2} \bigg)_i =
\begin{cases}
\frac{\displaystyle (11\phi_{i-1}-20\phi_i+6\phi_{i+1}+4\phi_{i+2}-\phi_{i+3})}{\displaystyle 12\Delta^2} & i = 2 \\[5pt] \frac{\displaystyle (-\phi_{i-2}+16\phi_{i-1}-30\phi_i+16\phi_{i+1}-\phi_{i+2})}{\displaystyle 12\Delta^2} & 2 < i < (N_y-1) \\[5pt] \frac{\displaystyle (-\phi_{i-3}+4\phi_{i-2}+6\phi_{i-1}-20\phi_i+11\phi_{i+1})}{\displaystyle 12\Delta^2} & i = N_y - 1
\end{cases}
\label{secondderivative}
\end{equation}

\noindent where the index $i$ denotes the node number in a uniformly discretised domain with $N_y$ elements. $\Delta$ is the distance between neighbouring cells, given as $\Delta = y_i - y_{i-1}$.  The (dimensional) velocity at the top boundary $u_{N_y}(t)$ is varied according to ${u_{N_y} = \gamma_0\omega H \mathrm{cos}(\omega t)}$ where $H$ is the gap height. $\gamma_0\omega$ represents the strain rate amplitude.  For non-dimensionalisation, $H$ is used for the length scale,  $\gamma_0\omega H$ is used for the velocity scale, and the time is still non-dimensionalised with $\omega$. $De$ and $Wi$ are then defined as they are for the 0D approach.

Assuming that the only nonzero velocity component is in the $x$-direction, and the flow is uniform in the $x$-direction, the resulting system of Partial Differential Equations (PDEs) to be solved can be expressed in dimensionless form as 

\begin{subequations}
\begin{gather}
\textit{Re} \bigg(\frac{\partial u}{\partial t} \bigg)_i = \bigg(\frac{\partial \tau_{p, 12}}{\partial y}\bigg)_i + \beta \bigg(\frac{\partial^2 u}{\partial y^2} \bigg)_i  \label{momentum}\\ \bigg(\frac{\partial}{\partial t} A_{11} \bigg)_i = (\mathcal{F}_{11})_i \\ \bigg(\frac{\partial}{\partial t} A_{12} \bigg)_i = (\mathcal{F}_{12})_i \\ \bigg(\frac{\partial}{\partial t} A_{22} \bigg)_i = (\mathcal{F}_{22})_i \\ \bigg(\frac{\partial}{\partial t} A_{33} \bigg)_i = (\mathcal{F}_{33})_i 
\end{gather}
\end{subequations}

\noindent where the Reynolds number $\textit{Re}$ is defined as $\textit{Re} = \rho H^2\gamma_0\omega/(\eta_s + \eta_p)$, and the tensor $\mathbfcal{F}$ is the right hand side of the constitutive model when expressed for the time derivative in conformation tensor form. First and second order derivatives are replaced with the discretised forms in Equations \eqref{firstderivative} and \eqref{secondderivative}, which turns the system of PDEs into a system of ODEs. For the momentum equation,$\tau_{p,12}$ is computed from $\mathsfbi{A}$ using Equation \eqref{cramersForm}, and then its gradient is discretised with Equation \eqref{firstderivative}. For the 1D MOL modelling, we could not totally omit the solvent contribution to the stress due to stability issues. We therefore used $\beta = 1/1001$, which was found to be large enough to stabilise the simulations but small enough so that the results were essentially insensitive to the value of $\beta$ in the range of $De$ and $Wi$ investigated. This is shown in the Supplementary Material. We also enforce true creeping flow so that inertia is neglected (i.e. the left hand side of Equation \eqref{momentum} is zero).

For the spacial resolution, we used $N_y = 128$ 
, which proved sufficiently accurate to ensure that the results were independent of $N_y$. This is also shown in the Supplementary Material. At the bottom wall the velocity was fixed at zero. The components $\mathsfbi{A}$ were linearly extrapolated at the top and bottom boundaries. Simulations were initiated with $u = 0$ and $\mathsfbi{A} = \mathsfbi{I}$.  We integrated the resulting system of equations in MATLAB using the adaptive-step ODE solver \textit{ode15s}, which can solve systems of DAEs using the Mass Matrix approach. We simulated the flow until the limit cycle was reached. Again, only data for the limit cycle is presented in Section \ref{results}. 

\section{Results and discussion}\label{results}

For all of the results except those presented in Section \ref{results5}, the 0D method is used (with $\beta = 0$) to obtain the solutions. In Sections \ref{results1} and \ref{results2}, we investigate the model responses in LAOS with the parameter substitutions $\epsilon = 1/L^2$,  $Wi_{\mathrm{UCM}} = Wi_{\mathrm{sPTT}} = Wi_{\mathrm{FP}}/a$, and $De_{\mathrm{UCM}} = De_{\mathrm{sPTT}} = De_{\mathrm{FP}}/a$. We present the results for various values of $De/a \ (De)$ and $Wi/a \ (Wi)$ for the FENE-P (sPTT or UCM) model. In Section \ref{results3}, we investigate the responses of "toy" models to help explain the observations from Sections \ref{results1} and \ref{results2}. In Section \ref{results4}, we analyse the model responses using the Sequence of Physical Processes methodology. Finally, in Section \ref{results5} we use the 1D MOL modelling to assess whether the constitutive models are prone to shear banding in LAOS. Throughout much of this study, we present the results by showing the Lissajous-Bowditch curves. For the shear stress, these are displayed as plots of $\tau_{p,12}$ \textit{versus} $\gamma$ and plots of $\tau_{p,12}$ \textit{versus} $\dot{\gamma}$. The former is referred to as the elastic projection and the latter is referred to as the viscous projection. The resulting patterns are often presented in Pipkin (or $De$ and $Wi$) space. For a more detailed overview of Lissajous-Bowditch plots, the reader is referred to the review by \citet{Hyun2011}.

\subsection{Scaling of the Lissajous curves}\label{results1}

\begin{figure}
\centering
\large 
$\tau_{p,12} \ $ \textit{vs} $\ \dot{\gamma}$
\includegraphics[width=0.9\textwidth]{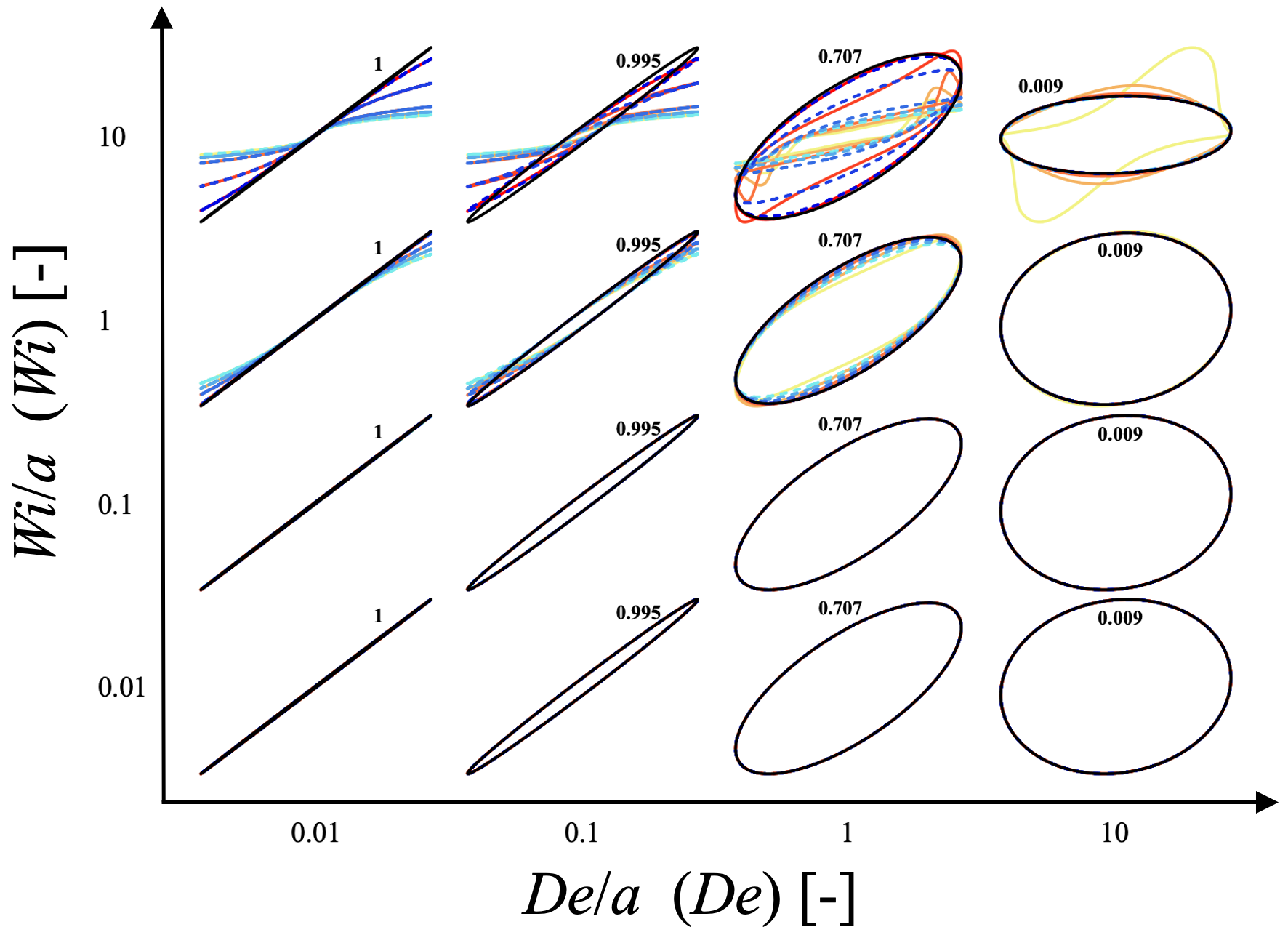}

\hspace{15mm}\includegraphics[width=0.85\textwidth]{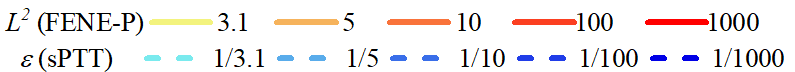}
\caption{Viscous Lissajous-Bowditch plots in $De/a \ (De) -Wi/a \ (Wi)$ space for the FENE-P (sPTT) model.  Black curves represent the UCM response. Black numbers in each plot represent the maximum value of $\tau_{p,12}$ in the UCM response,  since the $y$-axis is scaled differently in each plot.}
\label{viscousLBWi}
\end{figure}

\begin{figure}
\centering
\large 
$\tau_{p,12} \ $ \textit{vs} $\ \gamma $
\includegraphics[width=0.9\textwidth]{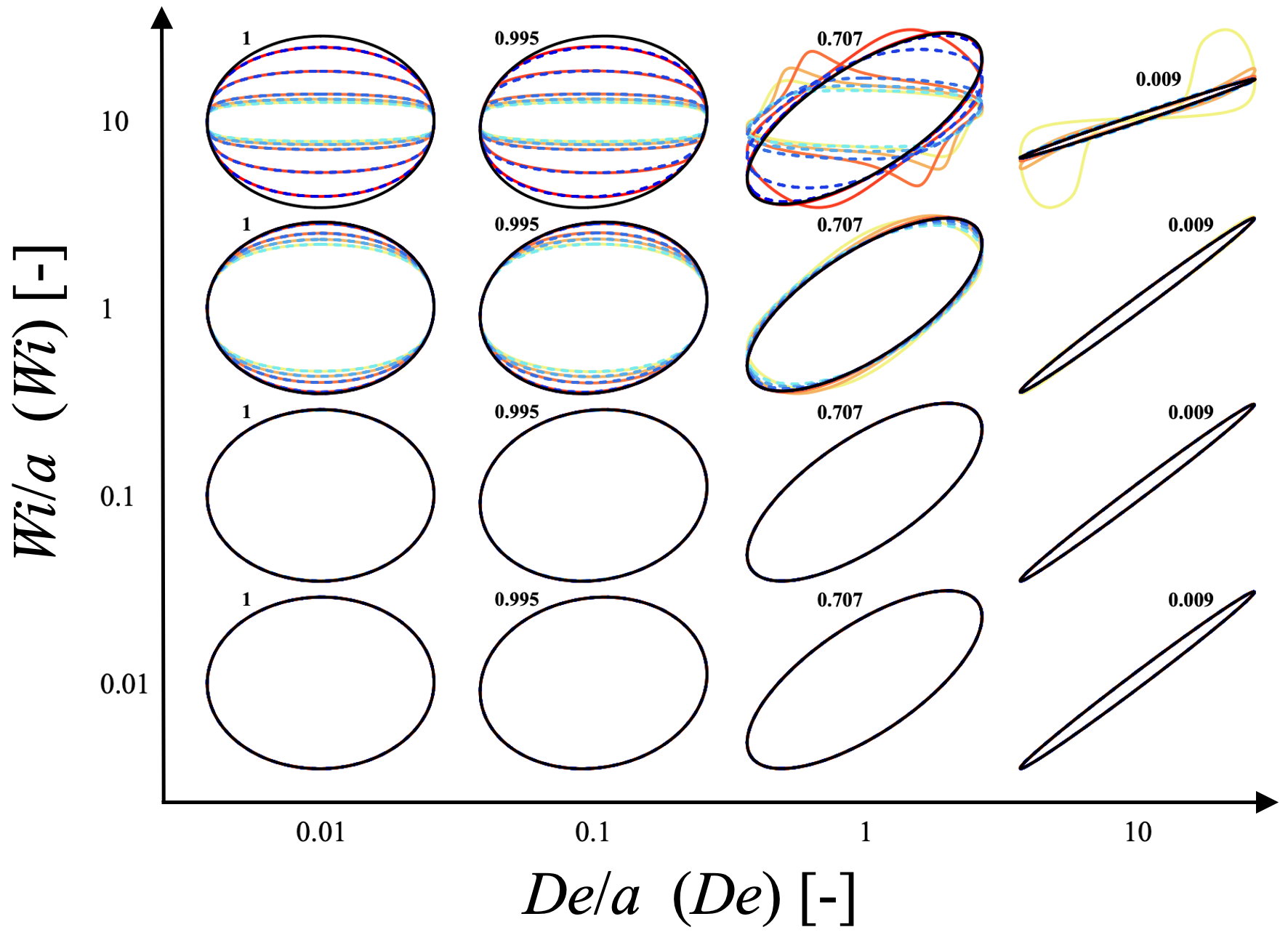}

\hspace{15mm}\includegraphics[width=0.85\textwidth]{Figures/LegendLogLog.PNG}
\caption{Elastic Lissajous-Bowditch plots in $De/a \ (De) -Wi/a \ (Wi)$ space for the FENE-P (sPTT) model.  Black curves represent the UCM response. Black numbers in each plot represent the maximum value of $\tau_{p,12}$ in the UCM response, since the $y$-axis is scaled differently in each plot.}
\label{elasticLBWi}
\end{figure}

Figures \ref{viscousLBWi} and \ref{elasticLBWi} show, respectively, the viscous and elastic projections of the Lissajous-Bowditch plots in the $De/a \ (De) -Wi/a \ (Wi)$ space for the UCM (black solid lines), FENE-P (yellow to red solid lines) and sPTT (cyan to blue dashed lines) models with varying values of $L^2 = 1/\epsilon$ for the FENE-P and sPTT models. The FENE-P and sPTT models deviate from the UCM model at high $Wi/a \ (Wi)$ and particularly at low (high) values of $L^2 \, (\epsilon)$. For the majority of the plots (except the four in the upper-right quadrant), the responses of the FENE-P and sPTT models to uniform oscillation are practically identical. However, for the upper-right quadrant the responses of the FENE-P and sPTT models differ from each other, particularly for the lower (higher) values of $L^2 \, (\epsilon)$, which matches the observations of \cite{Davoodi2022} for start-up shear flow.

In order to investigate how the responses of the models in LAOS scale with the model parameters, we look first at the way in which the model responses scale under Steady Simple Shear Flow (SSSF). For SSSF, the non-dimensional FENE-P model (Equation~\eqref{fenePNonDim}) yields the following solution for the (polymeric) shear stress

\begin{equation}
\frac{2}{(1-\beta)^2}\bigg(\frac{{Wi}}{aL}\bigg)^2  \, {\tau_{p,12}}^3 + \tau_{p, 12} = (1-\beta) \dot{\gamma} \label{fenepanalytical1}
\end{equation}

\noindent For constant $\beta$, the solution for $\tau_{p,12}$ in SSSF then evidently depends on the parameter $Wi / (aL)$. \citet{Oliveira1999} discussed this scaling, but for the sPTT model response instead, when they derived analytical solutions for fully-developed channel flow and showed that the solution scaled with the parameter $Wi \sqrt{\epsilon}$.  With the aforementioned substitution of model parameters, the scaling parameters of both the FENE-P and sPTT models are the same for SSSF. The existence of the scaling parameter $Wi/(aL) \ (Wi\sqrt{\epsilon})$ for the FENE-P (sPTT) models in SSSF was also shown by \cite{Oliveira2004} and \cite{Latreche2021}. A recent study by \citet{Yamani2022} showed, analytically, that the SSSF response of the FENE-P model scales with the dimensionless parameter $Wi / L$, which differs from the parameter used in this study, $Wi / (aL)$. However,  in the version of FENE-P model used in the study of \citet{Yamani2022} the value of $a$ was assumed to be unity. The different versions of the FENE-P model which appear in the literature have been presented and discussed by \citet{Alves2021} and \citet{Davoodi2022}.

\begin{figure}
\centering
\large
$\tau_{p,12} \ $ \textit{vs} $\ \dot{\gamma} $
\includegraphics[width=0.9\textwidth]{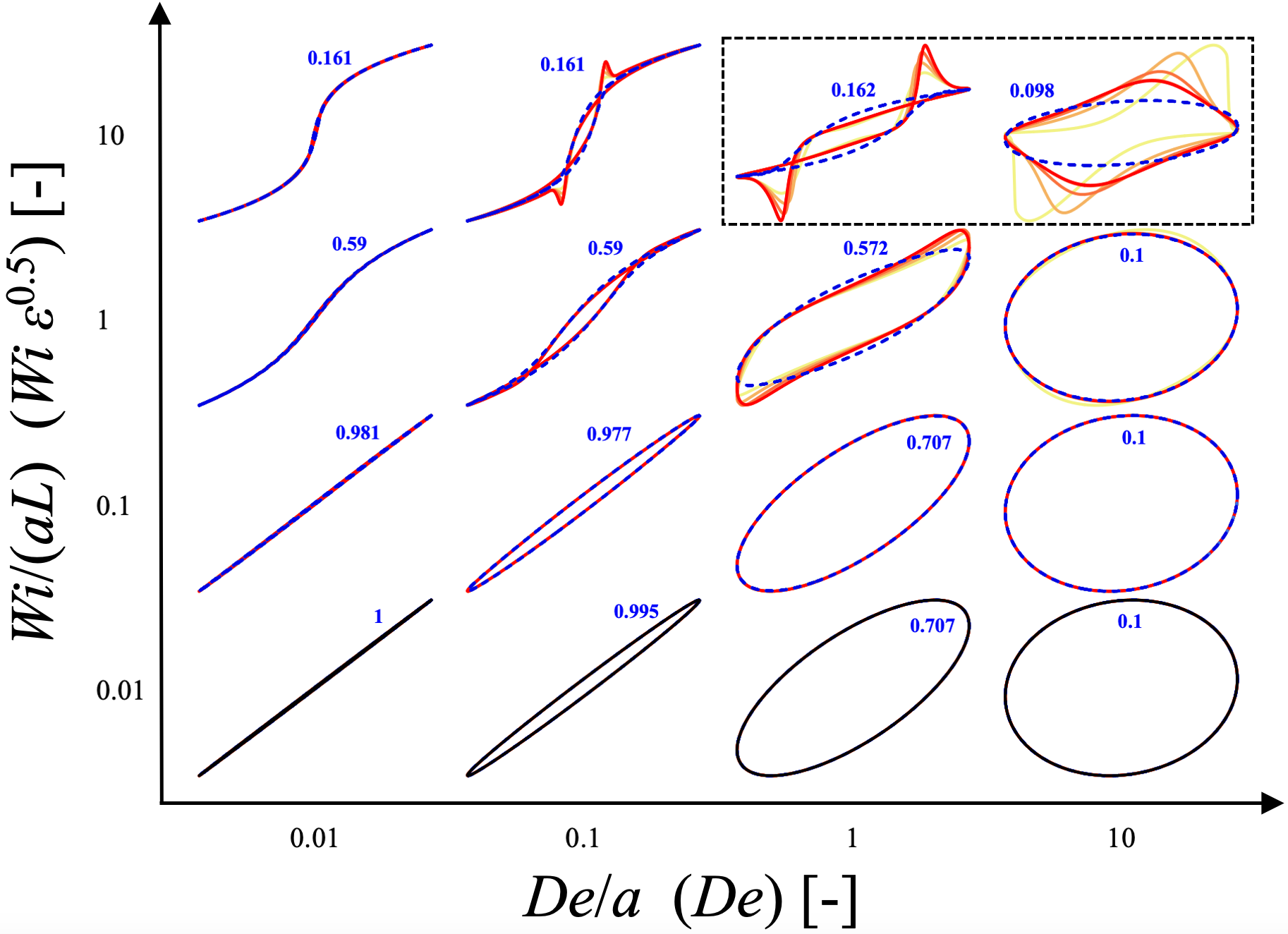}

\hspace{15mm}\includegraphics[width=0.85\textwidth]{Figures/LegendLogLog.PNG}
\caption{Viscous Lissajous-Bowditch plots in $Wi/(aL) \ (Wi\sqrt{\epsilon}) - De/a \ (De)$ space for the FENE-P (sPTT) model. Blue numbers in each plot represent the maximum value of $\tau_{p,12}$ in the sPTT response since the $y$-axis is scaled differently in each plot.  Plots in the black dashed box are shown at a larger scale below in Figure \ref{zoomedLissajous}.}
\label{viscousLBWi_La}
\end{figure}

\begin{figure}
\centering
\large 
$\tau_{p,12} \ $ \textit{vs} $\ \gamma $
\includegraphics[width=0.9\textwidth]{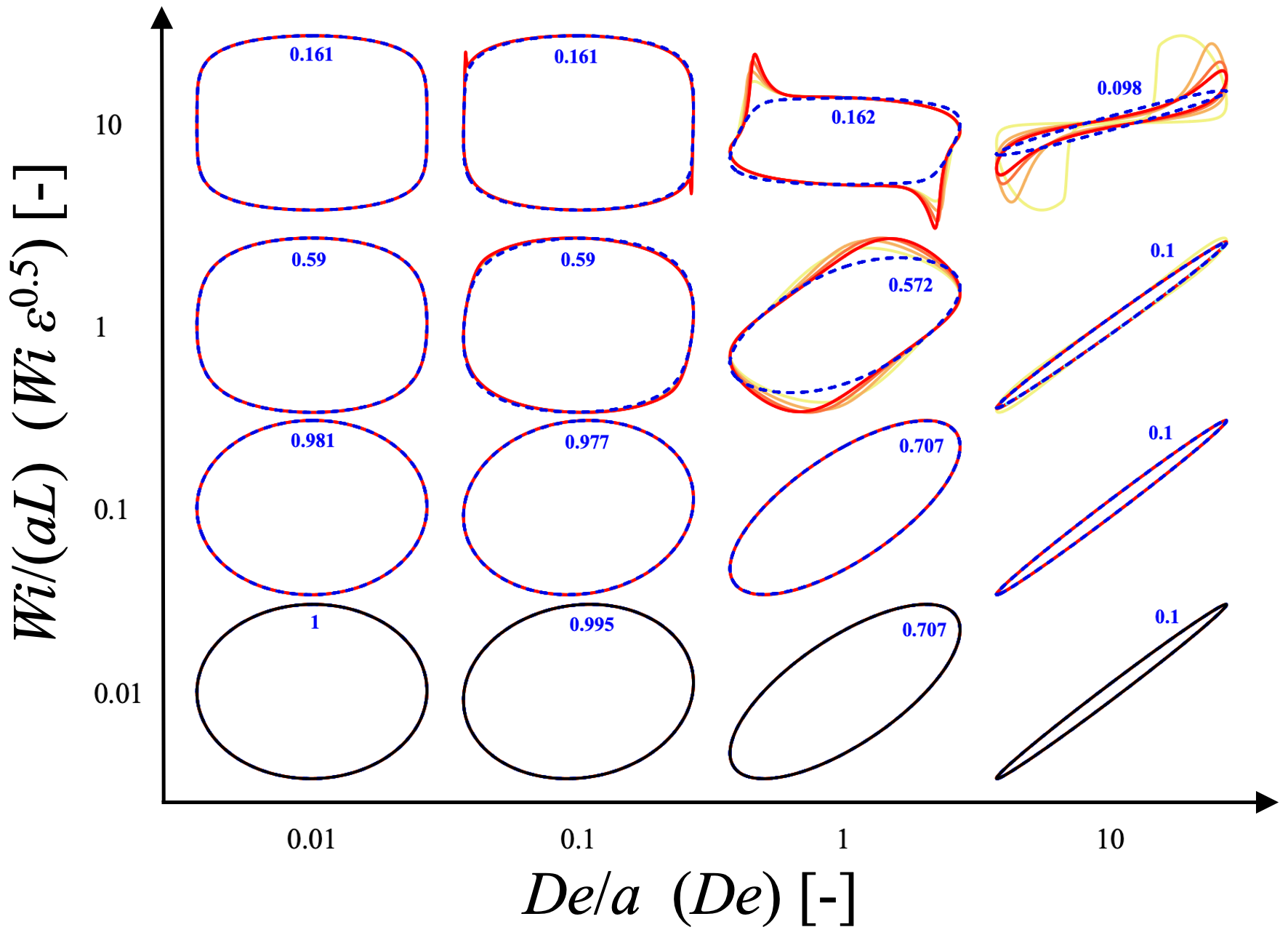}

\hspace{15mm}\includegraphics[width=0.85\textwidth]{Figures/LegendLogLog.PNG}
\caption{Elastic Lissajous-Bowditch plots in $Wi/(aL) \ (Wi\sqrt{\epsilon}) - De/a \ (De)$ space for the FENE-P (sPTT) model.  Black curves represent the UCM response. Blue numbers in each plot represent the maximum value of $\tau_{p,12}$ in the sPTT response, since the $y$-axis is scaled differently in each plot.}
\label{elasticLBWi_La}
\end{figure}

\begin{figure}
\centering
\subfloat[$~$]{\includegraphics[width=0.45\textwidth]{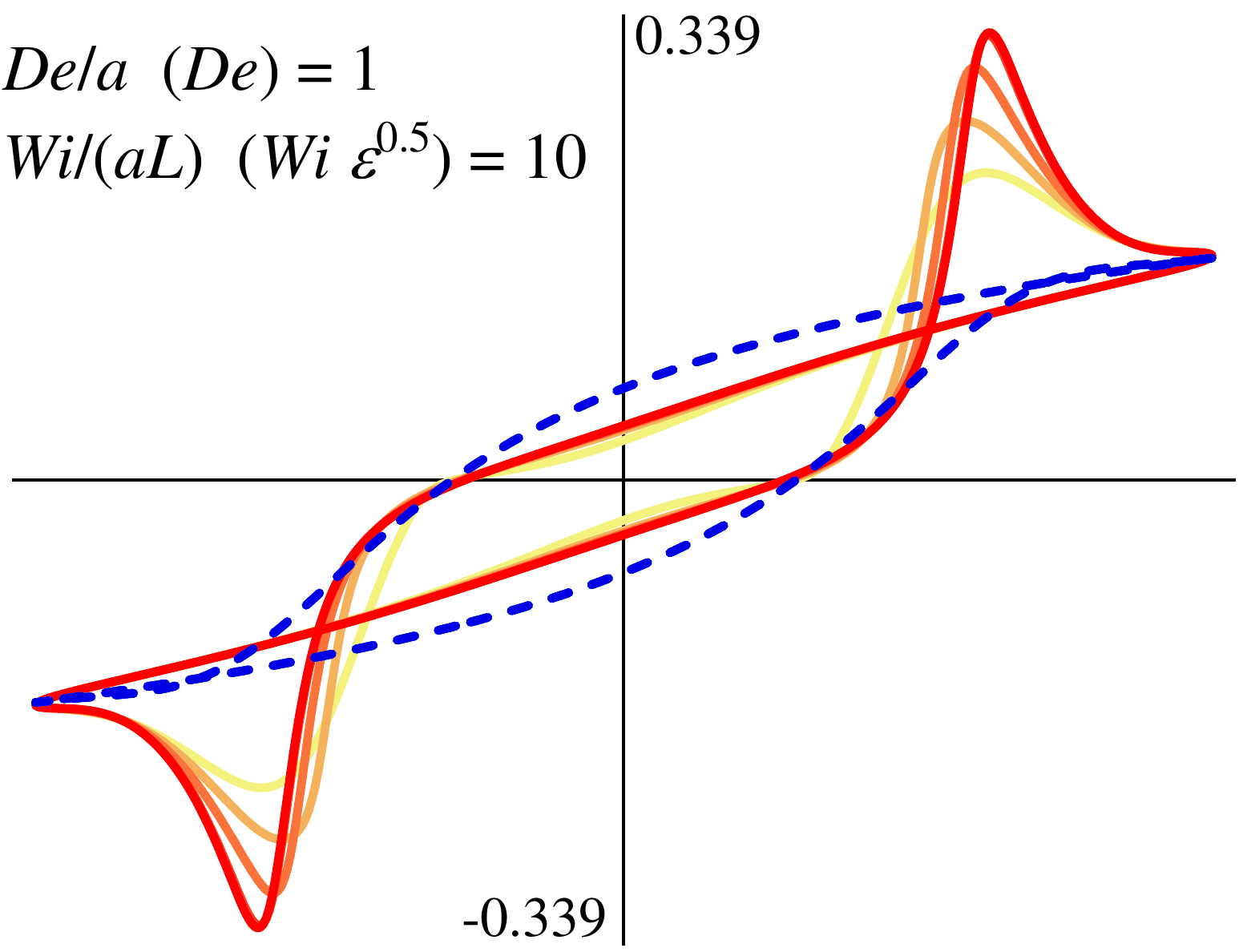}}
\subfloat[$~$]{\includegraphics[width=0.45\textwidth]{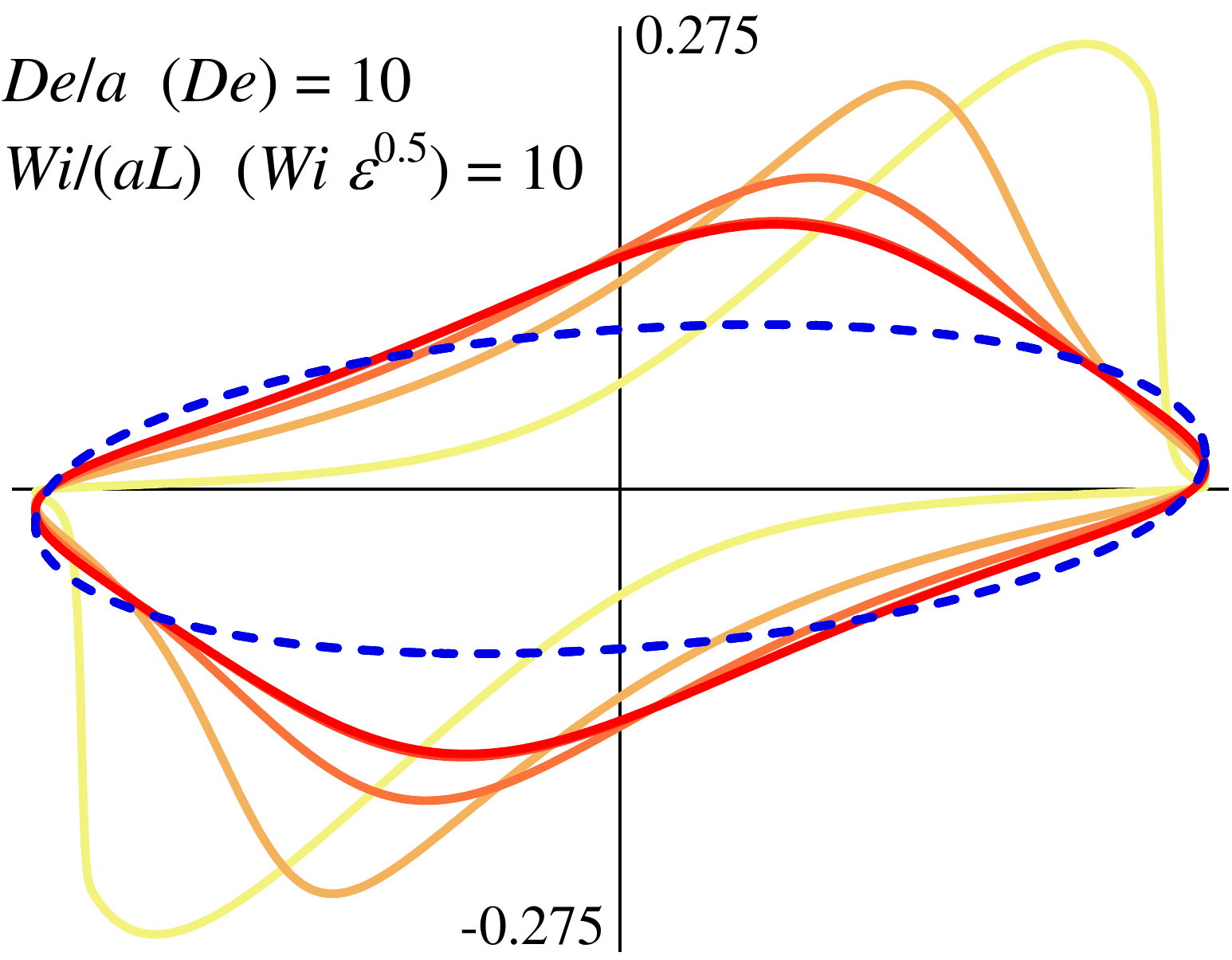}}
\caption{Zoomed Lissajous-Bowditch plots (viscous projection) for (a) $De/a \ (De) = 1$ and $Wi/(aL) \ (Wi\sqrt{\epsilon}) = 10$ and (b) $De/a \ (De) = 10$ and $Wi/(aL)\ (Wi\sqrt{\epsilon}) = 10$. See Figure \ref{viscousLBWi_La} for the legend.}
\label{zoomedLissajous}
\end{figure}

Figures \ref{viscousLBWi_La} and \ref{elasticLBWi_La} show the viscous and elastic Lissajous-Bowditch plots, respectively, in the $Wi/(aL) \ (Wi\sqrt{\epsilon}) - De/a \ (De)$ space for the FENE-P (sPTT) model. For the sPTT model, scaling the curves with $Wi\sqrt{\epsilon}$ causes the responses for the various values of $\epsilon$ to become universal (for constant $\beta$). Whilst this is expected in SSSF due to the form of the analytical solution (Equation \eqref{fenepanalytical1}), it may not be immediately obvious why this is also the case in LAOS. However, this can be shown by considering the following system of equations for the sPTT model (acknowledging that $A_{22} = A_{33} = 1$ due to the fact $F(A)$ is on the outside of the $(\mathsfbi{A}-\mathsfbi{I})$ term in the constitutive model)

\begin{subequations}
\begin{gather}
\frac{\dif {A}_{11}}{\dif t} = 2 \bigg(\frac{Wi}{De}\bigg) {A}_{12}  \ \mathrm{cos}(t) - \frac{1}{De} (1+\epsilon({A}_{11}-1))({A}_{11} - 1) \\
\frac{\dif {A}_{12}}{\dif t} = \bigg(\frac{Wi}{De}\bigg) \ \mathrm{cos}(t)
 - \frac{1}{De} (1+\epsilon({A}_{11}-1))({A}_{12}) \\
 \tau_{p,12} = \frac{(1-\beta)}{Wi}A_{12}
\end{gather}
\end{subequations}

\noindent and introducing new variables $x = \epsilon(A_{11} - 1)$ and $y = A_{12}/Wi$, which gives

\begin{subequations}
\begin{gather}
De \frac{\dif x}{\dif t} = 2 (Wi \sqrt{\epsilon})^2 y \ \mathrm{cos}(t) - (1+x)x \\
De \frac{\dif y}{\dif t} = \mathrm{cos}(t)
 - (1+x)y \\
 \tau_{p,12} = (1-\beta)y
\end{gather}
\label{pttscaling}
\end{subequations}

\noindent Thus, for constant values of $\beta$ (in our case $\beta = 0$), the system only depends on the parameters $De$ and $Wi \sqrt{\epsilon}$.

The FENE-P LAOS response does not scale universally with $Wi/(aL)$ under uniform oscillation, as its SSSF response does. For $De / a < 0. 1$ there is practically no difference in the curves for the various values of $L^2$ since the flow is approaching SSSF. For $De / a \geq 0.1$ and $Wi / (aL) \geq 1$, the difference in the FENE-P response with varying $L^2$ becomes significant. However, the FENE-P response does seem to become universal at constant values of $De/a$ and $Wi/(aL)$ for large enough values of $L^2$. This is highlighted in Figure \ref{zoomedLissajous} by the fact that the responses for $L^2 = 100$ and $L^2 = 1000$ are practically identical. There are at least two potential reasons that the FENE-P response is not universal for constant values of $De/a$ and $Wi/(aL)$ at low values of $L^2$. One is that the functional form of $F(A)$ in the FENE-P model is different to that in the sPTT model (see Equation \eqref{extensibilityFunctionConf}).  Another is that the position of $F(A)$ in the conformation tensor form of the FENE-P model is different to that in the sPTT model (i.e. on the inside of the brackets in the recoil term rather than on the outside of the brackets). If the latter is the cause for the difference in the scaling behaviour between the sPTT and FENE-P models, then it should likely be the case that a universal solution exists for the FENE-CR model, which is given under LAOS as

\begin{subequations}
\begin{gather}
De \frac{\partial}{\partial t} \mathsfbi{A} - Wi (\mathsfbi{A} \cdot \nabla \boldsymbol{u} + \nabla \boldsymbol{u}^{\mathrm{T}} \cdot \mathsfbi{A} - \boldsymbol{u} \bcdot \nabla \mathsfbi{A}) = - F(A)_{\mathrm{FP}} (  \mathsfbi{A} -  \mathsfbi{I} )  \\ 
\boldsymbol{\tau}_p = \frac{(1-\beta)}{Wi} F(A)_{\mathrm{FP}}(\mathsfbi{A} - \mathsfbi{I})
\end{gather}
\end{subequations}

\noindent Noting that $A_{22} = A_{33} = 1$ in the FENE-CR model due to the difference in the position of $F(A)$ compared to the FENE-P model, the system of equations for the FENE-CR model under LAOS is given as

\begin{subequations}
\begin{gather}
\frac{\dif {A}_{11}}{\dif t} = 2 \bigg(\frac{Wi}{De}\bigg) {A}_{12}  \ \mathrm{cos}(t) - \frac{1}{De} \bigg(\frac{L^2}{L^2 - A_{11} - 2}\bigg)(A_{11} - 1) \\
\frac{\dif {A}_{12}}{\dif t} = \bigg(\frac{Wi}{De}\bigg) \ \mathrm{cos}(t)
 - \frac{1}{De} \bigg(\frac{L^2}{L^2 - A_{11} - 2} \bigg) A_{12} \\
 \tau_{p,12} = \frac{(1-\beta)}{Wi} \bigg(\frac{L^2}{L^2 - A_{11} - 2} \bigg) A_{12}
\end{gather}
\end{subequations}

\noindent Introducing the new variable $x = (A_{11}-1)/L^2$,  the extensibility function can be rewritten as $(1-x-3/L^2)^{-1}$, which, for $L^2 \gg 3$, becomes $(1-x)^{-1}$. The system of equations for the FENE-CR model for $L^2\gg 3$ can then be rewritten, using also $y = A_{12}/Wi$, as

\begin{subequations}
\begin{gather}
De\frac{\dif x}{\dif t} = 2 \bigg(\frac{Wi}{L}\bigg)^2 y  \ \mathrm{cos}(t) - \bigg(\frac{1}{1-x}\bigg)x \\
De \frac{\dif y}{\dif t} = \mathrm{cos}(t)
 -\bigg(\frac{1}{1-x}\bigg)y \\
 \tau_{p,12} = (1-\beta) \bigg(\frac{1}{1-x}\bigg)y
\end{gather}
\label{fenecrscaling}
\end{subequations}

\noindent Therefore, the FENE-CR model has universal solutions for constant values of $De$ and $Wi/L$ only in the case that $L^2 \gg 3$, and the breakdown of the universality can be caused solely due to a change in the functional form of the extensibility function $F(A)$ without a change in its position in the constitutive model.  This result alone cannot explain the scaling of the FENE-P response due to the fact that both the form of $F(A)$ and its position in the model (i.e. $F(A)(\mathsfbi{A}-\mathsfbi{I})$ \textit{versus} $(F(A)\mathsfbi{A}-\mathsfbi{I})$) are different to those in the sPTT model. We will explore the effect of the position of $F(A)$ in the constitutive model on the scaling of the response further in Section \ref{results3}. We also note here that $(1-x)^{-1}$ can be expanded as $1 + x + \mathcal{O}(x^2)$. Therefore, for $\epsilon = 1/L^2$,  the evolution of $\mathsfbi{A}$ for the FENE-CR model becomes mathematically identical to that for the sPTT model in the MAOS regime in the case that $L^2 \gg 3$. In the MAOS regime, where the response is weakly non-linear, terms of $\mathcal{O}(x^2)$ can be neglected in the expansion of $F(A)$. There is, however, a difference in the stress response due to the presence of the extensibility function in the $\boldsymbol{\tau}_p$-$\mathsfbi{A}$ relationship in the FENE-CR model.  We highlight this further in Appendix \ref{appA}.

\subsection{Comparing the LAOS response of the FENE-P and sPTT models}\label{results2}

One of the primary aims of this study is to assess the difference between the sPTT and FENE-P models in oscillatory shear flow. As mentioned, in SSSF, both models exhibit identical responses in terms of $\boldsymbol{\tau}_p$ with $\epsilon = 1/L^2$ and $Wi_{\mathrm{sPTT}} = Wi_{\mathrm{FP}} / a$, and differences in the model responses arise only due to transients in the flow. 

In Figures \ref{viscousLBWi_La} and \ref{elasticLBWi_La}, it is observed that the sPTT and FENE-P models (with the aforementioned substitution of parameters) have, naturally, identical responses in oscillatory shear flow for $De / a  \ (De) \rightarrow 0$, which corresponds to the system approaching SSSF. Both models also exhibit identical responses for $Wi/(aL) \ (Wi\sqrt{\epsilon}) \rightarrow 0$, however in this case both models reduce to the UCM model. For large values of both $De/a \ (De)$ and $Wi/(aL) \ (Wi\sqrt{\epsilon})$, there is a significant difference between the responses of the two models.  One such difference is that the FENE-P response exhibits sharp overshoots in the shear stress, which are often observed in the responses of strongly non-linear viscoelastic models, such as those derived specifically for wormlike micelles. These can be observed in particular in the plots for $De / a = 1$ and $Wi / (aL) = 10$, which are shown at a larger scale in Figure \ref{zoomedLissajous}a.  Similar stress overshoots were also observed in the FENE-P model response during start-up shear flow \citep{Davoodi2022}. In LAOS, the pronounced shear stress overshoots manifest as self-intersecting secondary loops in the viscous Lissajous curves. The criterion for the presence of the self-intersecting secondary loops is that the gradient of the decomposed elastic stress with respect to the strain is negative at $\gamma = 0$ (or $\dot{\gamma} = 1$), indicating that elasticity is being relieved through recoil faster than it is being accumulated through increased rates of deformation \citep{Ewoldt2010}. These secondary loops are associated with strongly non-linear viscoelastic responses and have been observed both in experimental LAOS data and in the responses of several viscoelastic constitutive models, as well as, as already mentioned, from simulations of network models.

In order to quantify the deviation of the FENE-P and sPTT models from the UCM response, we define a function $G(A)$ given by

\begin{equation}
G(A) = 
\begin{cases}
\frac{\displaystyle F(A)_{\mathrm{FP}}}{\displaystyle a} -1 & \text{FENE-P} \\[6pt] 
F(A)_{\mathrm{sPTT}}- 1 & \text{sPTT}
\end{cases}
\label{ExtFunc}
\end{equation}

\noindent Note that with the relevant expression relating $\boldsymbol{\tau}_p$ to $\mathsfbi{A}$ for each model (Equation \eqref{cramersForm}) it is the case that $F(A)_{\mathrm{FP}} = F(\tau_p)_{\mathrm{FP}}$ and $F(A)_{\mathrm{sPTT}} = F(\tau_p)_{\mathrm{sPTT}}$. We point this out just to clarify that as $G(A) \rightarrow 0$, both the conformation tensor and stress tensor forms of the models asymptote towards the UCM model.

\begin{figure}
\centering
\large
$G(A) \ $ \textit{vs} $\ \dot{\gamma} $
\includegraphics[width=0.9\textwidth]{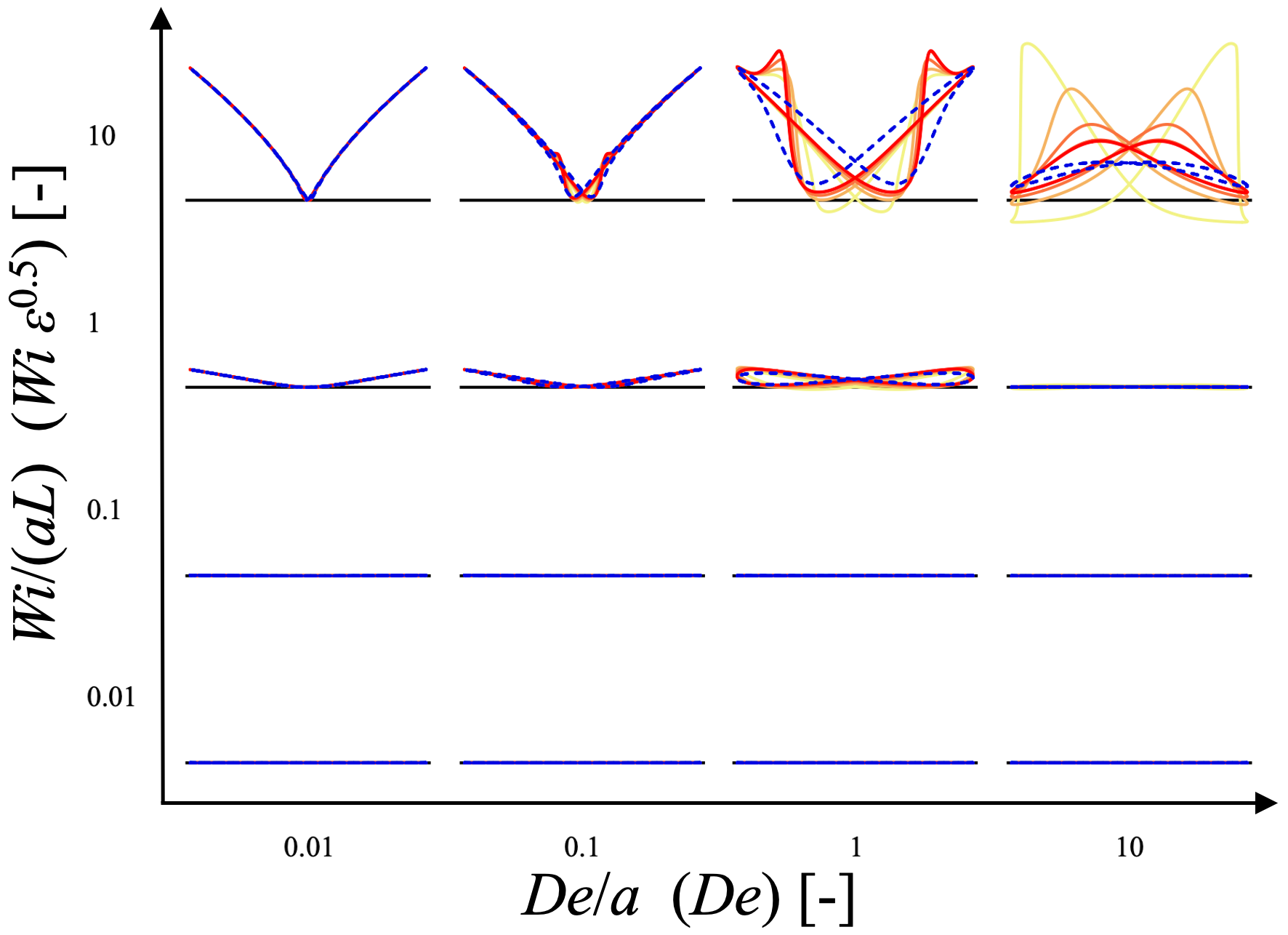}

\hspace{15mm}\includegraphics[width=0.85\textwidth]{Figures/LegendLogLog.PNG}
\caption{$G(A)$ \textit{vs} $\dot{\gamma}$ in $Wi/(aL) \ (Wi\sqrt{\epsilon}) - De/a \ (De)$ space for the FENE-P (sPTT) model. In each plot, the $y$-axis runs from -1 to 6.31. The $x$-axis runs from -1.04 to 1.04. The black solid line shows $y = 0$ on each plot.}
\label{ExtFuncLissajous}
\end{figure}

Figure \ref{ExtFuncLissajous} shows $G(A)$ against the dimensionless strain rate for the corresponding Lissajous curves presented in Figure \ref{viscousLBWi_La}. For both models, $G(A)$ is approximately 0 for values of $Wi/(aL) \ (Wi\sqrt{\epsilon}) \leq 0.1$ at any value of $De/a \ (De)$. Consequently, the dimensionless shear stress response shown in the corresponding Lissajous curves is essentially that of the UCM model. For $De/a \ (De) = 0.01$, even when $G(A)$ increases and the model responses become increasingly non-linear,  for each model, the solution is universal for various values of $L^2$ and $\epsilon$ at constant $Wi/(aL) \ (Wi\sqrt{\epsilon})$, which has been explained already by the fact that the steady-shear response of the dimensionless model contains only the scaling parameter $Wi/(aL) \ (Wi\sqrt{\epsilon})$.  Note here that the response of $G(A)$ is also the same for both the sPTT and FENE-P models, despite the fact that the functions $F(A)_{\mathrm{FP}}/a$ and $F(A)_{\mathrm{sPTT}}$ are not explicitly equivalent for $\epsilon = 1/L^2$, $Wi_{\mathrm{sPTT}} = Wi_{\mathrm{FP}} / a$, and $De_{\mathrm{sPTT}} = De_{\mathrm{FP}} / a$. Therefore, in SSSF, i.e., $De \rightarrow 0$, it is the case that $\mathrm{tr}(\mathsfbi{A})_{\mathrm{sPTT}}= (F(A)_{\mathrm{FP}}/a) \cdot \mathrm{tr}(\mathsfbi{A})_{\mathrm{FP}}$, which is also implied from Equations \eqref{cramersForm} if $(\boldsymbol{\tau}_p)_{\mathrm{FP}} = (\boldsymbol{\tau}_p)_{\mathrm{sPTT}}$. Thus, in steady and homogeneous flows, $\mathrm{tr}(\mathsfbi{A})$ for each model differs by a factor of $F(A)_{\mathrm{FP}}/a$, highlighting the difference in the physical interpretation of the polymeric stress from $\mathsfbi{A}$ in each model. For the higher values of $De/a \ (De)$ and $Wi/(aL) \ (Wi\sqrt{\epsilon})$, it is observed again that the sPTT solution for $G(A)$ is universal for constant $De$ and $Wi\sqrt{\epsilon}$ with varying $\epsilon$, but the FENE-P solution for $G(A)$ is only universal for constant $De/a$ and $Wi/(aL)$ at large values of $L^2$. It is also observed that there is significant correspondence between the results in Figures \ref{ExtFuncLissajous} and \ref{viscousLBWi_La}. Notably, the overshoots in $G(A)$ appear to correspond at least qualitatively with the overshoots in $\tau_p$. This will be discussed in more detail next.

\begin{figure}
\centering
\subfloat[$~$]{
\includegraphics[trim={0.35cm 0cm 0.4cm 0cm},clip,width=0.48\textwidth]{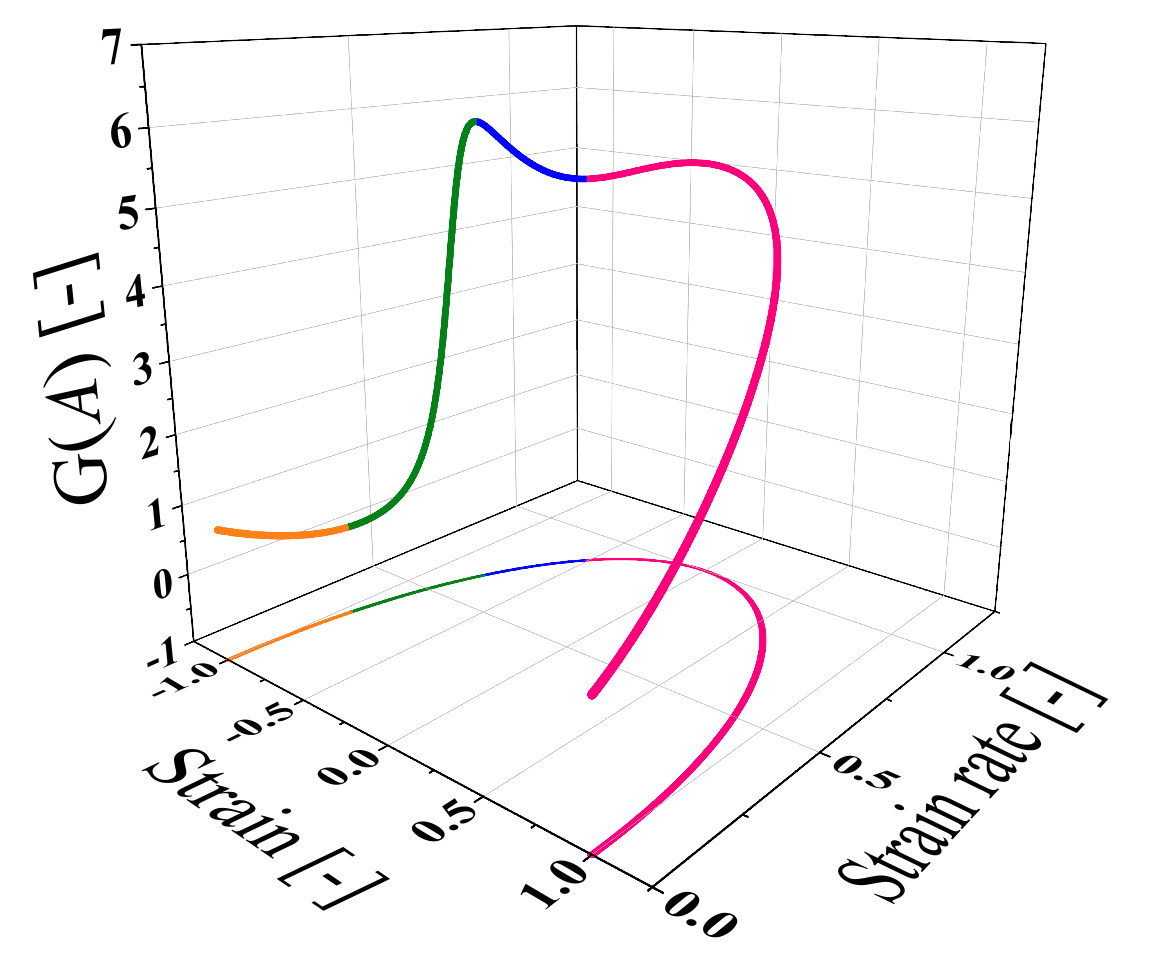}}
\subfloat[$~$]{
\includegraphics[trim={0.35cm 0cm 0.4cm 0cm},clip,width=0.48\textwidth]{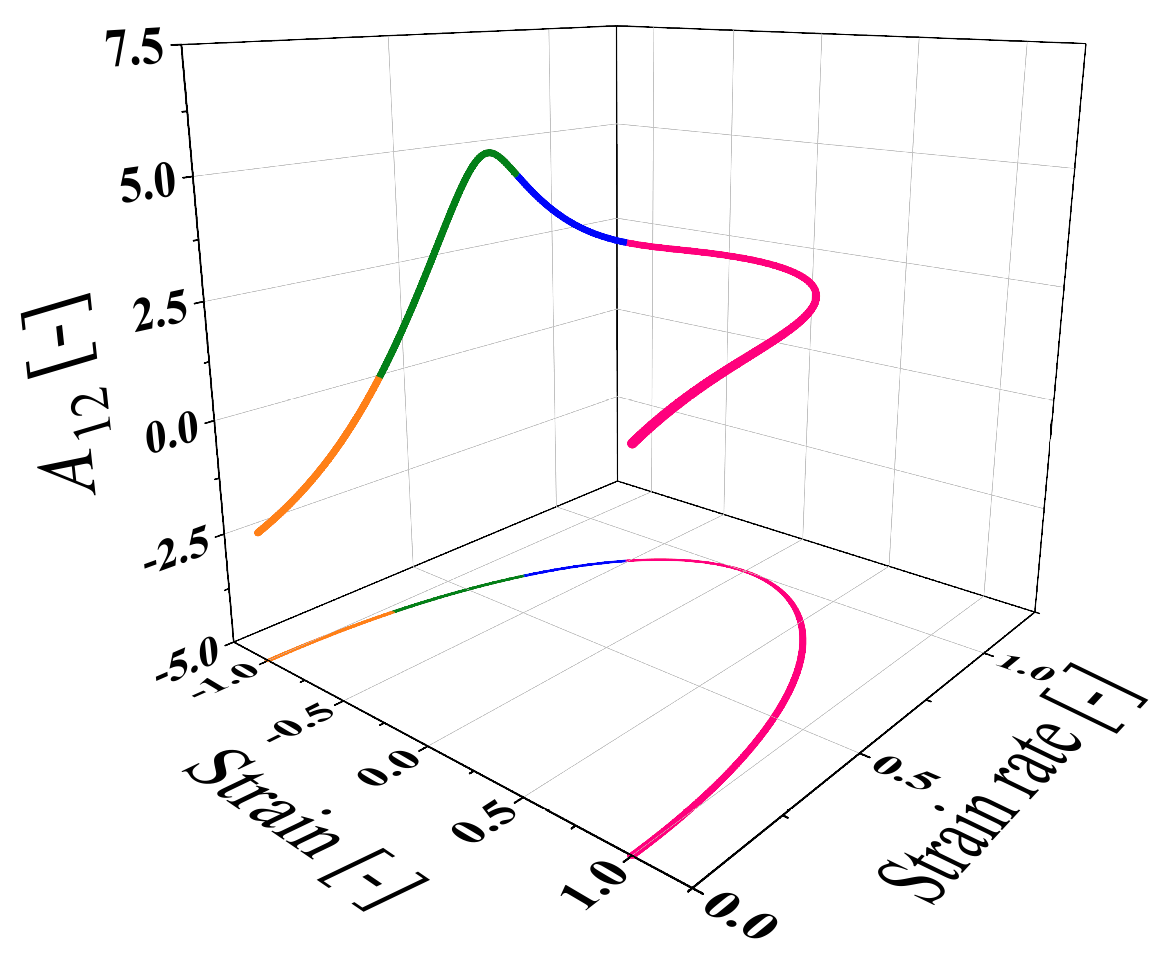}}

\subfloat[$~$]{
\includegraphics[trim={0.35cm 0cm 0.4cm 0cm},clip,width=0.48\textwidth]{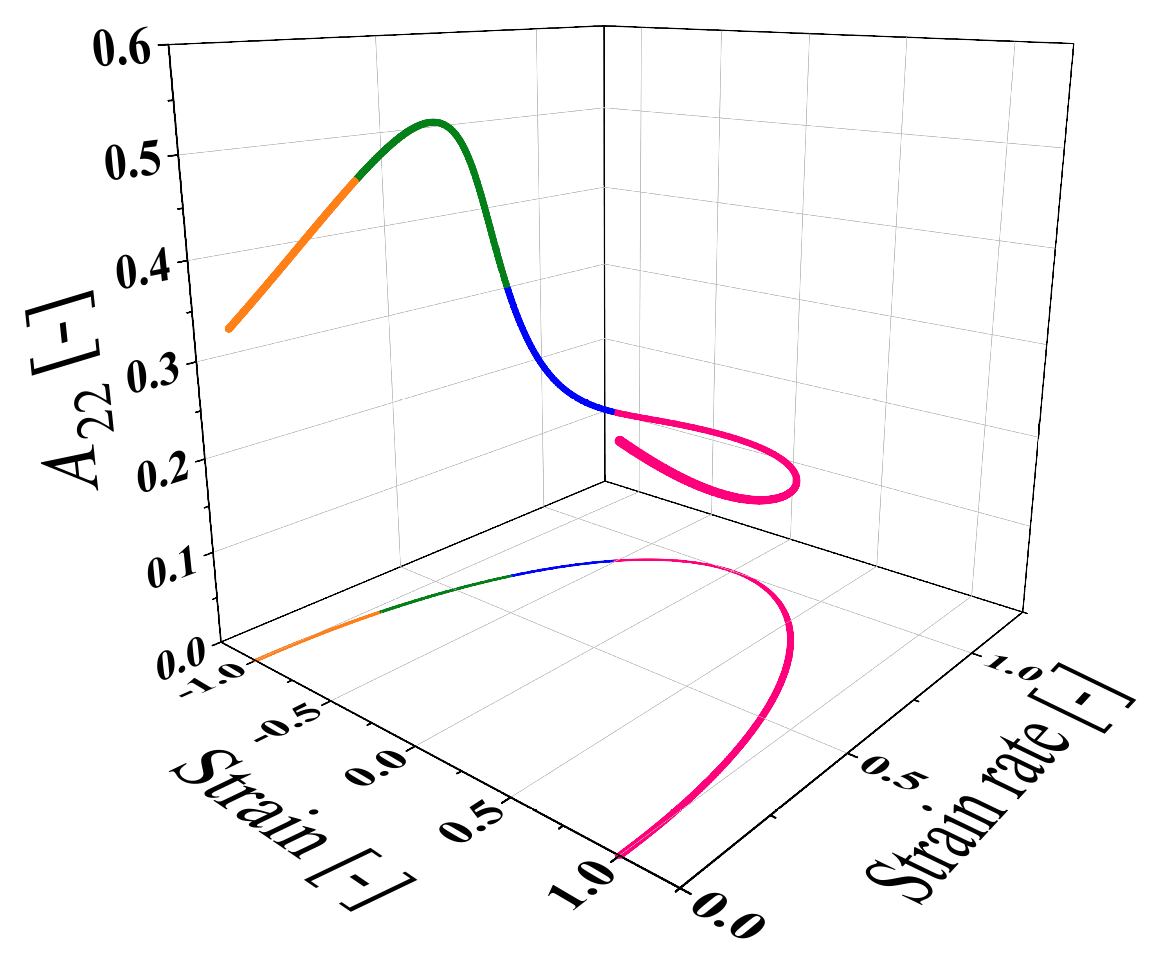}}
\subfloat[$~$]{
\includegraphics[trim={0.35cm 0cm 0.4cm 0cm},clip,width=0.48\textwidth]{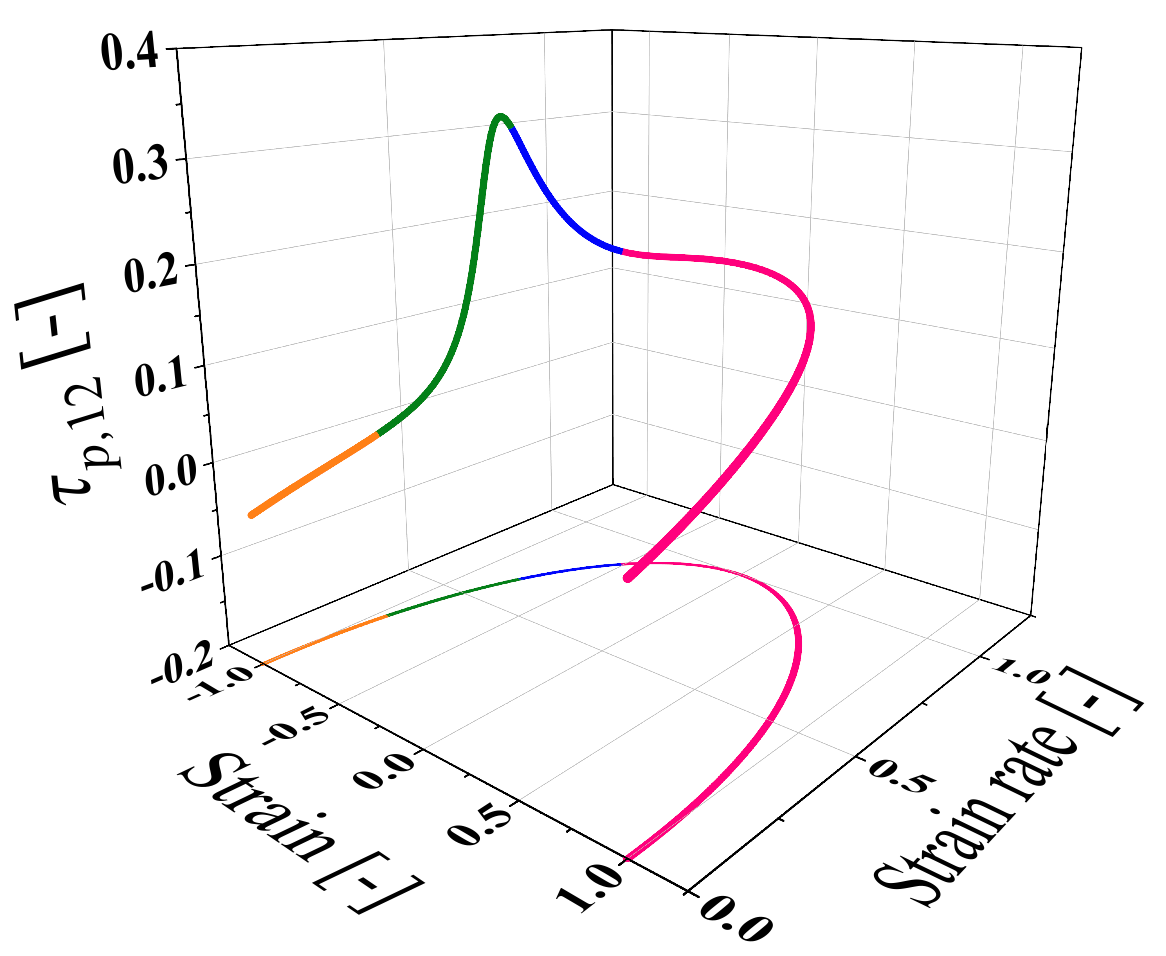}}

\centering
\includegraphics[width=0.75\textwidth]{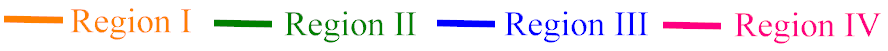}
\caption{3D plots showing (a) $G(A)$, (b) $A_{12}$, (c) $A_{22}$, (d) $\tau_{p, 12}$ in $\gamma-\dot{\gamma}$ space for the FENE-P response ($L^2 = 100$) for $De/a = 1$ and $Wi/(aL) = 10$. Only positive $\dot{\gamma}$ are displayed for convenience. Regions denoted with roman numerals are discussed in the text.}
\label{fenesplit}
\end{figure}

For  $De/a$ and $Wi/(aL) = 10$, we present in Figure \ref{fenesplit} 3D plots of $G(A)$, $A_{12}$,  $A_{22}$ and $\tau_{p, 12}$ against $\gamma$ and $\dot{\gamma}$ for the FENE-P model response in LAOS (with $L^2 = 100$). We present this to highlight more clearly the mechanisms responsible for the distinct behaviour exhibited by the FENE-P model in transient flows. We split the curve into 4 regions (note that only one half of the curve is shown in terms of the range of $\dot{\gamma}$). It is important here to refer back to Equations \eqref{pttoscillatoryshear} and \eqref{cramersForm} to explain the evolution of the shear stress. The regions are specified in chronological order (i.e. the system moves in time from Region \rom{1} to Region \rom{4}).

In Region \rom{1}, $A_{12}$ is negative and the rate of deformation is increasing, meaning that both the growth of $A_{12}$ due to deformation and the elastic recoil are acting in the same direction, causing a positive rate of change in time of $A_{12}$. The rate of change of $A_{12}$ is governed by a balance between growth due to deformation and reduction due to elastic recoil. In the sPTT model, $A_{22} = 1$ and so the growth of $A_{12}$ due to increasing rates of deformation is proportional to the strain rate. In the FENE-P model however, $A_{22}$ is time-dependent for large values of $Wi/(aL)$ and so the growth of $A_{12}$ due to increasing deformation rates is non-linear (this will be discussed in more detail in Section \ref{results3}). Whilst $G(A)$ is decreasing in Region \rom{1}, which reduces the degree of the elastic recoil, the growth of $A_{22}$ is relatively large and so ultimately $A_{12}$ grows non-linearly in Region \rom{1}. Despite the fact $A_{12}$ grows in Region \rom{1}, $\tau_{p,12}$ grows only slightly (and seemingly linearly) due to the presence of $F(A)$ in the $\boldsymbol{\tau}_p$-$\mathsfbi{A}$ relationship and the fact that $F(A)$ is decreasing.

Region \rom{2} is characterised by a sharp increase in $G(A)$ as $\mathrm{tr}(\mathsfbi{A}) \rightarrow L^2$, which causes $\tau_{p,12}$ to increase rapidly but also limits somewhat the growth of $A_{12}$ since the elastic recoil increases.  $A_{22}$ also goes through a maxima in Region \rom{2} and begins to decrease, which causes a reduction in the growth of $A_{12}$ with increasing rates of deformation, exacerbating the recoil effect.  In Region \rom{3}, $G(A)$ is large enough that the elastic recoil now exceeds the growth of $A_{12}$ due to the increasing rate of deformation, which drives a negative rate of change of $A_{12}$ for increasing strain rates. In this region, $G(A)$ also decreases as $\mathrm{tr}(\mathsfbi{A})$ decreases, and so $\tau_{p,12}$ decreases rapidly due to both decreasing $A_{12}$ and $G(A)$ (and the fact $F(A)$ is present in the $\boldsymbol{\tau}_p$-$\mathsfbi{A}$ relationship). 

At the start of Region \rom{4}, $A_{12}$ remains fairly constant, which is a consequence of the balance of the elastic recoil with the building of elasticity due to the rate of deformation. This might be thought of as being representative of the system approaching steady-state. Further into Region \rom{4}, there is a decrease in $G(A)$, which drives a reduction in the shear stress. However, as the deformation rate decreases, there is no significant decrease in $A_{12}$ due to the fact that $A_{22}$ is increasing and $G(A)$ is decreasing, which acts ultimately to slow the recoil of $A_{12}$ when the rate of deformation is decreased.

\begin{figure}
\centering
\subfloat[$~$]{
\includegraphics[trim={0.5cm 0cm 0.2cm 0cm},clip,width=0.48\textwidth]{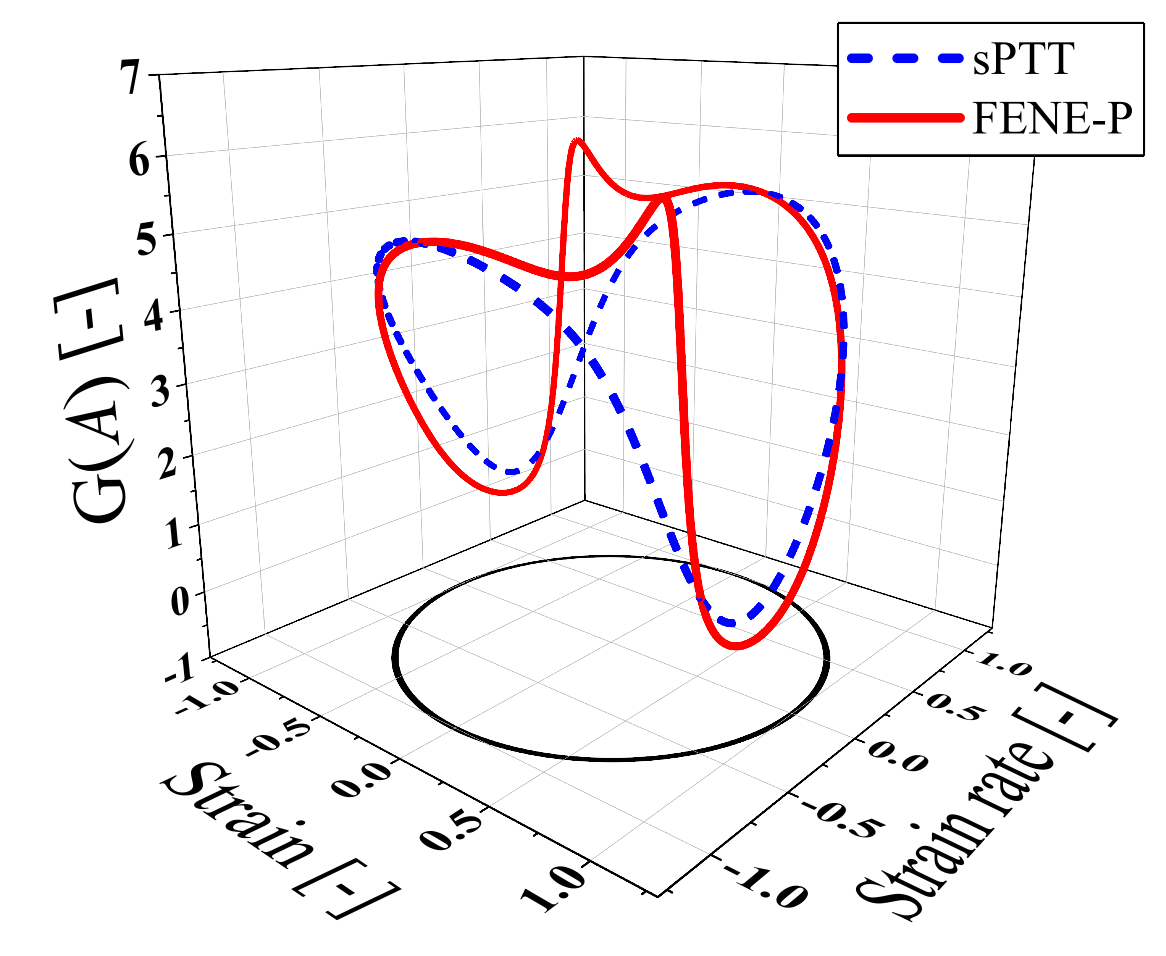}}
\subfloat[$~$]{
\includegraphics[trim={0.5cm 0cm 0.2cm 0cm},clip,width=0.48\textwidth]{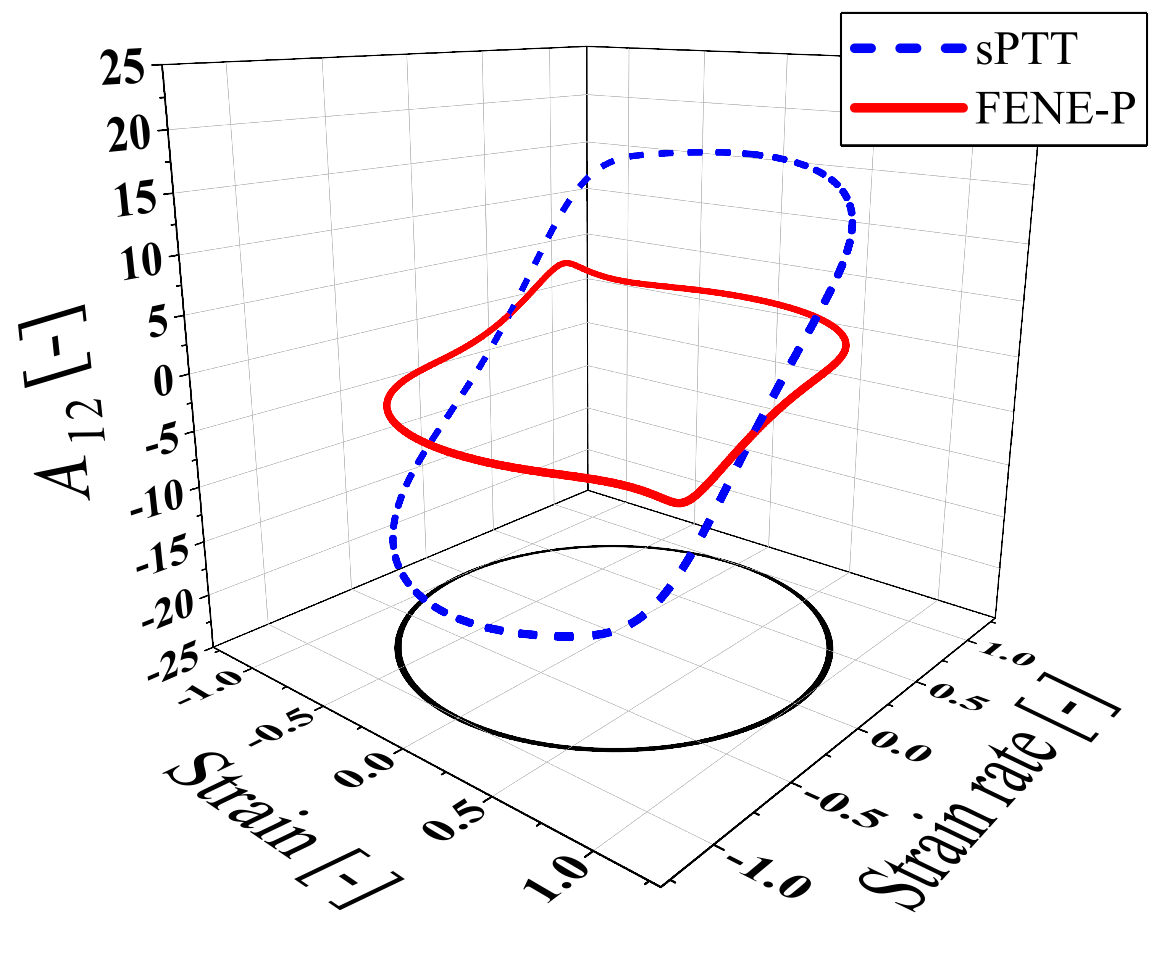}}

\subfloat[$~$]{
\includegraphics[trim={0.5cm 0cm 0.2cm 0cm},clip,width=0.48\textwidth]{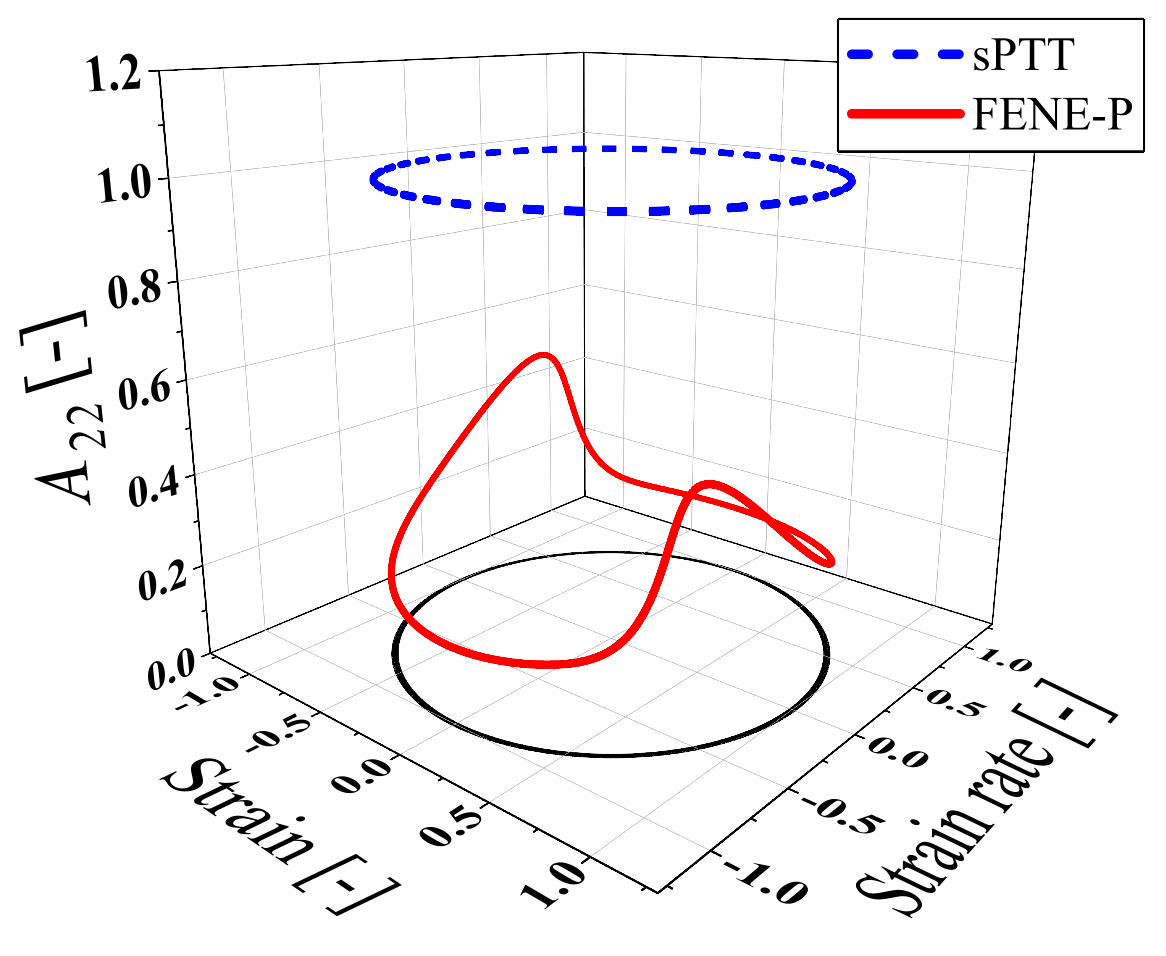}}
\subfloat[$~$]{
\includegraphics[trim={0.5cm 0cm 0.2cm 0cm},clip,width=0.48\textwidth]{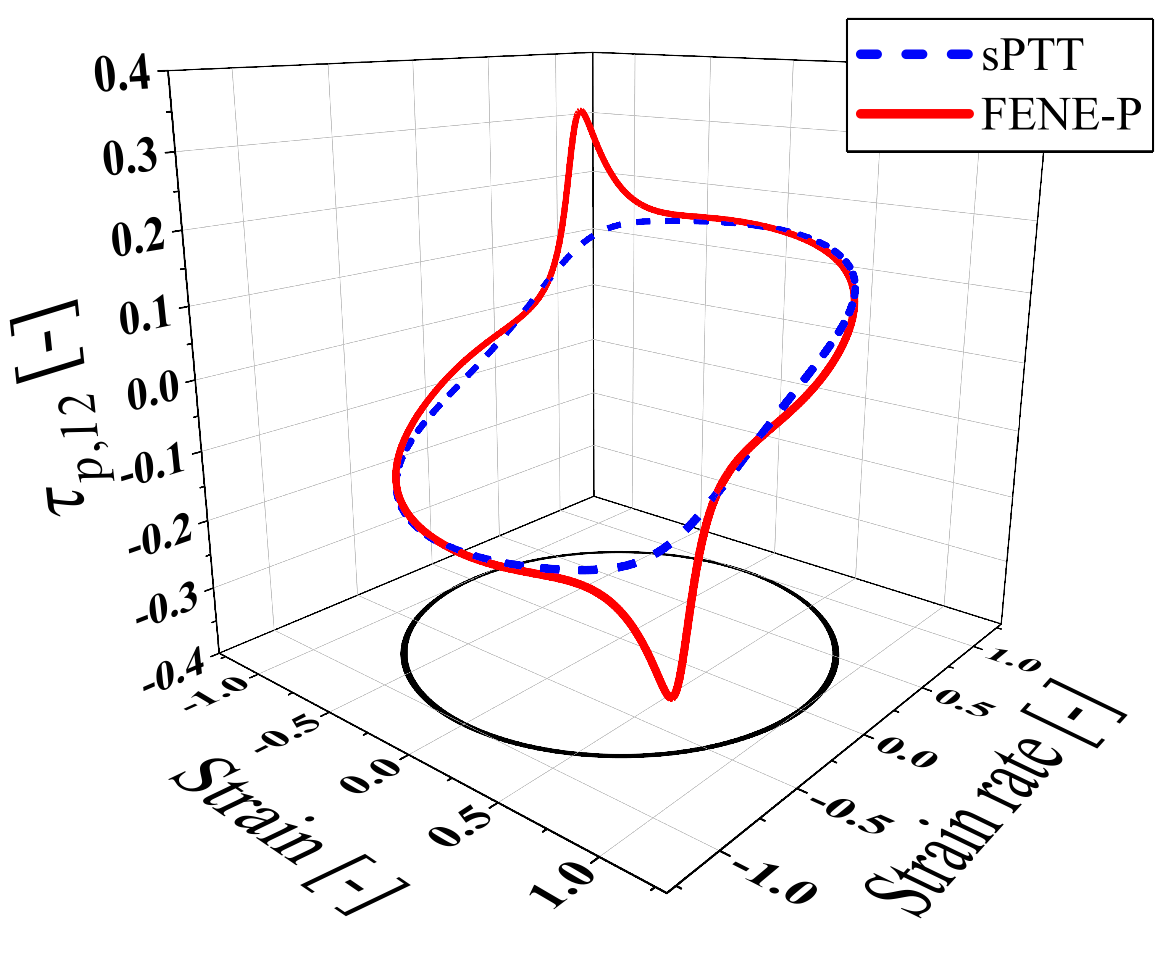}}
\caption{3D plots showing (a) $G(A)$, (b) $A_{12}$, (c) $A_{22}$, (d) $\tau_{p, 12}$ in $\gamma-\dot{\gamma}$ space for the FENE-P response ($L^2 = 100$) and sPTT response for $De/a \ (De) = 1$ and $Wi/(aL) \ (Wi\sqrt{\epsilon}) = 10$.}
\label{RegionsFene_sPTT}
\end{figure}

Considering the previous analysis, we then present 3D plots of $G(A)$, $A_{12}$,  $A_{22}$ and $\tau_{p, 12}$ against $\gamma$ and $\dot{\gamma}$ for the FENE-P ($L^2 = 100$) and sPTT model response for $De/a \ (De) = 1$ and $Wi/(aL) \ (Wi\sqrt{\epsilon}) = 10$ in Figure \ref{RegionsFene_sPTT}.  Note that the sharp overshoots in $G(A)$ are not observed for the sPTT model response and also note the significant differences in the $A_{22}$ responses of each model. $A_{12}$ is generally much larger for the sPTT model than for the FENE-P model, but the values of $\tau_{p, 12}$ are fairly similar for large parts of the oscillation due to Equation \eqref{cramersForm}. This analysis unravels exactly where the differences arise between the sPTT and FENE-P models in unsteady shear flows. It should be noted also that the unsteadiness here can be Eulerian or Lagrangian in nature, since the only difference between the two models written in stress tensor form (Equations \eqref{fenePNonDim} and \eqref{pttNonDim}) is a Lagrangian derivative term on the right hand side of the FENE-P model. This is not so easily observed when the models are expressed for the conformation tensor. This has significant consequences when these constitutive models are used to model flows in complex geometries which at first might seem steady-state due to the Eulerian steadiness, but might be Lagrangian unsteady (or inhomogeneous).

\subsection{Toy sPTT/FENE-P models}\label{results3}

In this subsection, we focus on identifying and investigating more closely the differences in the LAOS responses of the sPTT and FENE-P models, and particularly, the origins of the pronounced stress overshoots in the FENE-P response. To do this, we define "toy" models by manipulating slightly the standard sPTT and FENE-P models.  In this subsection, we do not focus on the aforementioned substitution of parameters to equate the FENE-P and sPTT responses for steady and homogeneous flows, as we just compare qualitatively the responses of each toy model. We only employ the 0D modelling approach to obtain the solutions, and we also use again $\beta = 0$ for all results.

For the toy sPTT model, we start with the generic network model given in Equation \eqref{pttrates} and adjust the rates of micro-structural destruction $D(A)$ and creation $C(A)$ as

\begin{subequations}
\begin{equation}
D(A) = 1 + \alpha \epsilon [\mathrm{tr}(\mathsfbi{A})-3]
\end{equation}

\begin{equation}
C(A) = 1 + (1-\alpha) \epsilon [\mathrm{tr}(\mathsfbi{A})-3]
\end{equation}
\end{subequations}

\begin{figure}
\centering
\large
$\tau_{p, 12} \ $ \textit{vs} $\ \dot{\gamma} $

\subfloat[$~$]{
\includegraphics[width=0.328\textwidth]{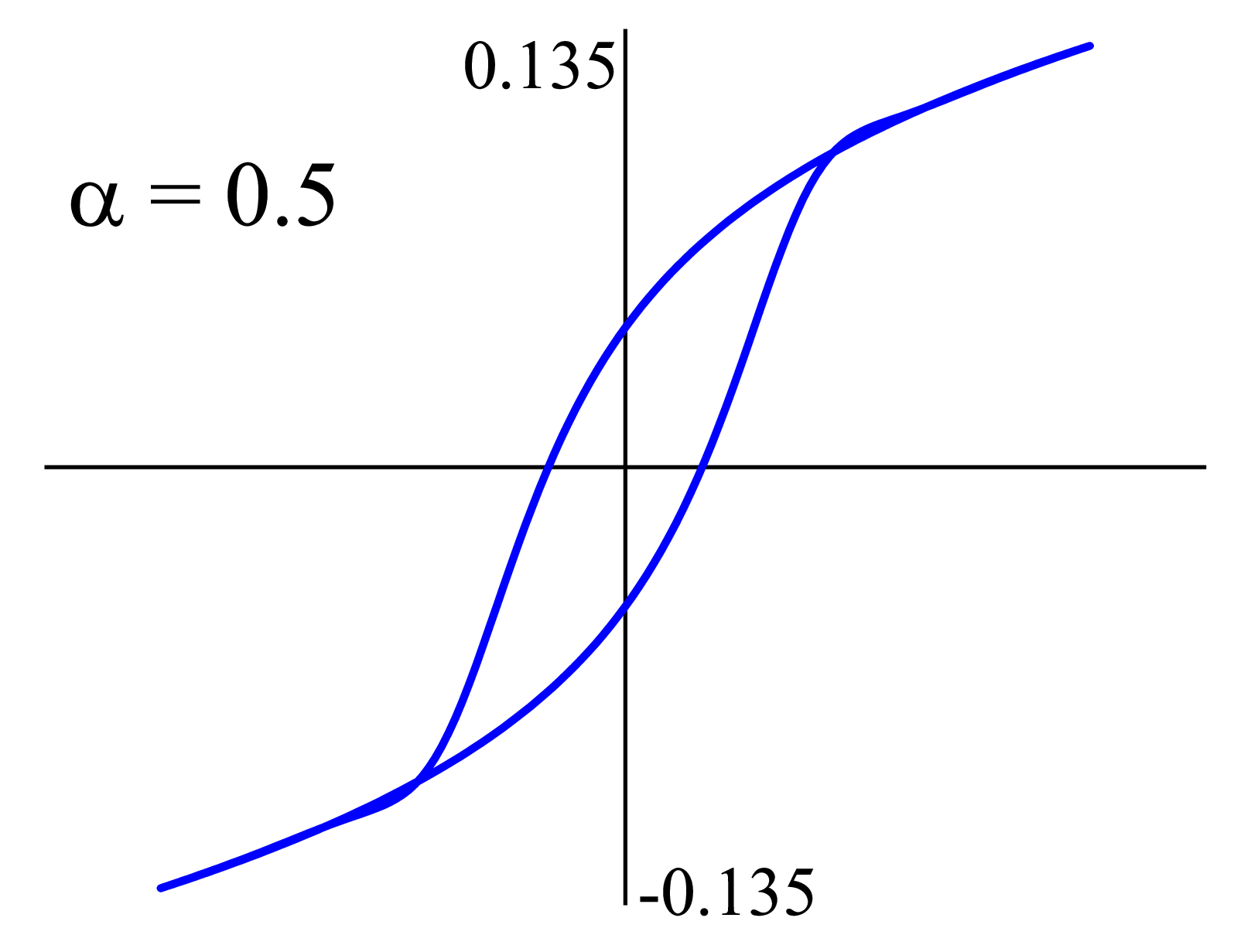}}
\subfloat[$~$]{
\includegraphics[width=0.328\textwidth]{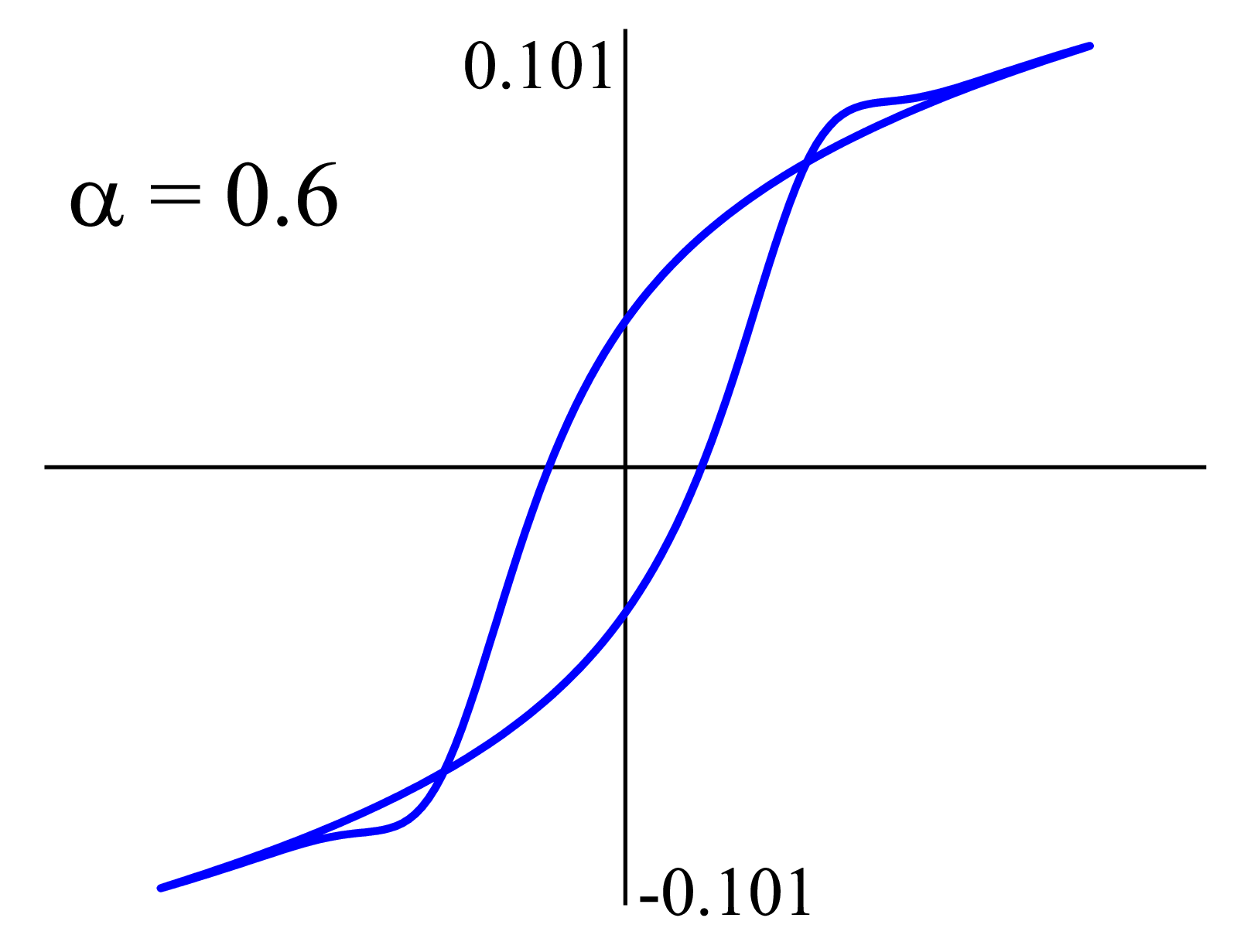}}
\subfloat[$~$]{
\includegraphics[width=0.328\textwidth]{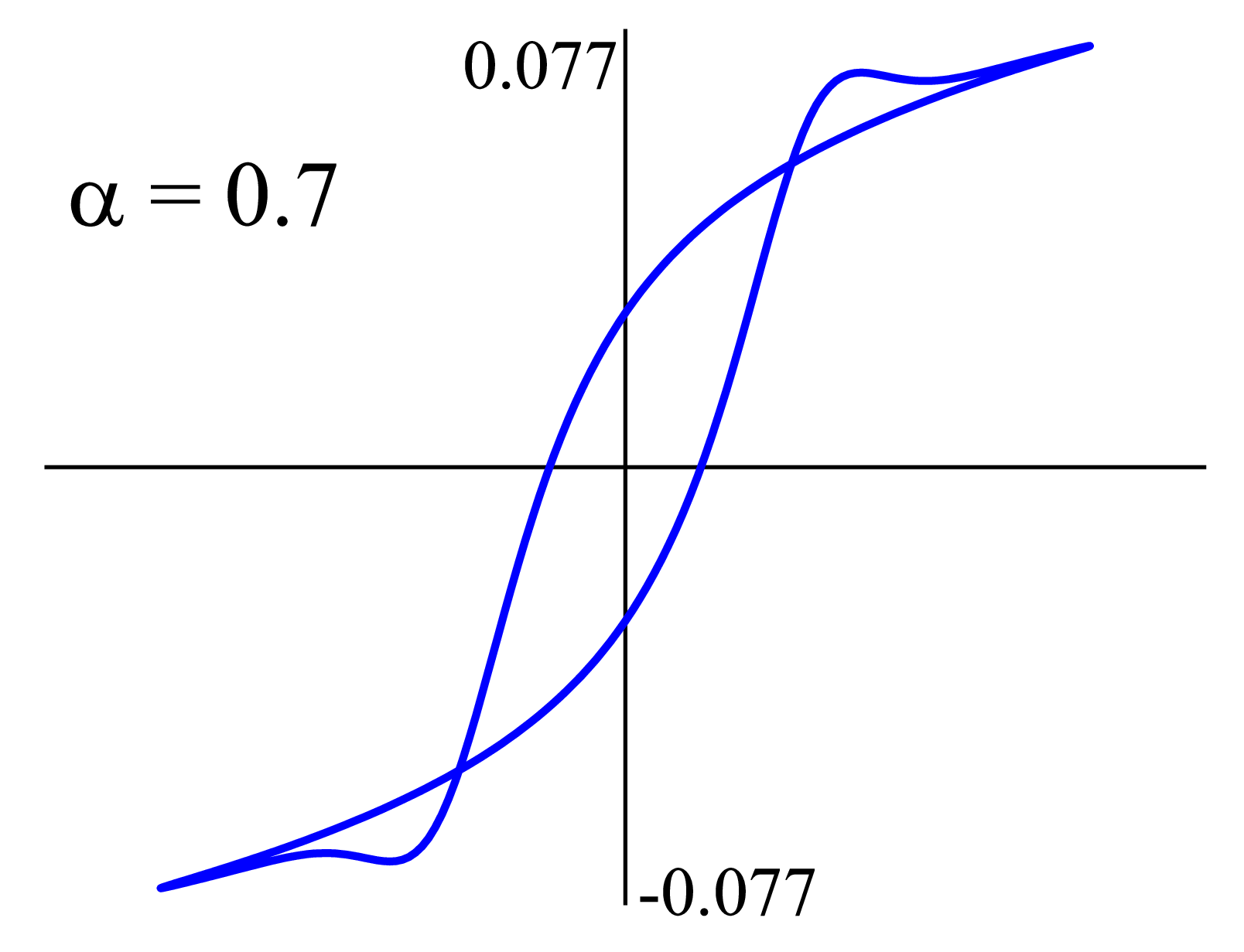}}

\subfloat[$~$]{
\includegraphics[width=0.328\textwidth]{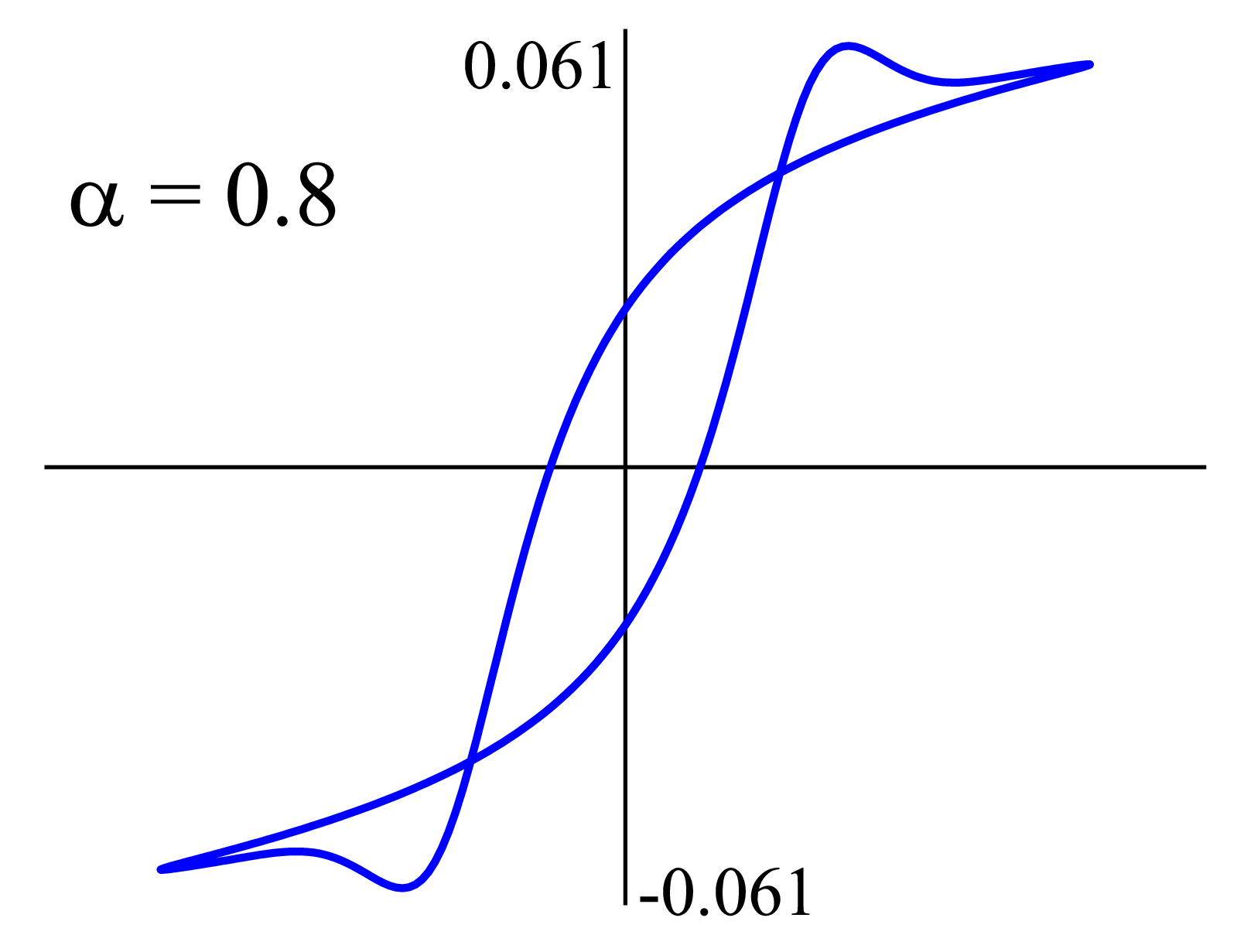}}
\subfloat[$~$]{
\includegraphics[width=0.328\textwidth]{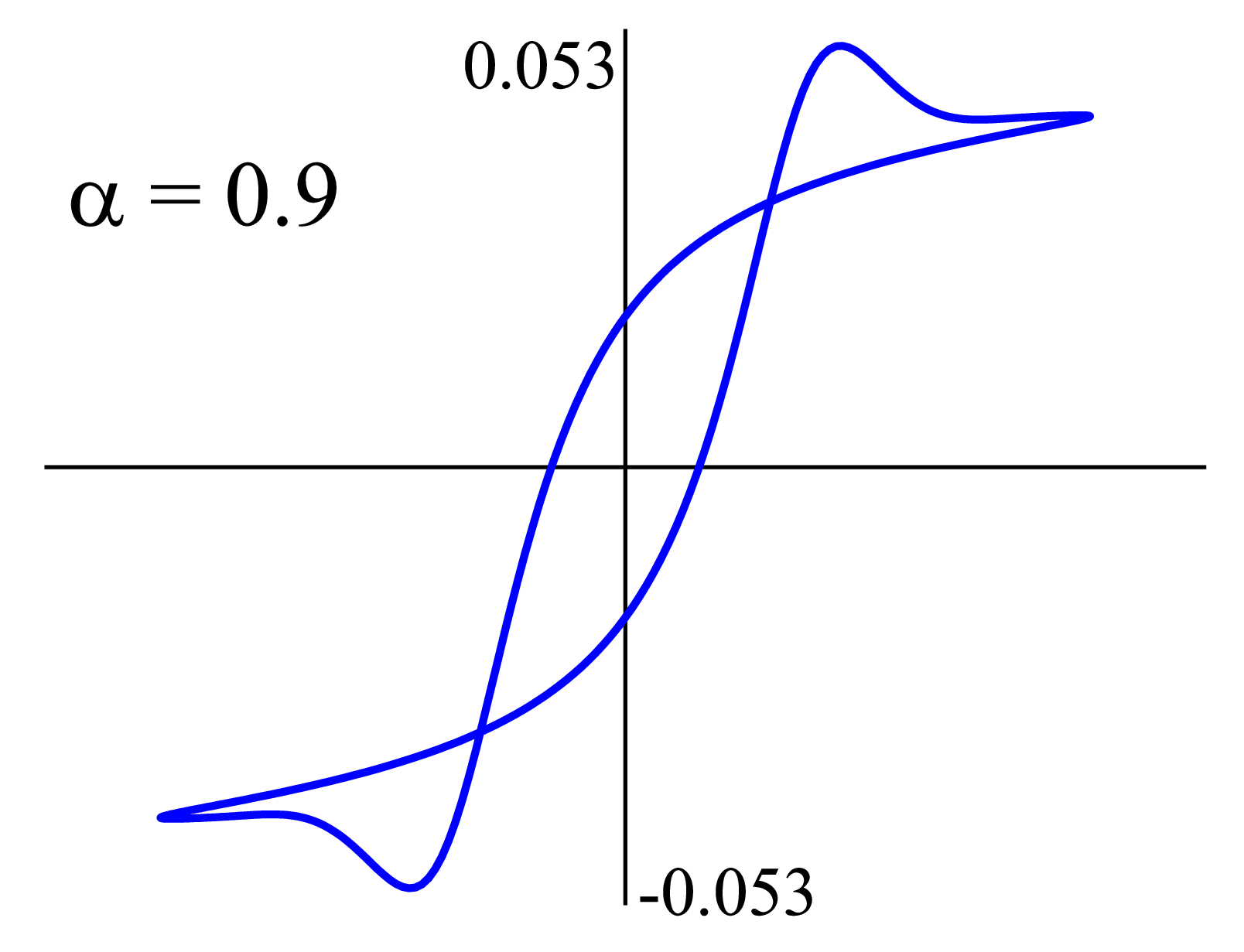}}
\subfloat[$~$]{
\includegraphics[width=0.328\textwidth]{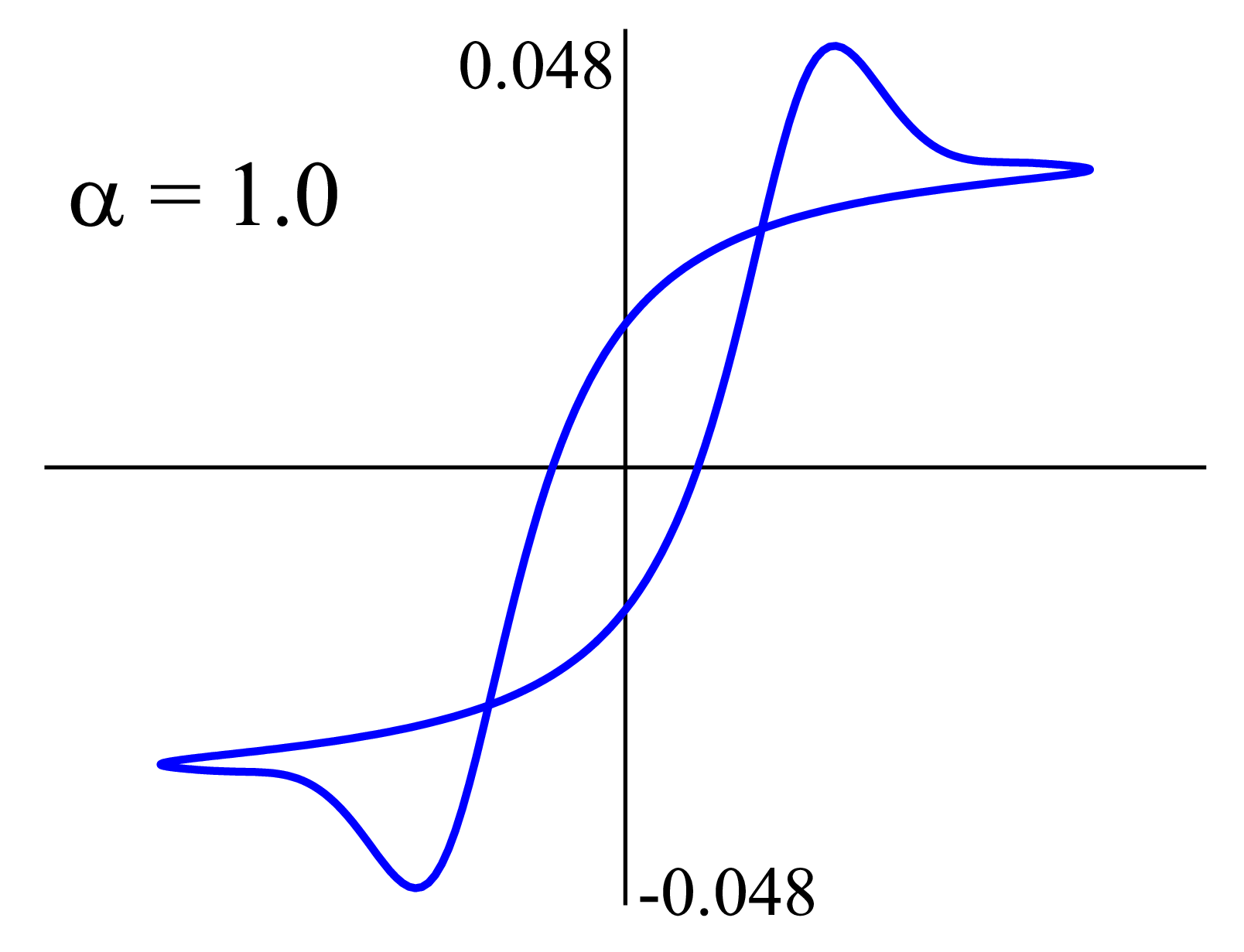}}
\caption{Viscous Lissajous curves for the toy sPTT model ($\epsilon = 1/100$) with various values of $\alpha$ between 0.5 (for which $C(A) = D(A)$) and 1 (for which $C(A) = 0$).  $De = 0.5$ and $Wi = 200$. $y$-axis limits are shown by the numbers adjacent to the ends of the axes. $x$ axes run from -1.04 to 1.04.}
\label{ToysPTT}
\end{figure}

\begin{figure}
\centering
\large
$\tau_{p, 12} \ $ \textit{vs} $\ \dot{\gamma} $

\subfloat[$~$]{
\includegraphics[width=0.328\textwidth]{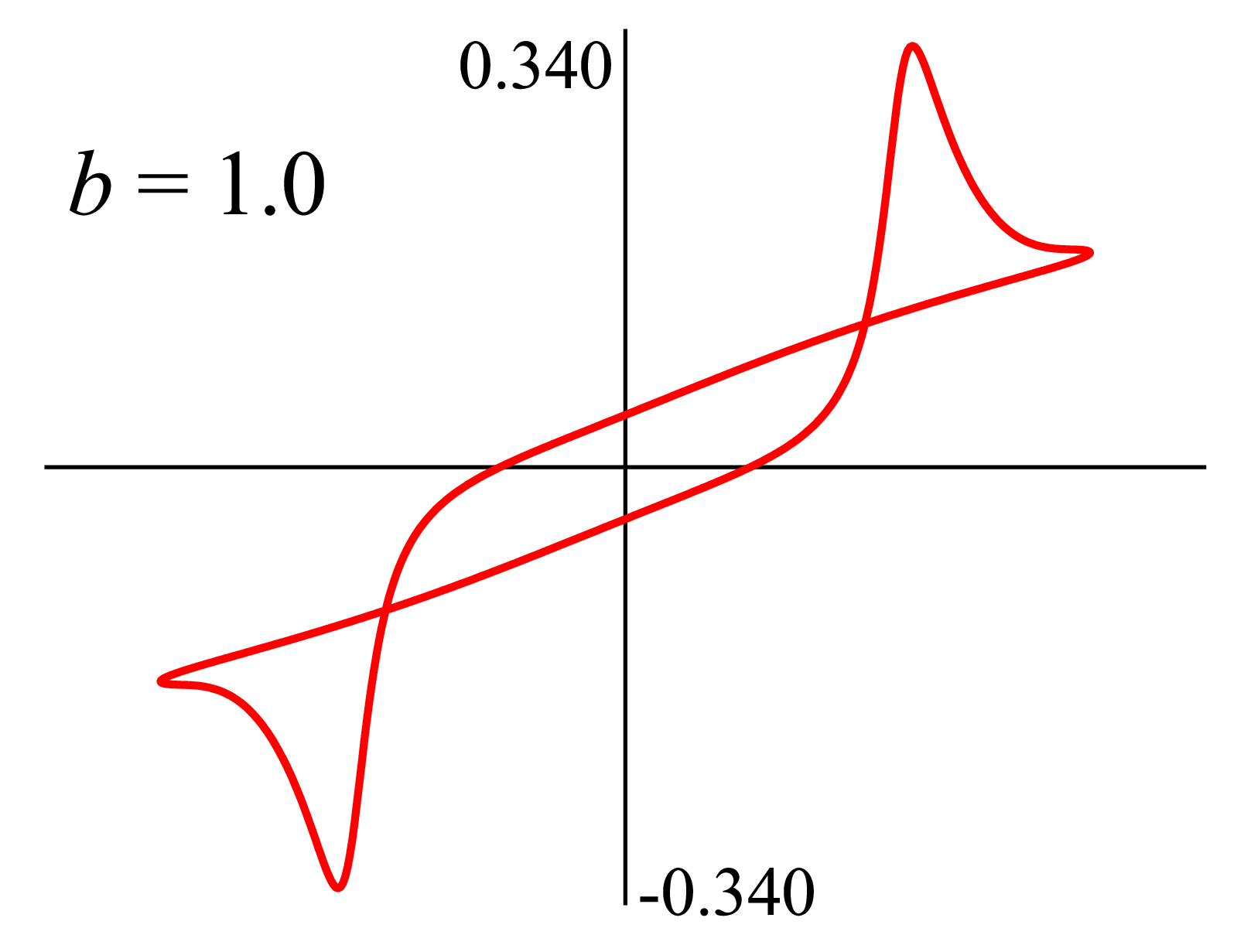}}
\subfloat[$~$]{
\includegraphics[width=0.328\textwidth]{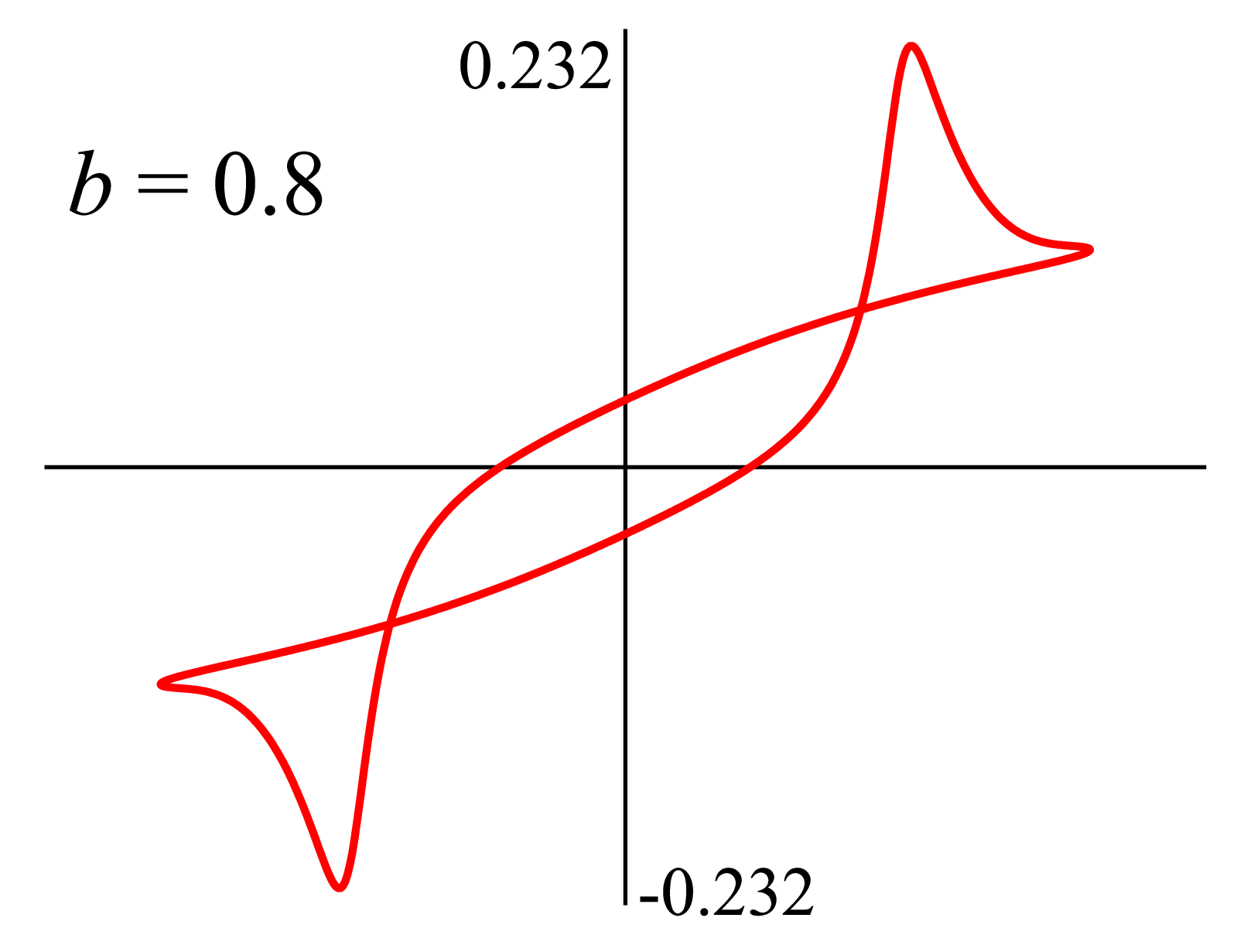}}
\subfloat[$~$]{
\includegraphics[width=0.328\textwidth]{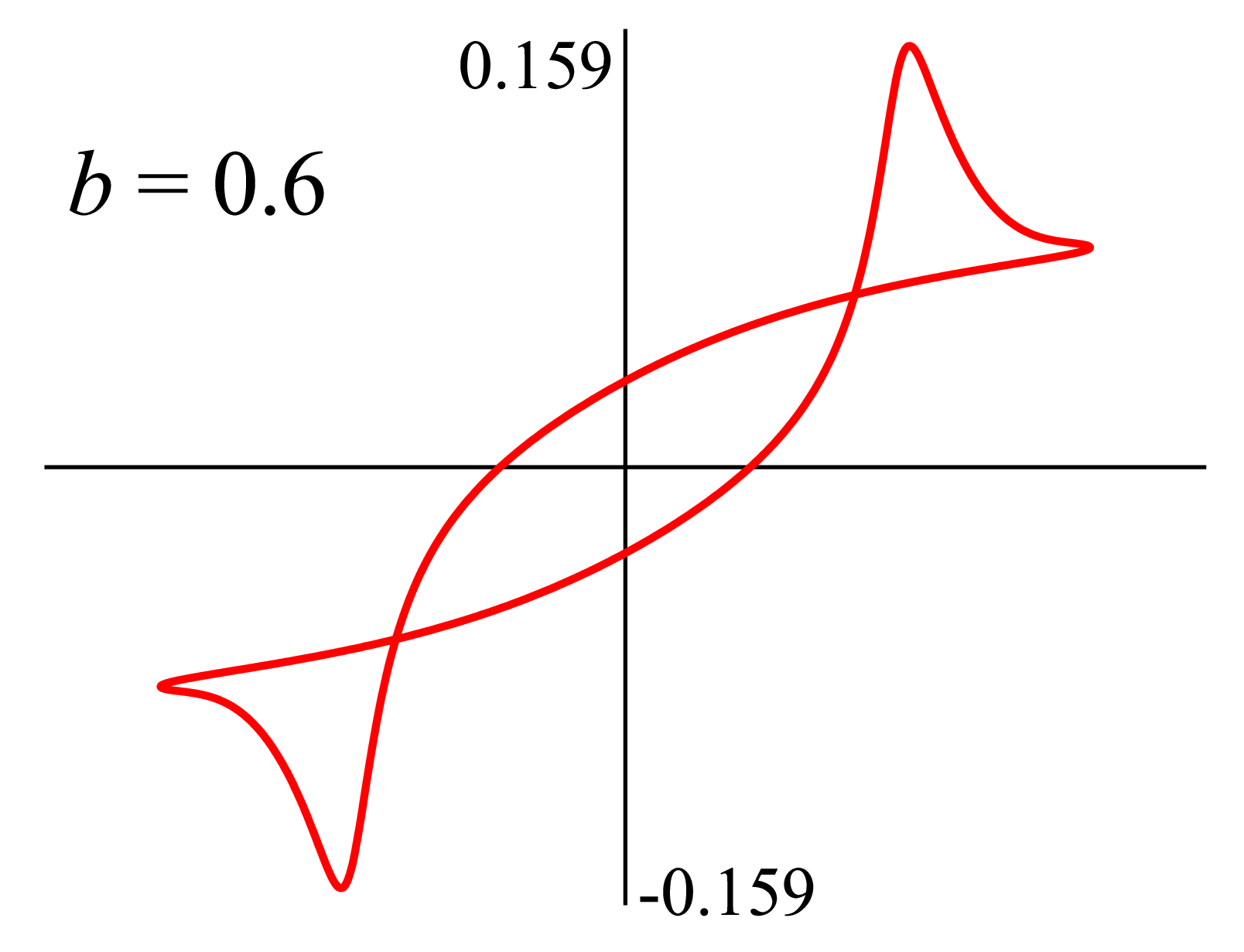}}

\subfloat[$~$]{
\includegraphics[width=0.328\textwidth]{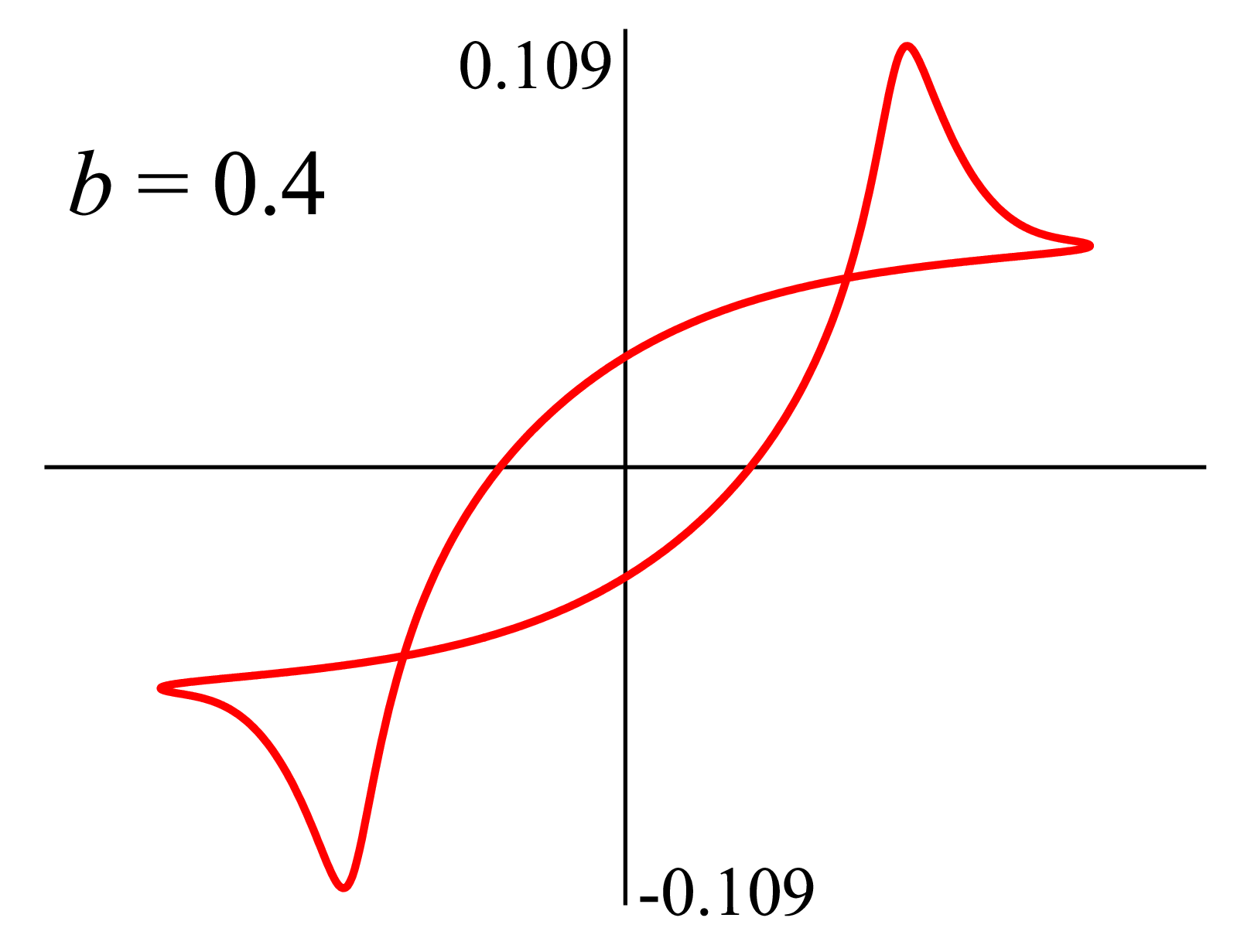}}
\subfloat[$~$]{
\includegraphics[width=0.328\textwidth]{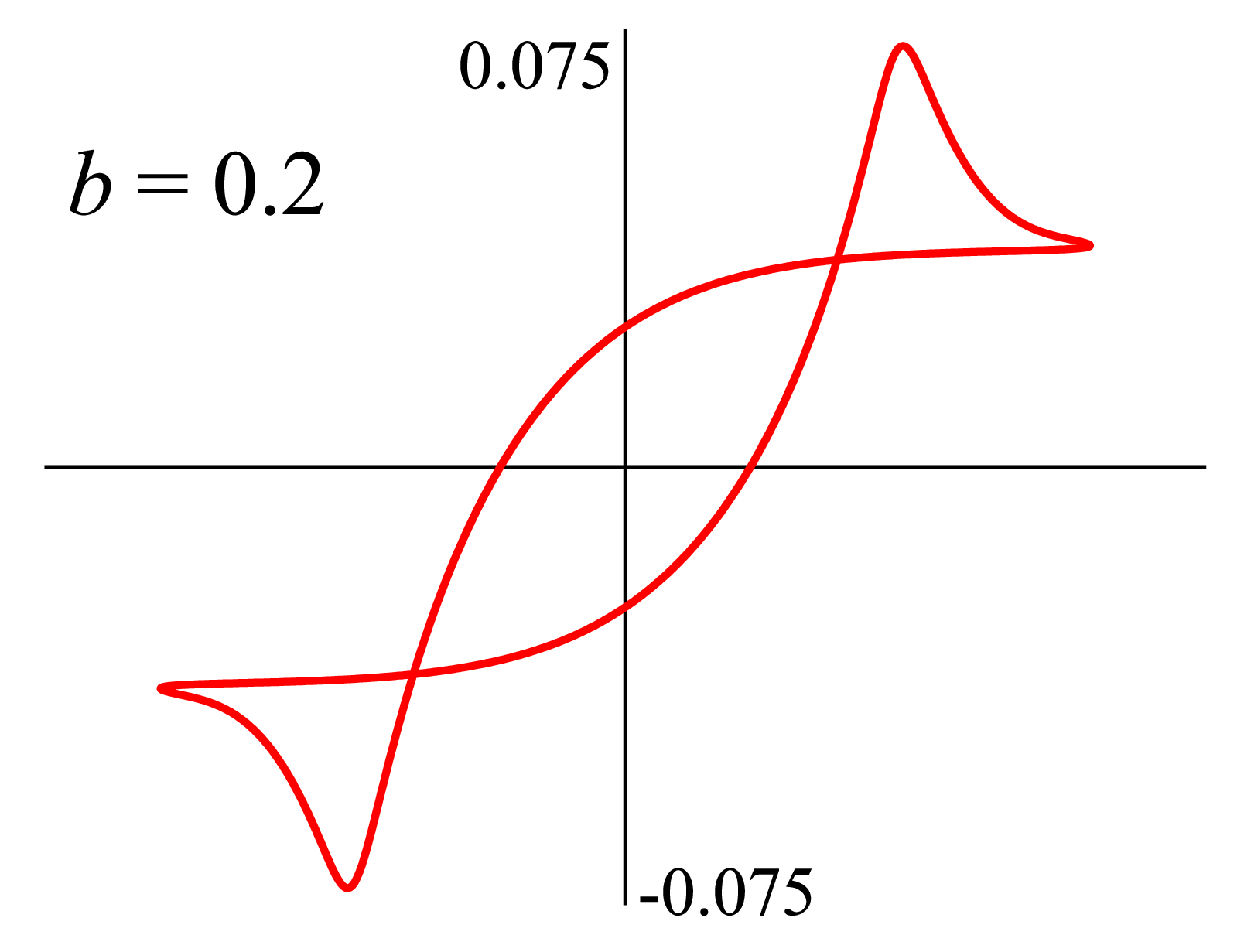}}
\subfloat[$~$]{
\includegraphics[width=0.328\textwidth]{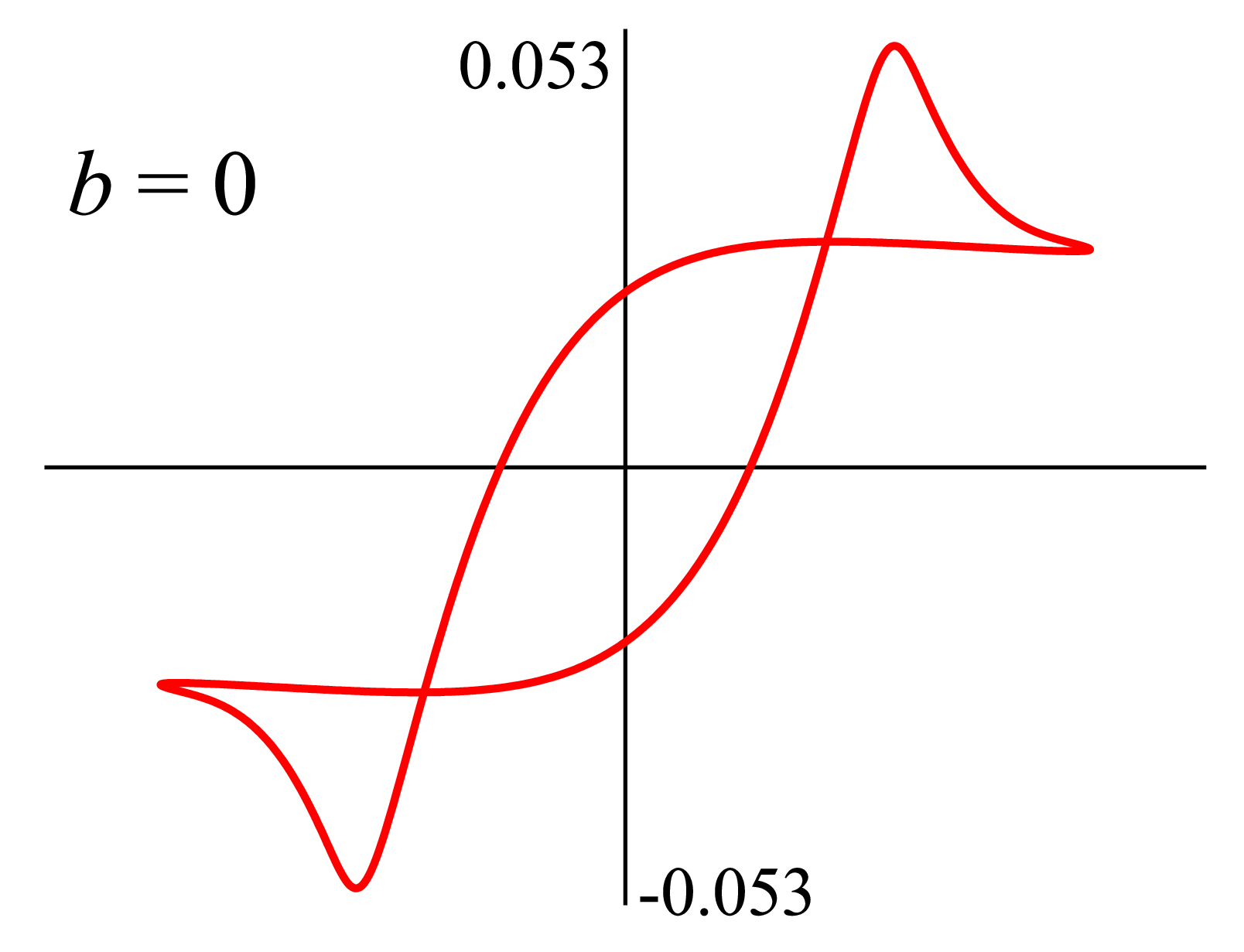}}
\caption{Viscous Lissajous curves for the toy FENE-P model ($L^2 = 100$) with various values of $b$ between 1 and 0.  $De = 1$ and $Wi = 100$. $y$-axis limits are shown by the numbers adjacent to the ends of the axes. $x$ axes run from -1.04 to 1.04.}
\label{ToysFENE}
\end{figure}

\noindent Thus, for $\alpha = 0.5$ we recover the sPTT model (with the value for $\epsilon$ halved), and for $\alpha = 1$ we recover a model with a similar form to the FENE-P model but without $F(A)$ in the $\boldsymbol{\tau}_p$-$\mathsfbi{A}$ relationship.  $\alpha$ therefore, in a way, controls the position of $F(A)$ in the recoil term of the constitutive model. The LAOS response for the toy sPTT model with $\epsilon = 1/100$ at $De = 0.5$ and $Wi = 200$ is presented for various values of $\alpha$ in Figure \ref{ToysPTT}. As $\alpha$ is increased, the stress overshoots and self-intersecting secondary loops become significantly more pronounced, as is expected for systems which exhibit large rates of micro-structural destruction \citep{Davoodi2022, Vargas2023}, and the response begins to appear qualitatively similar to the FENE-P response. This suggests that the primary reason for the FENE-P response exhibiting large stress overshoots in transient flows is that the extensibility function is multiplied by $\mathsfbi{A}$ instead of $(\mathsfbi{A}-\mathsfbi{I})$, and not due to the fact that the extensibility function appears in the $\boldsymbol{\tau}_p$-$\mathsfbi{A}$ relationship, or due the difference in the natures of $F(A)_{\mathrm{FP}}$ and $F(A)_{\mathrm{sPTT}}$. We also show this more explicitly using a toy FENE-P model where the evolution equation for $\mathsfbi{A}$ is unchanged (given by Equation \eqref{feneConf2}), but $\boldsymbol{\tau}_p$ is given by 

\begin{equation}
\boldsymbol{\tau}_p = \frac{a(1-\beta)}{Wi} \bigg(\bigg( \frac{(F(A)_{\mathrm{FP}}}{a} \bigg)^b \mathsfbi{A} - \mathsfbi{I}\bigg)
\end{equation}

\noindent where $0 \leq b \leq 1$. When $b = 1$, the original FENE-P model is obtained, and when $b = 0$, the $\boldsymbol{\tau}_p$-$\mathsfbi{A}$ relationship reverts to the original form given by the Kramers' relation \citep{Kramers1944}, and that used for the sPTT model. The viscous Lissajous curves for the toy FENE-P model with $L^2 = 100$ at $De = 1$ and $Wi = 100$ with varying values of $b$ are shown in Figure \ref{ToysFENE}. The stress overshoots and self-intersecting loops are observed for all values of $b$, indicating that the presence of $F(A)$ in the $\boldsymbol{\tau}_p$-$\mathsfbi{A}$ relationship does not explicitly cause pronounced shear stress overshoots in the transient response. As $b \rightarrow 0$,  the region directly after the stress overshoot becomes significantly flatter, and the Lissajous curves resemble more those of constitutive models derived for wormlike micellar systems. We also note for both Figures \ref{ToysFENE} and \ref{ToysPTT} that the scale of the $y$ axis in each plot is not fixed. The parameters $\alpha$ and $b$ do affect significantly the maximum stresses reached in the oscillation. In this sense, the stress overshoots that we are discussing here are relative.

\begin{figure}
\centering
\includegraphics[width=0.65\textwidth]{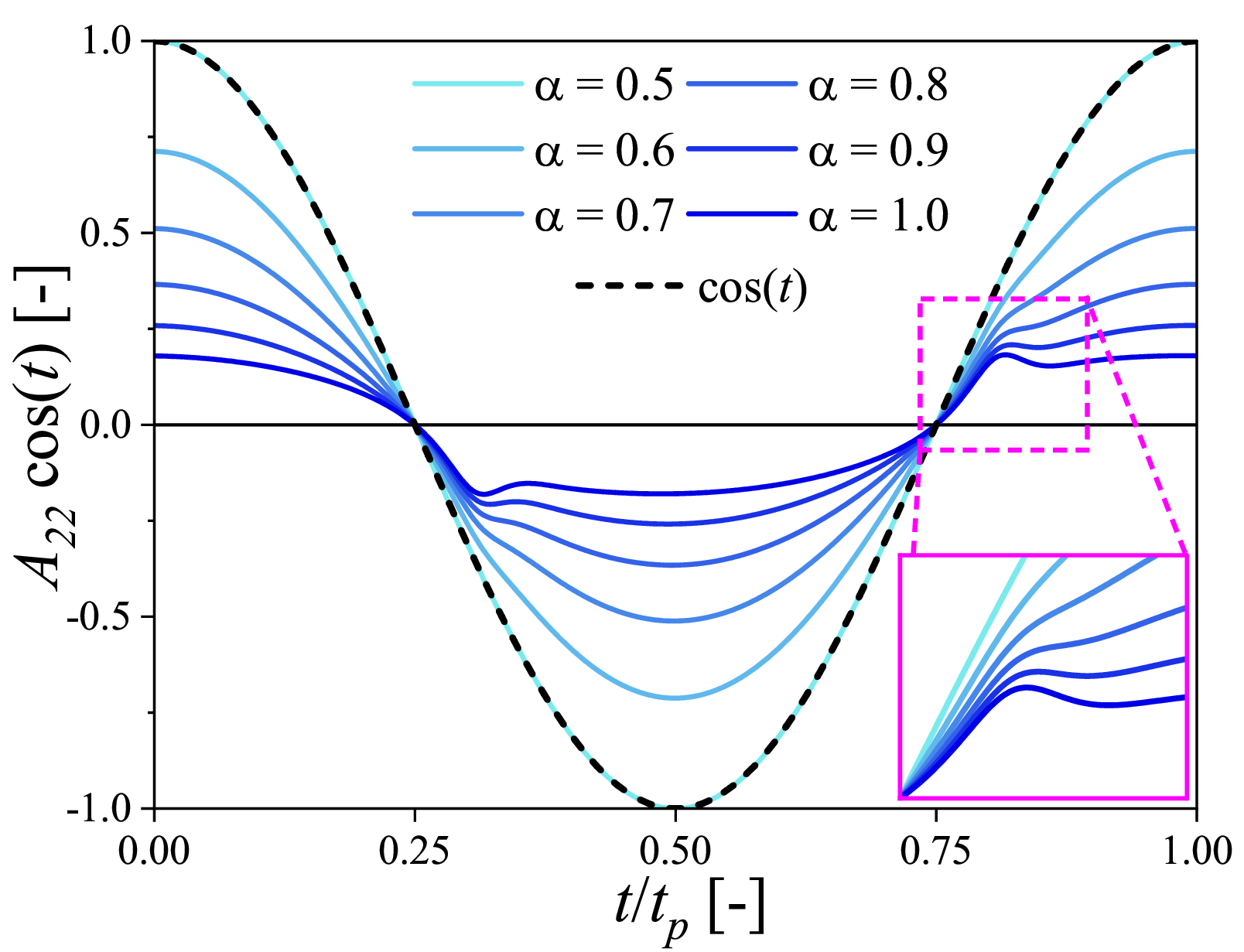}
\caption{$A_{22} \mathrm{cos}(t)$ \textit{vs} time for the toy sPTT model ($\epsilon = 1/100$) for $De = 0.5$ and $Wi = 200$ during one oscillation with varying $\alpha$.  Zoomed inset highlights the point of the overshoot in $A_{22}$ when $\alpha \rightarrow 1$. }
\label{toypttA22}
\end{figure}

In the standard form of the sPTT model ($\alpha = 0.5$), the growth term for $A_{12}$ due to the rate of deformation is equal to $(Wi/De) \mathrm{cos}(t)$ since $A_{22} = 1$. As $\alpha \rightarrow 1$, there is significant deviation of $A_{22}$ from unity during the oscillation, and hence deviation of the growth term from a pure cosine wave. Since the time rate of change of $A_{12}$ is governed by a balance between the growth and the elastic recoil, the deviation of $A_{22}$ from unity contributes significantly to the occurrence of pronounced stress overshoots and self-intersecting secondary loops in LAOS. Ultimately, the origin of this behaviour lies in the use of the upper-convected time derivative combined with the specific positioning of $F(A)$ within the constitutive model (i.e. $F(A)(\mathsfbi{A}-\mathsfbi{I})$ \textit{versus} $(F(A)\mathsfbi{A}-\mathsfbi{I})$). Figure \ref{toypttA22} shows $A_{22} \mathrm{cos}(t)$ during one oscillation for the toy sPTT model with $\epsilon = 1/100$ at $Wi = 200$ and $De = 0.5$ for various values of $\alpha$. Clear overshoots of $A_{22}\mathrm{cos}(t)$ are observed at roughly $t/t_p = 0.3$ and $t/t_p = 0.8$ as $\alpha \rightarrow 1$. 

To highlight more clearly the role of $A_{22}$ in the generation of the pronounced stress overshoots and self-intersecting secondary loops, we define the growth and recoil terms for the evolution of $A_{12}$ as

\begin{equation}
Q_g = \bigg(\frac{Wi}{De}A_{22}\bigg) \mathrm{cos}(t) \label{growth}
\end{equation} 

\begin{equation}
Q_r = \bigg(\frac{1}{De}\bigg)F(A) A_{12} \label{recoil}
\end{equation} 

\begin{figure}
\centering
\subfloat[$~$]{\includegraphics[width=0.48\textwidth]{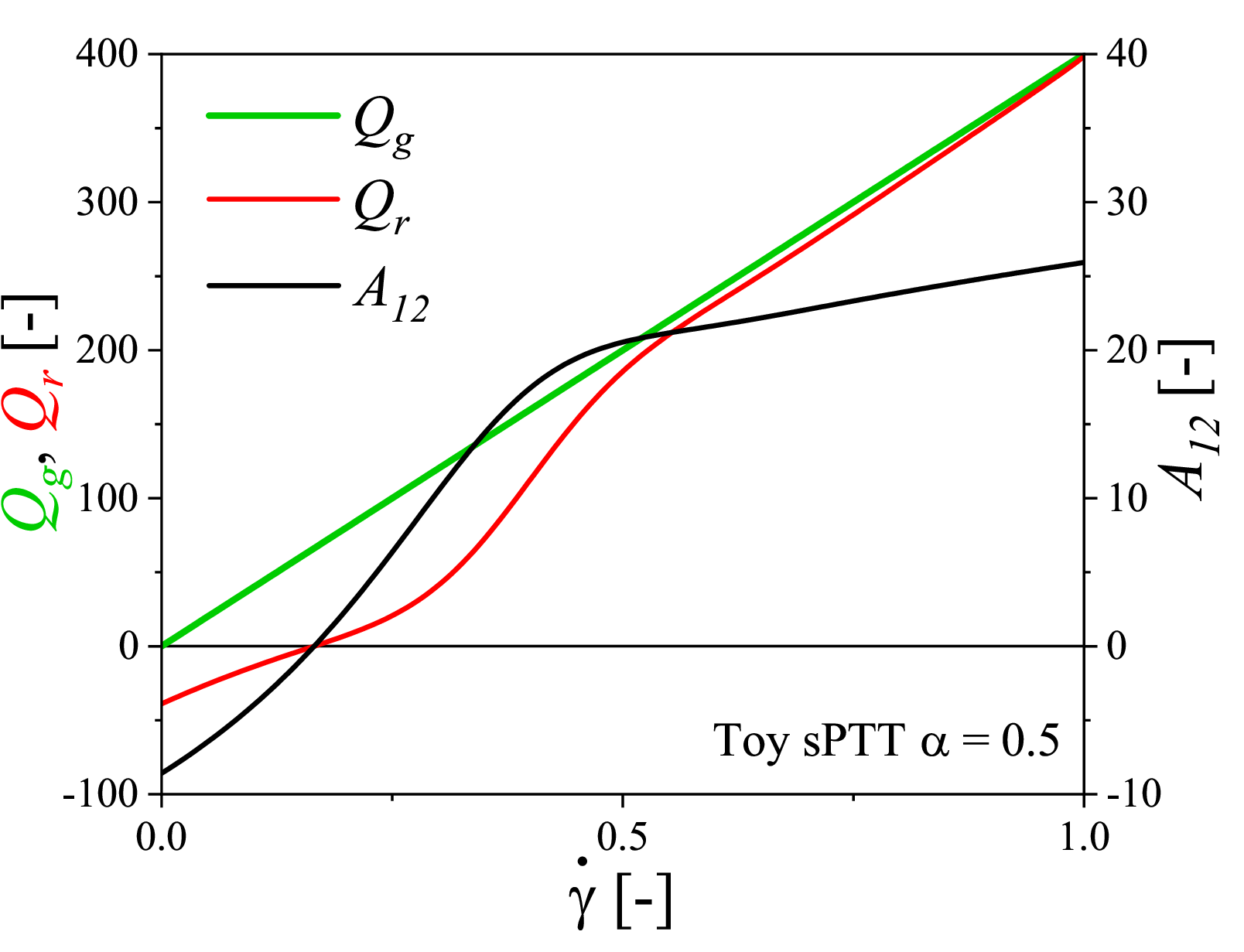}}
\subfloat[$~$]{\includegraphics[width=0.48\textwidth]{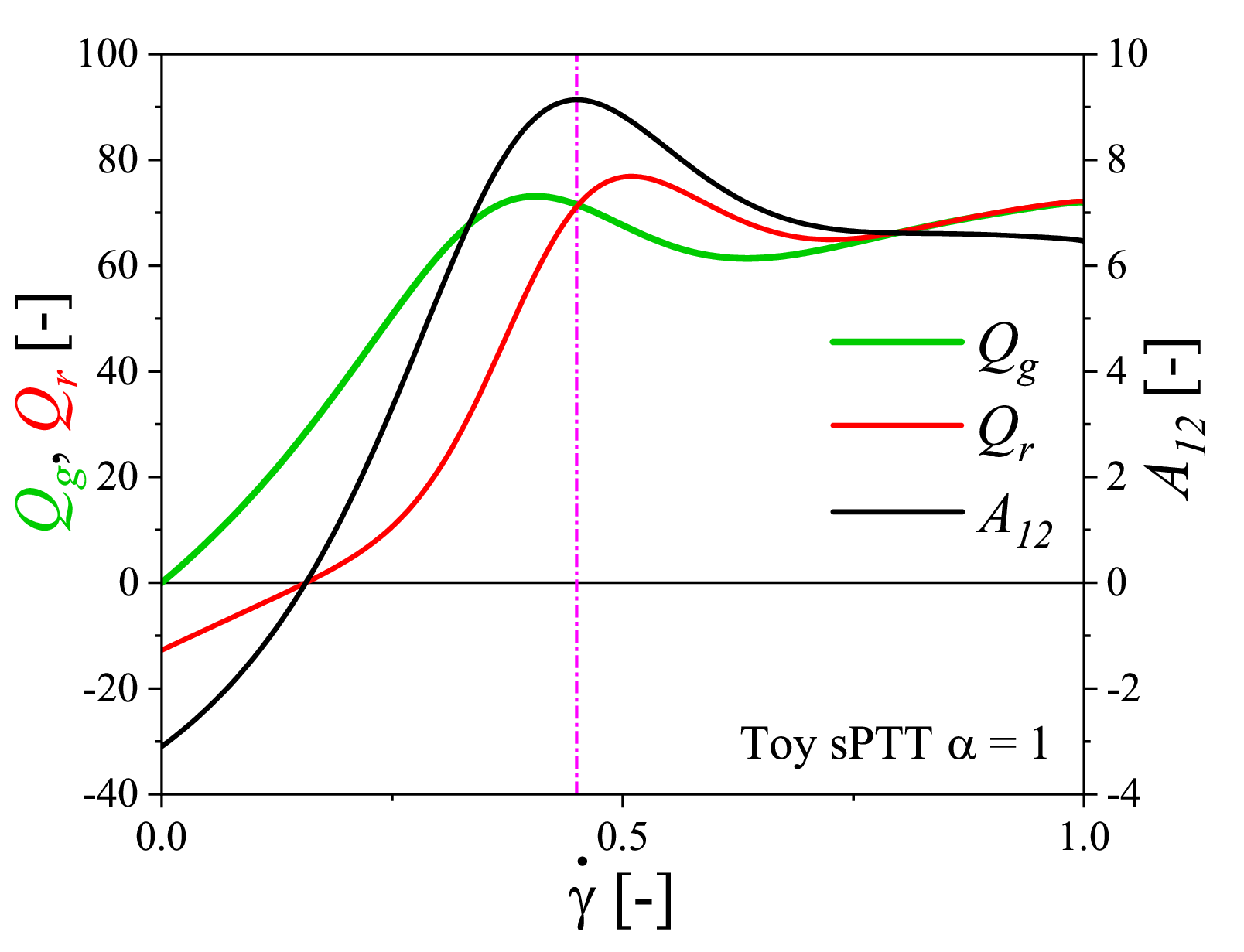}}
\caption{$Q_g$, $Q_r$, and $A_{12}$ \textit{vs} $\dot{\gamma}$ in one quarter of an oscillation for the toy sPTT model ($\epsilon = 1/100$) for $De = 0.5$ and $Wi = 200$ with (a) $\alpha = 0.5$ and (b) $\alpha = 1$. Dashed vertical line shows the cross-over point of $Q_g$ and $Q_r$, and hence also the maxima of $A_{12}$.}
\label{fig:pttRates}
\end{figure}

\noindent Note that these are just, respectively, the first and second terms on the right hand side of Equations \eqref{feneposcillatoryshear}b and \eqref{pttoscillatoryshear}b. Figure \ref{fig:pttRates} shows $Q_g$, $Q_r$, and $A_{12}$ for the toy sPTT model with $\epsilon = 1/100$ when $\alpha = 0.5$ and $\alpha = 1$, respectivly, during a quarter of the oscillation period. During this quarter of the period, $\dot{\gamma}$ is increasing from 0 to 1. Naturally, stress overshoots are observed in this region for the case when $Q_r > Q_g$, which is seemingly the case for the toy sPTT model when $\alpha \rightarrow 1$ and thus $C(A)=1$.  Whilst it is difficult to assess explicitly the role that $A_{22}$ plays in the evolution of $Q_g$ and $Q_r$ due to the strong coupling in the equations, it is clearly the case that the sharp decrease in $Q_g$, which can only be caused by a change in $A_{22}$, occurs before any observed decrease in $Q_r$, which allows for a region where $Q_r$ is significantly larger than $Q_g$. In Figure \ref{fig:pttRates}(b), the point of intersection of $Q_g(\dot{\gamma})$ and $Q_r(\dot{\gamma})$ of course defines the exact position for the maxima of $A_{12}$ (and hence $\tau_{p, 12}$) associated with the stress overshoot. As $\dot{\gamma} \rightarrow 1$, there is a large region where $Q_g \approx Q_r$, which as mentioned might represent the system approaching steady-shear. 

\begin{figure}
\centering
\includegraphics[width=0.48\textwidth]{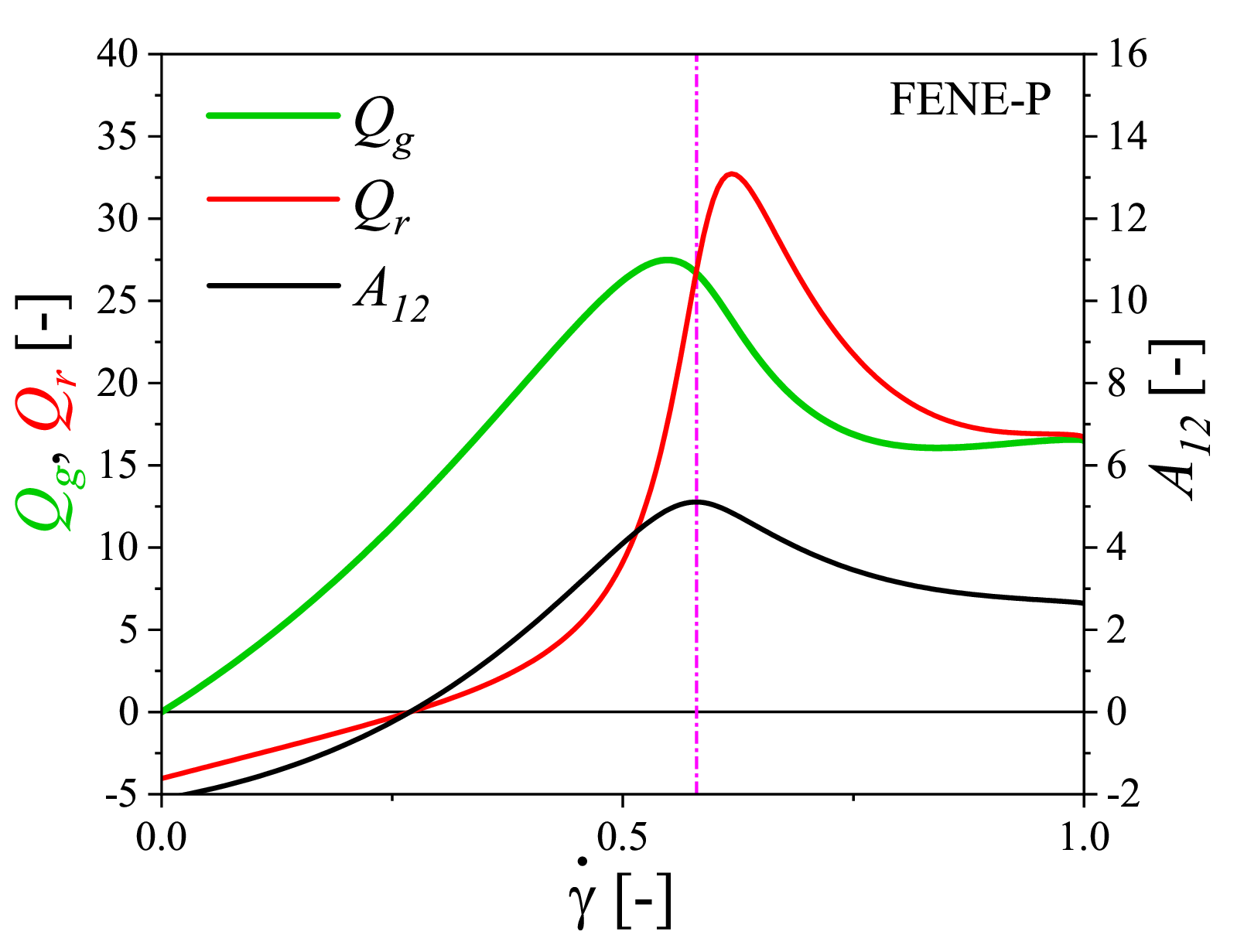}
\caption{$Q_g$, $Q_r$, and $A_{12}$ \textit{vs} $\dot{\gamma}$ in one quarter of an oscillation for the FENE-P model ($L^2 = 100$) for $De = 1$ and $Wi = 100$. Dashed vertical line shows the cross-over point of $Q_g$ and $Q_r$, and hence also the maxima of $A_{12}$.}
\label{FENEPRates}
\end{figure}

Figure \ref{FENEPRates} shows $Q_g$, $Q_r$, and $A_{12}$ for the FENE-P model (or toy FENE-P model with $b=1$) with $L^2 = 100$ at $De = 1$ and $Wi = 100$. Again, the decrease in $Q_g$, caused by the time-dependence of $A_{22}$, is significant and is primarily responsible for the generation of the pronounced stress overshoots. $Q_r$ grows much more sharply for the FENE-P model than for the toy sPTT model due to the difference between $F(A)_{\mathrm{FP}}$ and $F(A)_{\mathrm{sPTT}}$. The overshoots in the FENE-P response are likely exacerbated somewhat by this. We should note that if $F(A)_{\mathrm{sPTT}}$ is replaced by $F(A)_{\mathrm{FP}}$ in the original sPTT model, stress overshoots can still occur even when $A_{22} = 1$ solely due to the increased non-linearity of $Q_r$. However, these overshoots are significantly smaller than those observed when $Q_g$ is non-linear. This replacement of $F(A)_{\mathrm{sPTT}}$ by $F(A)_{\mathrm{FP}}$ in the sPTT model essentially corresponds to a toy FENE-CR model, which we explore in the Appendix \ref{appA}. 

\begin{figure}
\centering
\large
$\tau_{p, 12} \ $ \textit{vs} $\ \dot{\gamma} $

\includegraphics[width=0.65\textwidth]{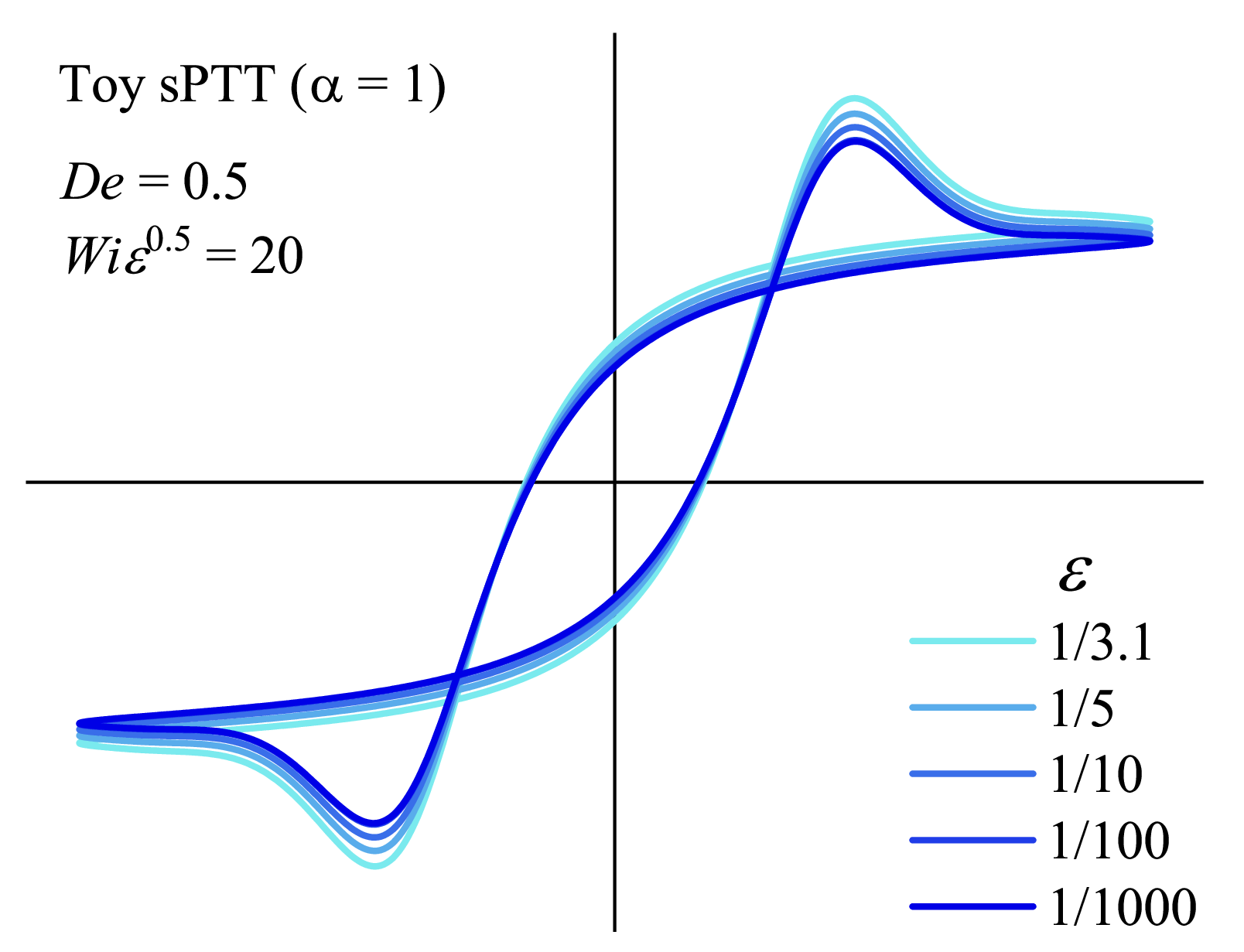}
\caption{Viscous Lissajous curves for the toy sPTT model for $\alpha = 1$, $De = 0.5$ and $Wi\sqrt{\epsilon} = 20$, with varying $\epsilon $. }
\label{Scalingtest}
\end{figure}

As discussed in Section \ref{results1}, the standard sPTT model has  universal solutions both in steady-shear and in LAOS (for constant values of $De$) for constant values of $Wi\sqrt{\epsilon}$.  However, the FENE-P model LAOS response only has universal solutions for constant values of $Wi/(aL)$ for large enough values of $L^2$.  In Section \ref{results1}, we show that replacing $F(A)_{\mathrm{sPTT}}$ with $F(A)_{\mathrm{FP}}$ in the sPTT model causes a breakdown of this universality, and, therefore, that the existence of universal solutions is dependent on the specific form of $F(A)$ used. Here, we use the toy models to also investigate the effect of the positioning of $F(A)$ on the scaling of the LAOS response (i.e. $F(A)(\mathsfbi{A}-\mathsfbi{I})$ \textit{versus} $(F(A)\mathsfbi{A}-\mathsfbi{I})$). In Figure \ref{Scalingtest}, we show the LAOS response of the toy sPTT model with $\alpha = 1$ (such that $C(A) = 1$) for $De = 0.5$ and $Wi\sqrt{\epsilon} = 20$ for various values of $\epsilon$. The solution seems to only be universal for small enough values of $\epsilon$, which is strongly reminiscent of the FENE-P model response presented in Section \ref{results1}.  Note that $\alpha = 1$ makes the toy sPTT model appear in a similar form to the FENE-P model in the network model framework. As highlighted by \cite{Tanner2006} and \cite{Davoodi2022}, the generic network model (Equation \eqref{pttrates}) can be written for steady and homogeneous flows in stress tensor form as

\begin{equation}
D(\tau_p) \boldsymbol{\tau}_p  + Wi \, \overset{\kern0em\triangledown}{\boldsymbol{\tau}}_p =2 (1-\beta)\mathsfbi{D} - \frac{D(\tau_p)(1-\beta)}{Wi}\bigg(1 - \frac{C(\tau_p)}{D(\tau_p)}\bigg) \mathsfbi{I} \label{toypttstress}
\end{equation}

\noindent Note here that both the time and the velocity gradient are made dimensionless by $\dot{\gamma}$, and so the entire upper-convected time derivative is multiplied by $Wi$.  For $D(A) \neq C(A)$, the last term on the right hand side of Equation \eqref{toypttstress} is nonzero.  For the toy sPTT model with $\alpha = 1$, $D(\tau_p) = F(\tau_p)_{\mathrm{sPTT}}$ and $C(\tau_p) = 1$. In this case, the solution of Equation \eqref{toypttstress} in SSSF is given by the following system

\begin{subequations}
\begin{equation}
\bigg[ 1 + \frac{Wi \epsilon}{(1-\beta)}(\tau_{p,11}+\tau_{p,22}+\tau_{p,33}) \bigg] \tau_{p,11} -2Wi\dot{\gamma}\tau_{p,12}=-\epsilon(\tau_{p,11}+\tau_{p,22}+\tau_{p,33})
\end{equation}

\begin{equation}
\bigg[ 1 + \frac{Wi \epsilon}{(1-\beta)}(\tau_{p,11}+\tau_{p,22}+\tau_{p,33}) \bigg] \tau_{p,12} -Wi\dot{\gamma}\tau_{p,22}= (1-\beta)\dot{\gamma}
\end{equation}

\begin{equation}
\bigg[ 1 + \frac{Wi \epsilon}{(1-\beta)}(\tau_{p,11}+\tau_{p,22}+\tau_{p,33}) \bigg] \tau_{p,22} =-\epsilon(\tau_{p,11}+\tau_{p,22}+\tau_{p,33})
\end{equation}

\begin{equation}
\bigg[ 1 + \frac{Wi \epsilon}{(1-\beta)}(\tau_{p,11}+\tau_{p,22}+\tau_{p,33}) \bigg] \tau_{p,33} =-\epsilon(\tau_{p,11}+\tau_{p,22}+\tau_{p,33})
\end{equation}
\label{networksteadyshear}
\end{subequations}

\noindent For finite values of $Wi \,\epsilon$ (or $Wi\sqrt{\epsilon}$), but in the limit that $\epsilon \rightarrow 0$, the term $-\epsilon(\tau_{p,11}+\tau_{p,22}+\tau_{p,33})$ on the right hand side of the diagonal components approaches zero, in which case, ${\tau_{p, 22} = \tau_{p, 33} \rightarrow 0}$ and Equation \ref{networksteadyshear} reduces to

\begin{subequations}
\begin{equation}
\bigg[ 1 + \frac{Wi \epsilon}{(1-\beta)}(\tau_{p,11}) \bigg] \tau_{p,11}-2Wi\dot{\gamma}\tau_{p,12}=0
\end{equation}

\begin{equation}
\bigg[ 1 + \frac{Wi \epsilon}{(1-\beta)}(\tau_{p,11}) \bigg] \tau_{p,12} = (1-\beta)\dot{\gamma}
\end{equation}
\label{networksteadyshear2}
\end{subequations}

\noindent which is the solution for the original sPTT model. As discussed in Section \ref{results1}, the solution to Equation \ref{networksteadyshear2} only depends on the parameter $Wi\sqrt{\epsilon}$ (something which is also seen by introducing the new variable $x = Wi \, \epsilon \, \tau_{p,11}/(1-\beta)$ into Equation \eqref{networksteadyshear2}). It highlights that the breakdown of the universal scaling with $Wi\sqrt{\epsilon}$ can be caused, at least in SSSF, by setting $D \neq C$ even when $F(A)_{\mathrm{sPTT}}$ is used for the extensibility function. Therefore,  considering also the scaling analysis for the FENE-CR model presented in Section \ref{results1}, the breakdown of the universal scaling for low values of $L^2$ in the FENE-P LAOS response is likely a combined effect of the functional form of $F(A)$ and its position in the conformation tensor form of the constitutive model (i.e. $F(A)(\mathsfbi{A}-\mathsfbi{I})$ \textit{versus} $(F(A)\mathsfbi{A}-\mathsfbi{I})$).  We note briefly that, for the FENE-P model, the extra term on the right hand side of Equation \eqref{toypttstress} is multiplied by the Lagrangian derivative of $1/F(\tau_p)$ due to the presence of $F(A)$ in the $\boldsymbol{\tau}_p$-$\mathsfbi{A}$ relationship, which is why this term does not affect the SSSF solution for the FENE-P model.

For clarity, we will briefly summarise this subsection. When $D(A) \neq C(A)$ in the network model framework, temporal changes in $A_{22}$ cause both non-linear growth and non-linear recoil of $A_{12}$ simultaneously. This causes a region where $Q_r$ is significantly larger than $Q_g$ which manifests as pronounced stress overshoots in the Lissajous curves.  The presence of $F(A)$ in the $\boldsymbol{\tau}_p$-$\mathsfbi{A}$ relationship does not seem to have much effect on the relative stress overshoots. The sPTT model response scales universally due to both the specific functional form of $F(A)$ used in the model and the fact that $F(A)$ is on the outside of the brackets in the recoil term (or $D(A) = C(A)$ in the network model framework). The FENE-P response does not scale universally in LAOS due to the specific form of $F(A)$ used and the fact that $F(A)$ is on the inside of the brackets in the recoil term (or $D(A) \neq C(A)$ in the network model framework).

\subsection{Sequence of Physical Processes (SPP)}\label{results4}

Previously, we have discussed the responses of the FENE-P and sPTT models predominantly in terms of the nature of the underlying ODEs being solved. Here, we analyse the responses of the models in terms of the physical phenomena being represented by the models. To do this, we employ the SPP analysis  \citep{Rogers2011}, which will now be briefly outlined. 

Whilst the moduli $G^{'}$ and $G^{''}$ are time-independent during SAOS, and their values can be found from simple Fourier analysis, these moduli are transient during LAOS for a non-linear viscoelastic material or constitutive model. The SPP framework identifies, at each instant, these transient moduli, denoted as $G^{'}_t(t)$ and $G^{''}_t(t)$, by utilising the Frenet-Serret theorem. In this approach, each point in the oscillation is given by a position vector in the strain-rate-stress space ${\boldsymbol{P}(t)~=~\langle \gamma(t),\ \dot{\gamma}(t),\ \tau_{p,12}(t)\rangle}$. Three additional vectors are then defined at each position on the Lissajous curve, which are used to identify the transient moduli. The tangent vector ${\boldsymbol{T}(t) = \dot{\boldsymbol{P}}(t)/|\dot{\boldsymbol{P}}(t)|}$ points in the direction of the instantaneous trajectory (the overdot here denoting total differentiation with respect to time). The normal vector $\boldsymbol{N}(t) = \dot{\boldsymbol{T}}(t)/|\dot{\boldsymbol{T}}(t)|$ points to the center of the curvature of the instantaneous trajectory. Finally, the bi-normal vector $\boldsymbol{B}(t) = \boldsymbol{T}(t)\times\boldsymbol{N}(t)$ points in the direction normal to the plane in which the instantaneous trajectory sits (the osculating plane). The transient moduli are then defined (noting that $\omega=1$ in our case due to the equations being solved in non-dimensional form) using the components of $\boldsymbol{B}$ as

\begin{subequations}
\begin{gather}
G^{'}_t(t) = \frac{\dif \tau_{p,12}}{\dif \gamma} = - \frac{B_{\gamma}(t)}{B_{\tau_{p,12}}(t)} \\
G^{''}_t(t) = \frac{\dif \tau_{p,12}}{\dif \dot{\gamma}} = - \frac{B_{\dot{\gamma}}(t)}{B_{\tau_{p,12}}(t)}
\end{gather}
\end{subequations}

\noindent The stress response is then reconstructed as

\begin{equation}
\tau_{p,12}(t) = G^{'}_t\gamma + G^{''}_t\dot{\gamma} + \tau_{p,12}^d
\end{equation}

\noindent where $\tau_{p,12}^d$ is the displacement stress. For a more detailed explanation of the SPP framework, the reader is referred to the works of \citet{Rogers2011, Rogers2012, Rogers2017, Lee2017}. 

The time-dependent behaviour of $G^{'}_t$ and $G^{''}_t$ during the (half) oscillation inform us of underlying physical phenomena occurring in the stress response. Increasing values of $G^{'}_t$ represents intra-cycle strain-hardening whilst decreasing values of $G^{'}_t$ represents intra-cycle strain-softening. Similarly, for the viscous modulus, increasing values of $G^{''}_t$ represents intra-cycle shear-thickening whereas decreasing values of $G^{''}_t$ represents intra-cycle shear-thinning. Negative instantaneous values of $G^{'}_t$ can be thought of as representing elastic recoil, and negative instantaneous values of $G^{''}_t$ can be thought of as representing viscous backflow \citep{Rogers2017, Choi2019, Donley2022}.  In this subsection, we perform the SPP analysis for the sPTT (FENE-P) responses for a fixed $De \ (De/a) = 0.5$ with $L^2 = 1/\epsilon = 100$. We also investigate the responses of the toy models outlines in Section \ref{results4}. The SPP freeware\footnote{\url{https://publish.illinois.edu/rogerssoftmatter/freeware/}} for MATLAB was used for the analysis, which was kindly provided to us by the developers. Central differencing was used for differentiation of the stress response, and a single, limit cycle, period with 401 time points was used for the analysis.

\begin{figure}
\centering
\subfloat[$~$]{
\includegraphics[width=0.475\textwidth]{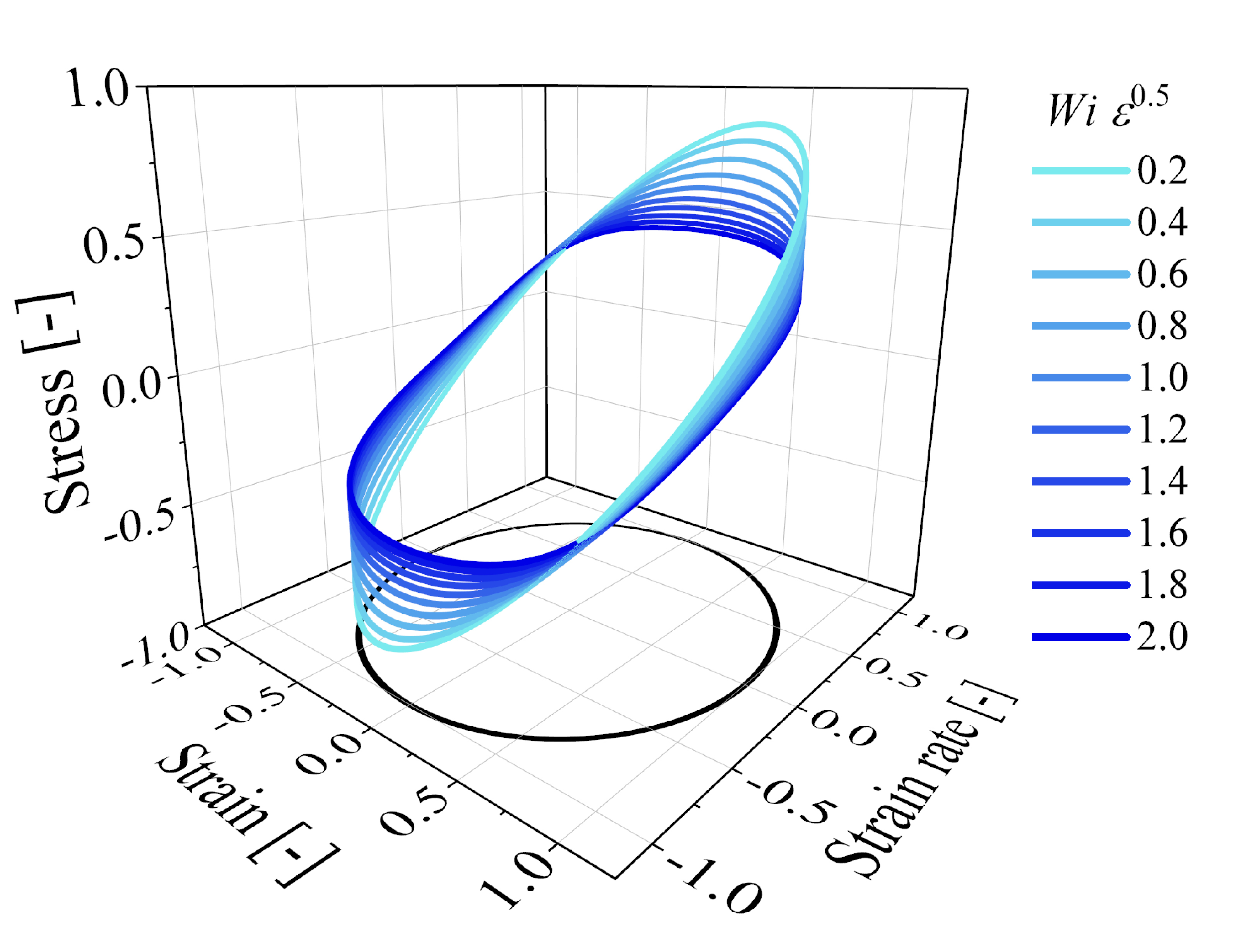}}
\subfloat[$~$]{
\includegraphics[width=0.475\textwidth]{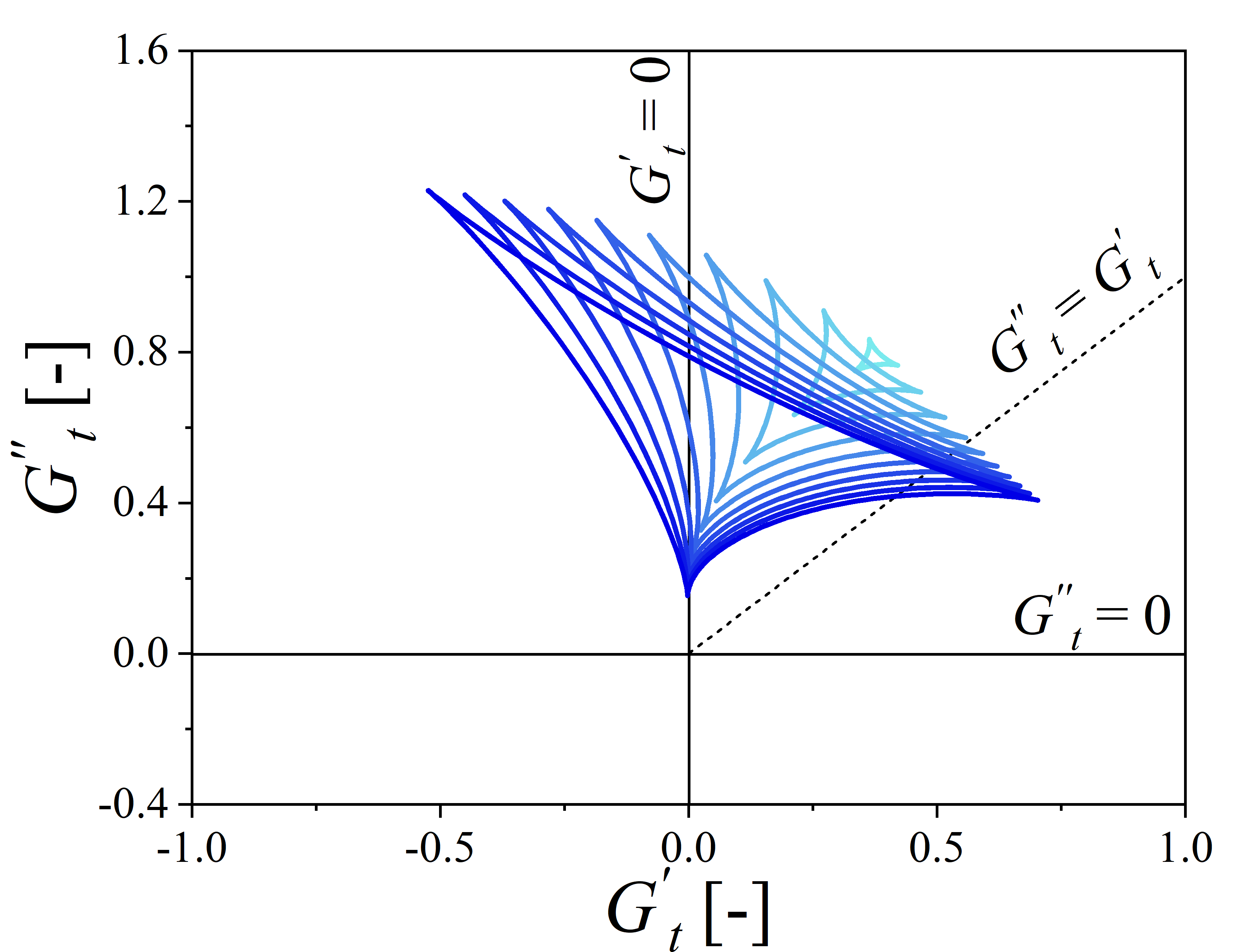}}

\subfloat[$~$]{
\includegraphics[width=0.475\textwidth]{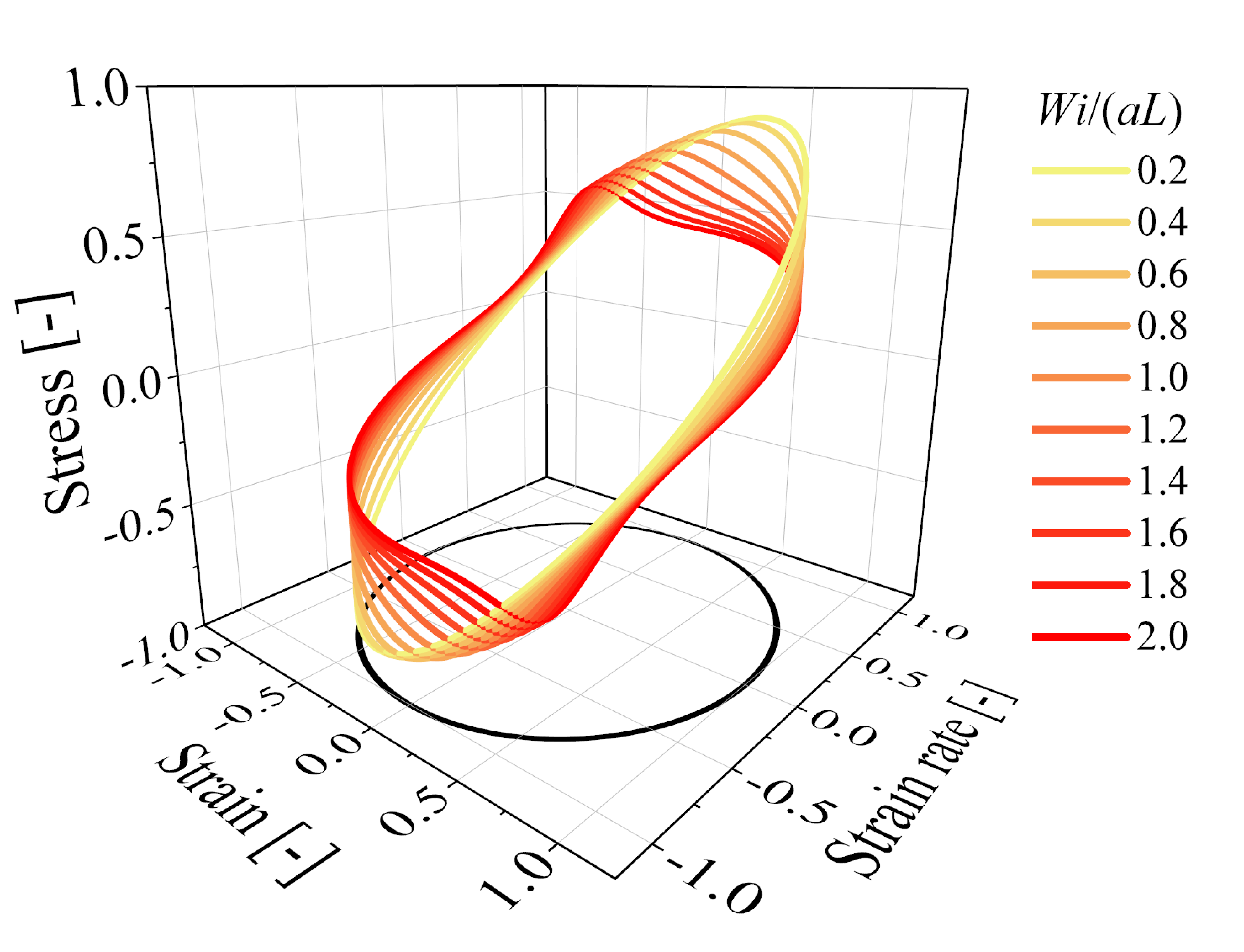}}
\subfloat[$~$]{
\includegraphics[width=0.475\textwidth]{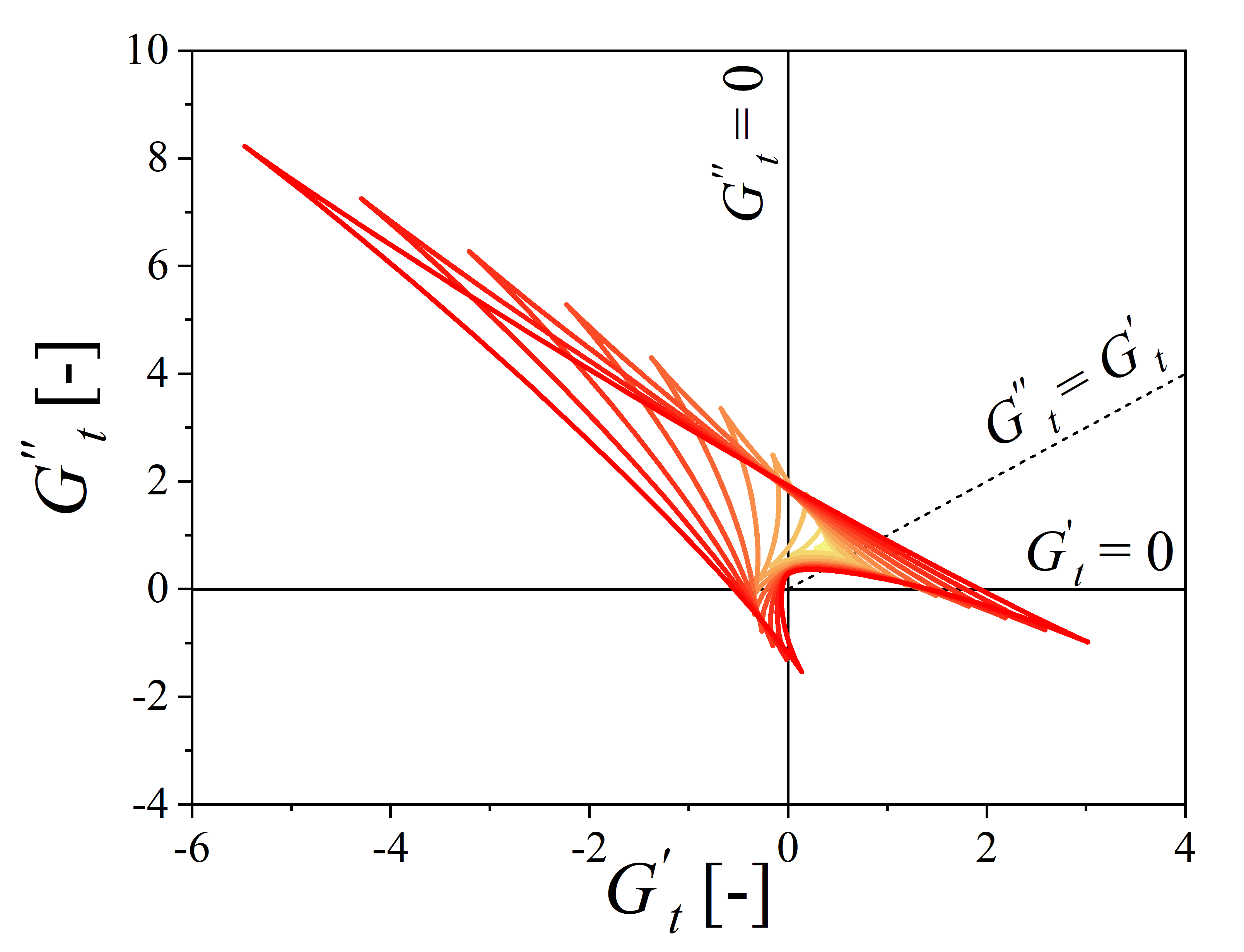}\label{De_a_0_5_FENEElasticStress}}
\caption{SPP analysis for the sPTT and FENE-P model responses at $De \ (De/a) = 0.5$ and $L^2 = 1/\epsilon = 100$. (a) and (c) show the 3D Lissajous curves for the sPTT and FENE-P models, respectively, whilst (b) and (d) show the respective Cole-Cole plots.}
\label{ColeColePlots}
\end{figure}

Figure \ref{ColeColePlots} shows the 3D Lissajous curves for  the sPTT and FENE-P models as well as the Cole-Cole plots ($G^{''}_t$ \textit{versus} $G^{'}_t$). For the sPTT model, the response manifests as deltoids in the Cole-Cole plots, which are commonly observed experimentally for a range of viscoelastic materials, including doughs \citep{Erturk2022, Park2020}.  As $Wi\sqrt{\epsilon}$ is increased the deltoids increase in size, which physically might represent an increase in the degree of micro-structural change during the oscillation \citep{Park2020}. For the FENE-P response, as $Wi$ is increased (note $L^2$ is fixed at $100$ and we do not assume the solution truly scales with $Wi/(aL)$) the Cole-Cole plots resemble instead arrow-head shapes which are significantly larger in size than the deltoids observed in the sPTT response (see the scales of the axes).

\begin{figure}
\centering
\subfloat[$~$]{
\includegraphics[width=0.475\textwidth]{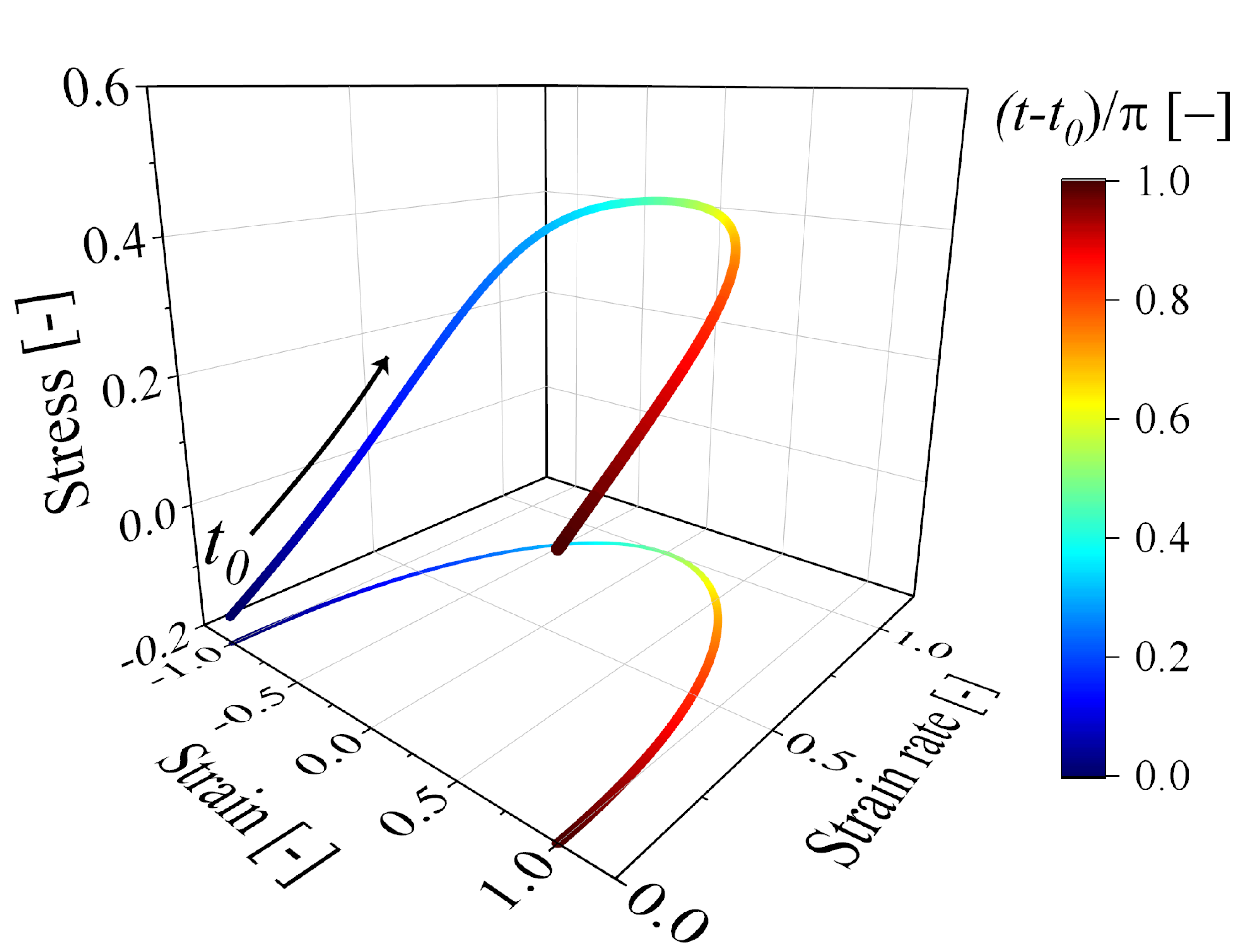}}
\subfloat[$~$]{
\includegraphics[width=0.475\textwidth]{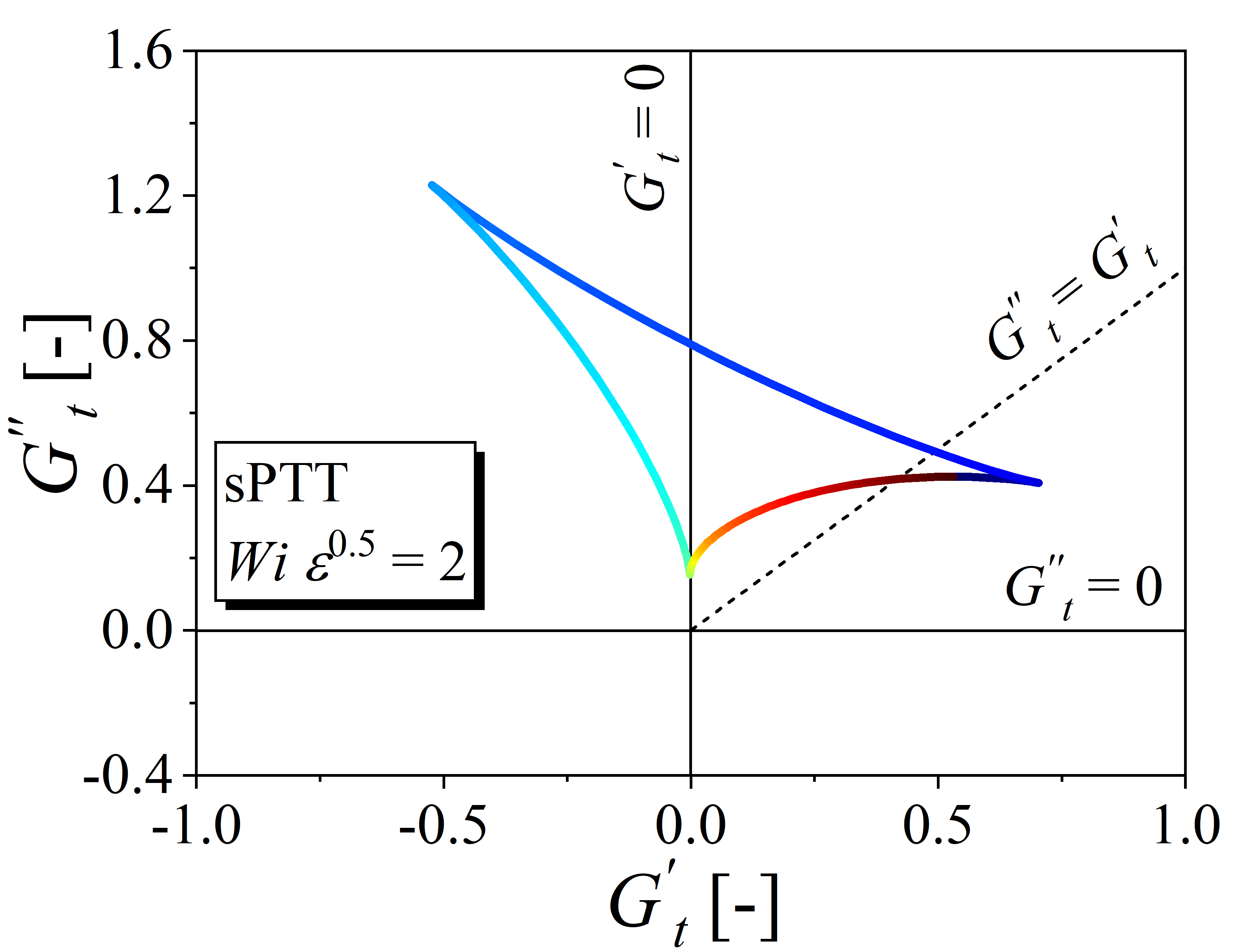}}

\subfloat[$~$]{
\includegraphics[width=0.475\textwidth]{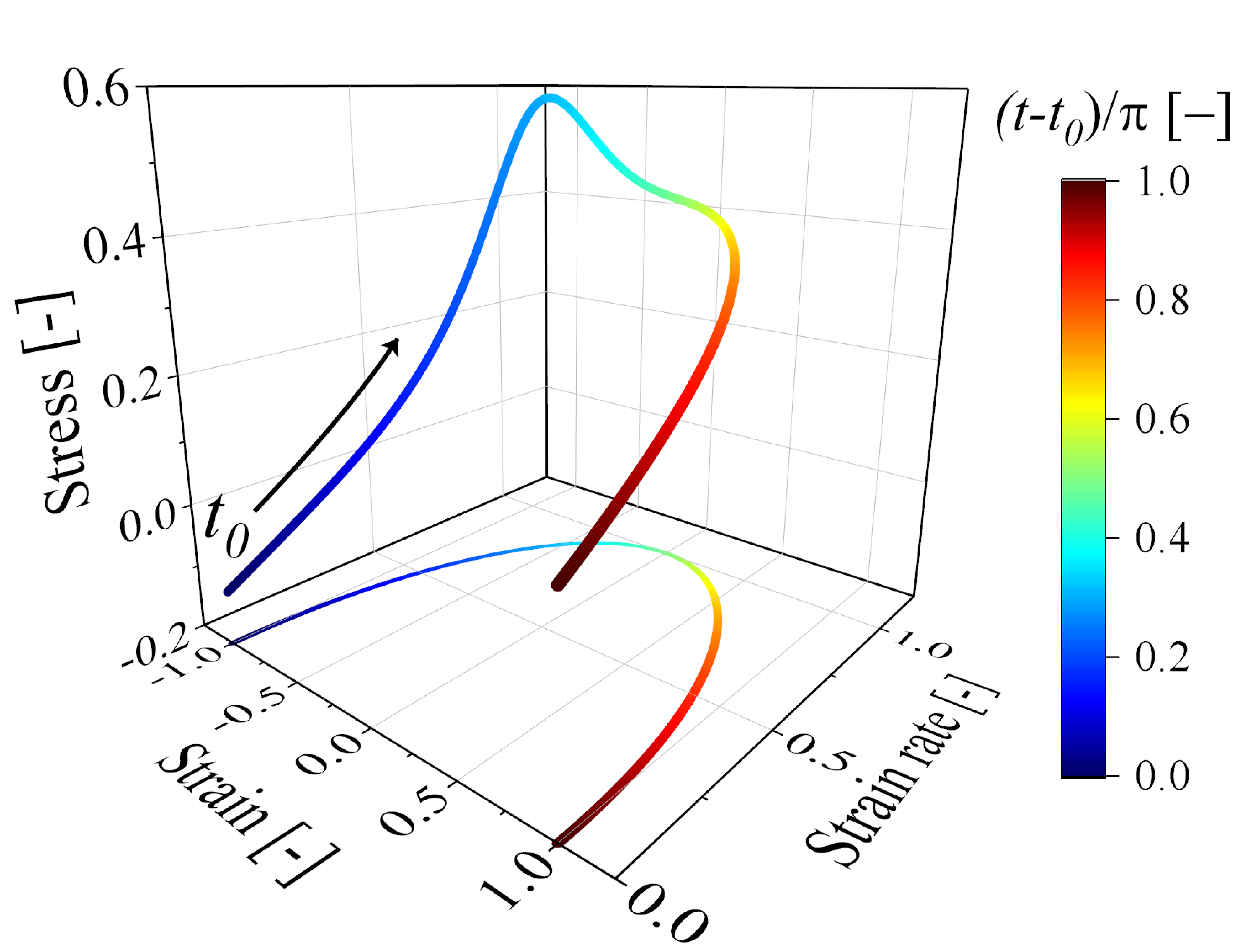}}
\subfloat[$~$]{
\includegraphics[width=0.475\textwidth]{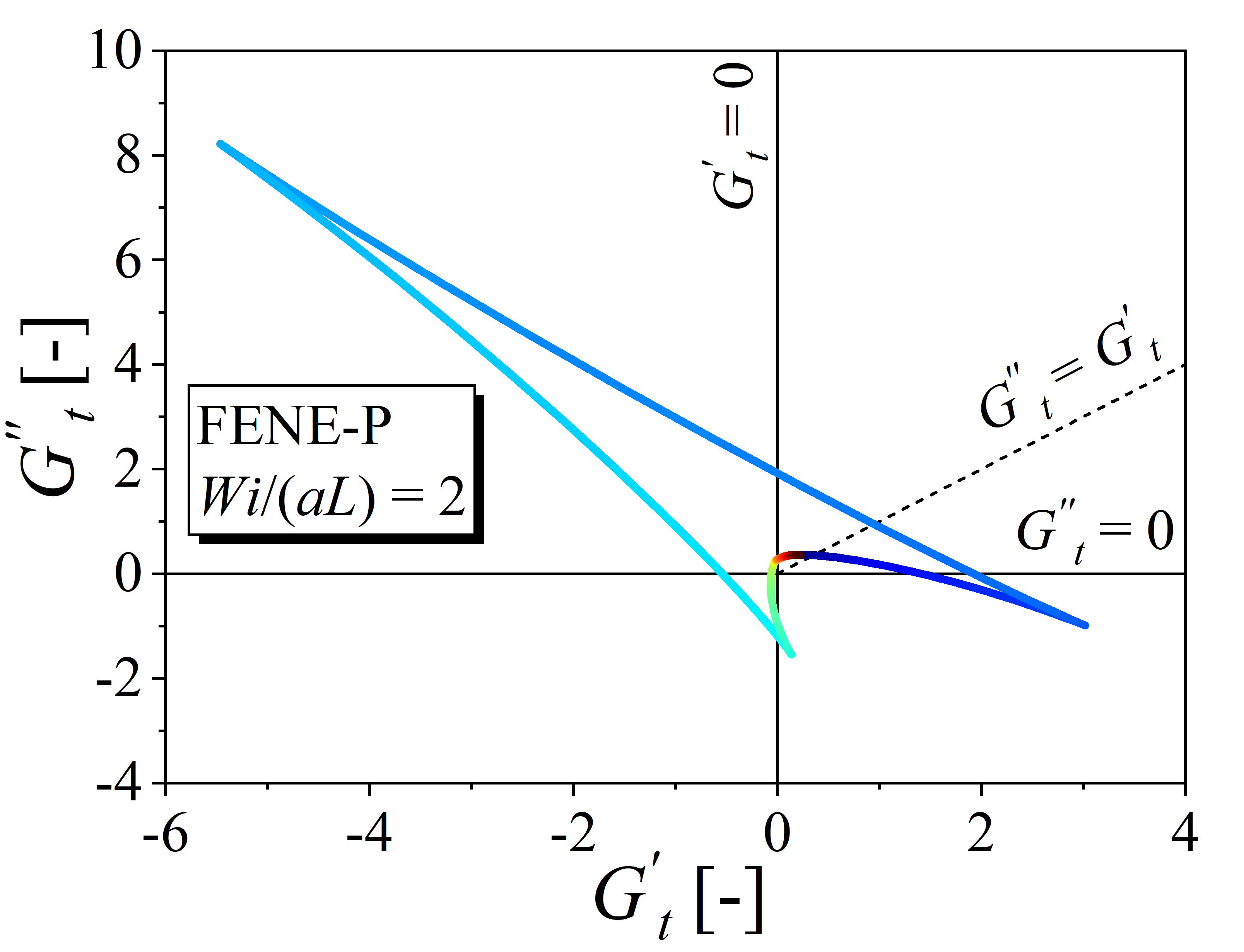}\label{De_a_0_5_FENEElasticStress}}
\caption{SPP analysis for the sPTT [(a) and (b)] and FENE-P [(c) and (d)] models  for $De \ (De/a) = 0.5$ and $Wi\sqrt{\epsilon} \ (Wi/(aL)) = 2$ with $\epsilon = 1/L^2 = 1/100$. The color bar indicates the normalised time in one half of the oscillation from the point indicated by $t_0$. (a) and (c) show the 3D Lissajous curves and (b) and (d) show the Cole-Cole plots.}
\label{SPPwithtime}
\end{figure}

Figure \ref{SPPwithtime} highlights more clearly the temporal evolution of $G^{'}_t$ and $G^{''}_t$ during the (half) oscillation for the sPTT and FENE-P models for the largest value of $Wi$ shown in Figure \ref{ColeColePlots}. (a) and (c) show the 3D Lissajous curves for the sPTT and FENE-P models, respectively, whilst (b) and (d) show the respective Cole-Cole plots. The color bar indicates the normalised time between the start point $t_0$ (chosen arbitrarily as $\gamma = -1$ and $\dot{\gamma}=0$) and the end point $t_0 +\pi$ (i.e. half a period after $t_0$).  For both the sPTT and FENE-P model responses, the majority of the temporal change in $G^{'}_t$ and $G^{''}_t$ takes place between $t_0$ and $(t-t_0)/\pi \approx 0.4$, which corresponds to the region directly before and after the stress overshoot in the FENE-P model. In the sPTT model response, initially (i.e. at $t_0$) both  $G^{'}_t$ and $G^{''}_t$ are positive with $G^{'}_t$ being slightly larger than $G^{''}_t$. Then, in the period between $t_0$ and $(t-t_0)/\pi \approx 0.4$, there is firstly a decrease in $G^{'}_t$ with an increase in $G^{''}_t$ (which corresponds to strain-softening and shear-thickening) followed by an increase in $G^{'}_t$ with a decrease in $G^{''}_t$ (which corresponds to strain-stiffening and shear-thinning).  For a significant portion of the region between between $t_0$ and $(t-t_0)/\pi \approx 0.4$,  $G^{'}_t$ is negative, indicating the presence of elastic recoil despite there being practically no self-intersecting loops in the viscous Lissajous curve.  In the FENE-P response, both $G^{'}_t$ and $G^{''}_t$ are smaller at $t_0$ than for the sPTT model, and $G^{''}_t$ is slightly larger than $G^{'}_t$.  In the period between $t_0$ and $(t-t_0)/\pi \approx 0.4$, the FENE-P response is drastically different to the sPTT response. Initially, there is a sharp increase in $G^{'}_t$ with a decrease in $G^{''}_t$. This decrease in $G^{''}_t$ is sharp enough that it becomes negative in this region before the stress overshoot. Then, similarly to the sPTT response, a decrease in $G^{'}_t$ with an increase in $G^{''}_t$ is observed. This behaviour is, however, more extreme for the FENE-P response than for the sPTT response. Then, an increase in $G^{'}_t$ with a decrease in $G^{''}_t$ is observed as the trajectory passes through the sharp stress overshoot. The decrease in $G^{''}_t$ in this region for the FENE-P response is so large that $G^{''}_t$ is negative after the stress overshoot.  In the supplementary animations Movie 1 and Movie 2, we show the evolution of the Frenet-Serret frame along the Lissajous curves displayed in Figure \ref{SPPwithtime}, along with the projections of $\boldsymbol{B}$ in the $\dot{\gamma}$ and $\tau_{p,12}$ plane (highlighting the sign of $G^{''}_t$) and the current position in the Cole-Cole plots.

\begin{figure}
\centering
\subfloat[Toy FENE-P model]{
\includegraphics[trim={1.7cm 0cm 3cm 0cm},clip,width=0.46\textwidth]{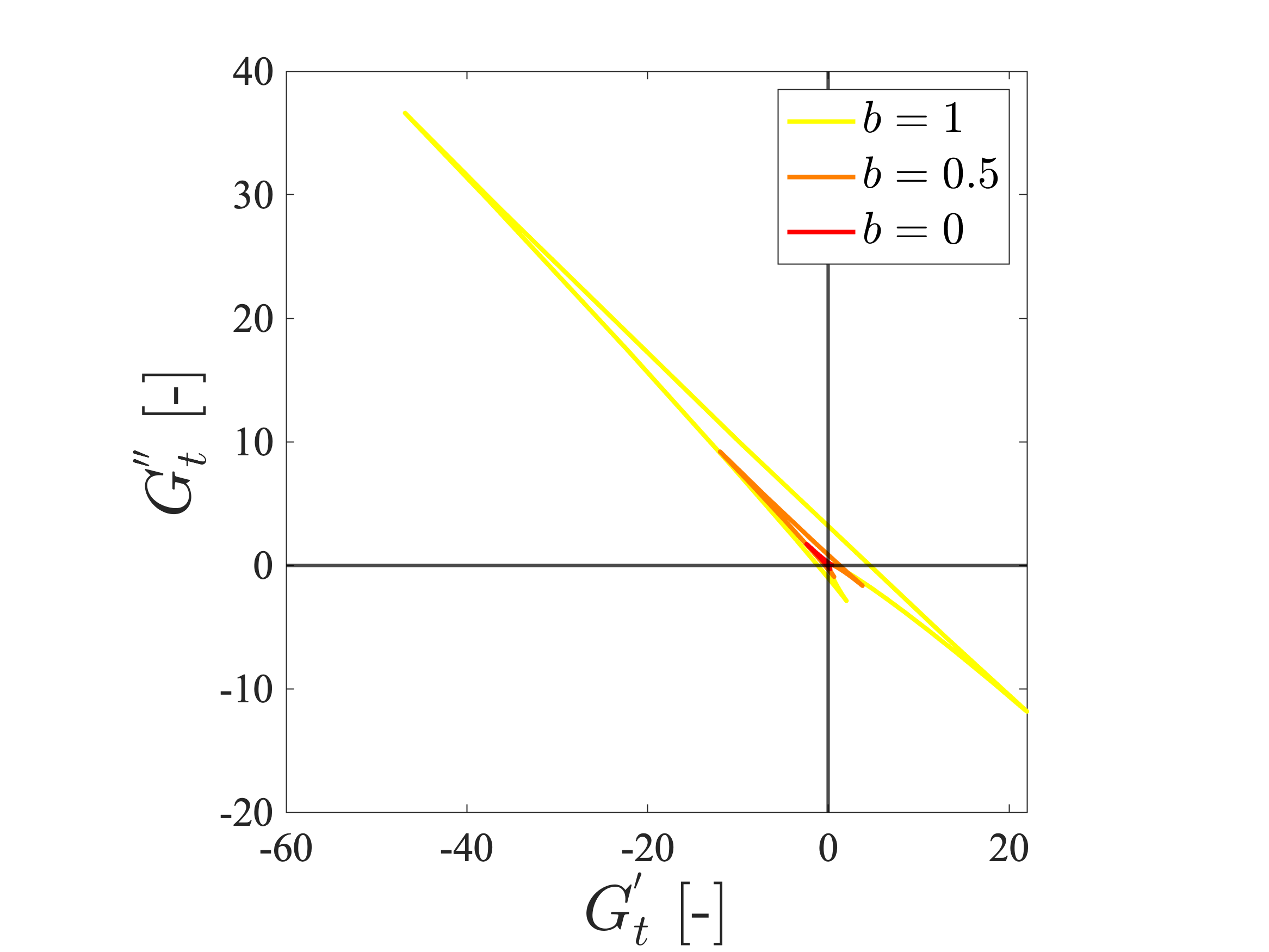}}
\subfloat[Toy sPTT model]{
\includegraphics[trim={1.7cm 0cm 3cm 0cm},clip,width=0.46\textwidth]{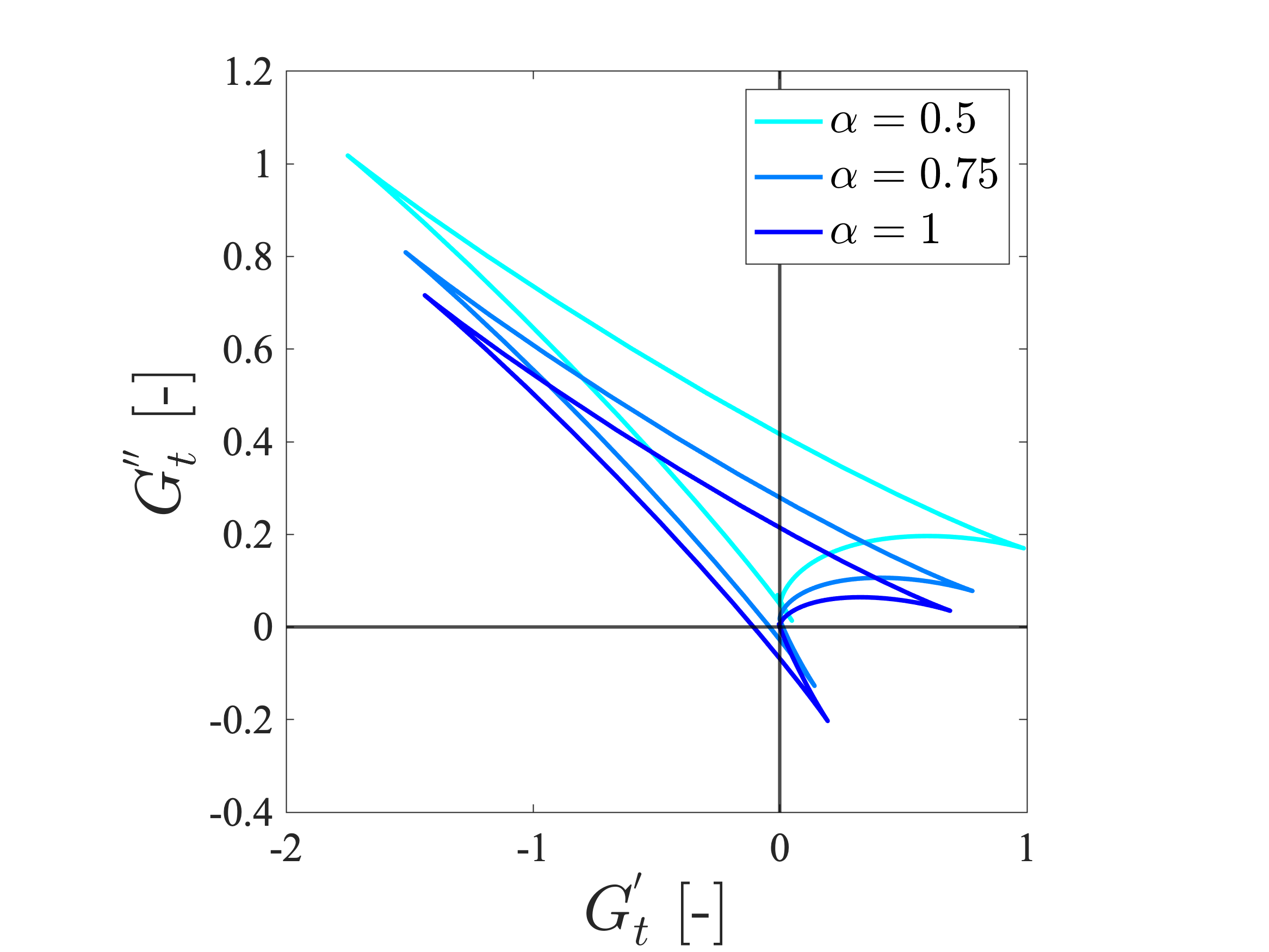}}
\caption{Cole-Cole plots for (a) the toy FENE-P model at $De = 1$, $Wi = 100$, $L^2 = 100$ and (b) the toy sPTT model at $De = 0.5$, $Wi = 200$, $\epsilon = 1/100$. The corresponding Lissajous curves can be seen in Figure \ref{ToysFENE} and \ref{ToysPTT}.}
\label{ToyModelsSPP}
\end{figure}

\begin{figure}
\centering
\includegraphics[trim={1.7cm 0cm 3cm 0cm},clip,width=0.46\textwidth]{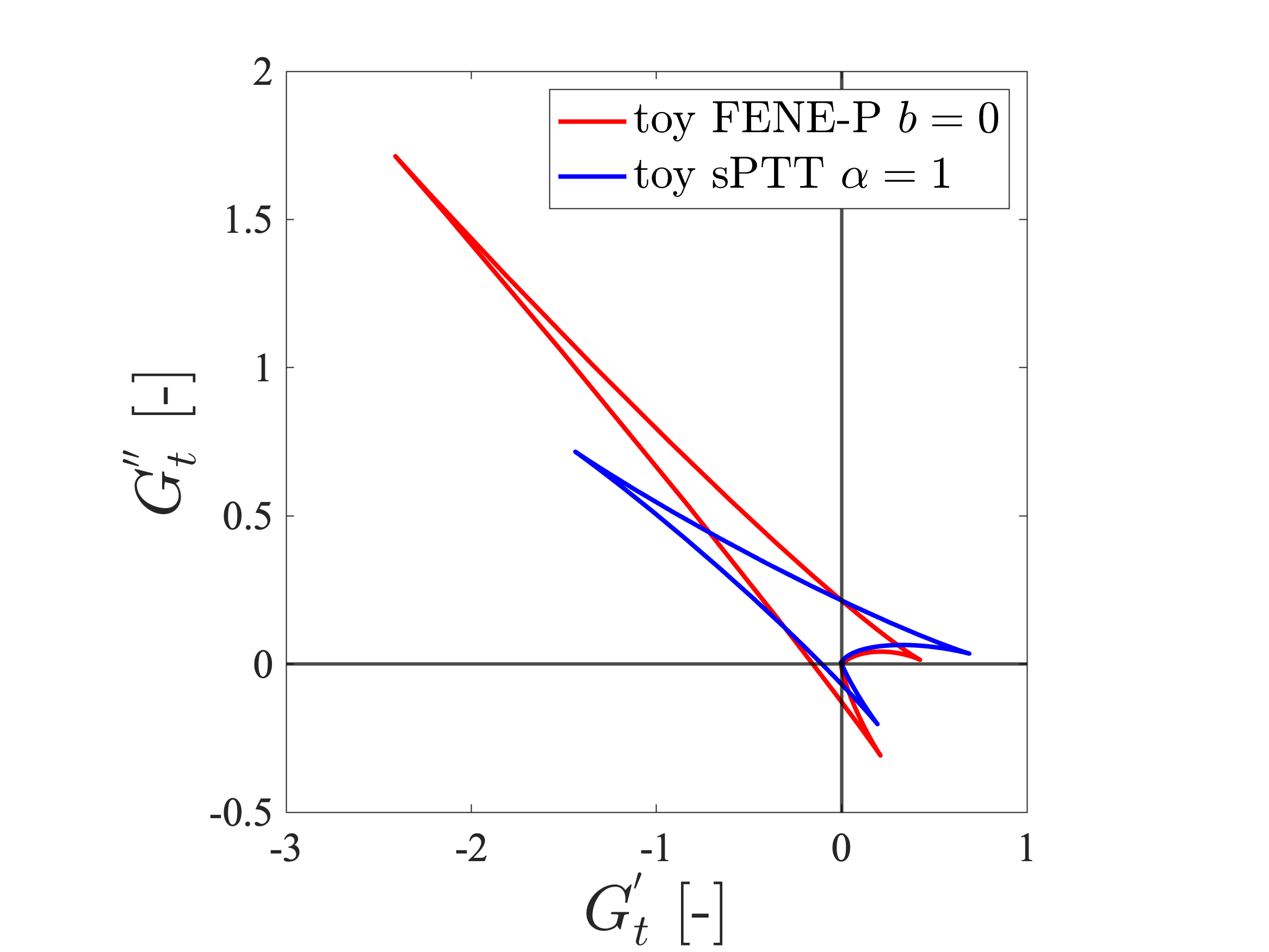}
\caption{Cole-Cole plots comparing the toy sPTT model response with $\alpha = 1$ and the toy FENE-P model response with $b=0$. }
\label{ToyColeColeComp}
\end{figure}

We now use the toy models outlined in Section \ref{results4} to further highlight the link between the nature of the constitutive model and its behaviour in terms of the SPP analysis. Figure~\ref{ToyModelsSPP} shows the Cole-Cole plots obtained from the SPP framework for the (a) toy FENE-P and (b) toy sPTT models. Note that $De = 0.5$ and $Wi = 200$ for the toy sPTT model, whilst $De = 1$ and $Wi = 100$ for the toy FENE-P model. This corresponds to the conditions for the Lissajous curves shown in Figures \ref{ToysPTT} and \ref{ToysFENE}.  For the toy sPTT model, increasing the value of $\alpha$ has a seemingly minor effect on the Cole-Cole plots. The general shape remains fairly constant, however, a region of negative $G^{''}_t$ develops which corresponds to the developing stress overshoots (seen in Figure \ref{ToysPTT}). The link between this negative region of $G^{''}_t$ and the stress overshoot is seen clearly in the supplementary animation (Movie 1) for the FENE-P model response. Essentially, after the stress overshoot, the normal vector $\boldsymbol{N}$ points in the positive stress direction as the recoil fades, which points the bi-normal vector $\boldsymbol{B}$ towards the negative strain rate direction. For the toy FENE-P model, the value of $b$ has a significant effect on the Cole-Cole plot. For $b \rightarrow 1$,  the extremities of $G^{'}_t$ and $G^{''}_t$ become significantly large in magnitude. This indicates that the SPP analysis is highly sensitive to the presence of $F(A)$ in the $\boldsymbol{\tau}_p$-$\mathsfbi{A}$ relationship (which relates physically to explicit finite extensibility of the polymer chains) despite the fact that the primary features of the Lissajous curves do not, at least qualitatively, appear to change significantly (see Figure \ref{ToysFENE}).  More quantitatively, the presence of $F(A)$ in the $\boldsymbol{\tau}_p$-$\mathsfbi{A}$ relationship appears to give rise to a region of backflow (i.e. negative $G^{''}_t$) before the onset of the stress overshoot. This behaviour cannot be easily explained in terms of the evolution of $\mathsfbi{A}$, as the region of backflow after the stress overshoot can (i.e. $Q_r < Q_g$). The region of backflow before the stress overshoot is observed only in the toy FENE-P and toy FENE-CR (see Appendix \ref{appA}) models when $b>0$. In the supplementary animation for the FENE-P response (Movie 1), it is evident that this region of backflow in the FENE-P response arises due to the exceptionally sharp increase in the stress before the overshoot, which is explained by the fact that the stress grows non-linearly with $\mathsfbi{A}$ according to Equation \eqref{cramersForm}. Figure \ref{ToyColeColeComp} shows the Cole-Cole plots for the toy sPTT model with $\alpha = 1$ and the toy FENE-P model with $b = 0$, such that the only difference between the models is the functional form of $F(A)$. The qualitative features of the Cole-Cole plots are very similar. The size of the deltoids are similar, both responses exhibit backflow after the stress overshoot (owing to the position of $F(A)$ in the model and the transient nature of $A_{22}$), and neither of the responses exhibit backflow before the stress overshoot (owing to the absence of $F(A)$ in the $\boldsymbol{\tau}_p$-$\mathsfbi{A}$ relationship). In this sense, the SPP framework can be used to quickly identify the presence of finite extensibility effects in LAOS responses, and differentiate them from other non-linear effects such as micro-structural destruction. 

\subsection{1D modelling of LAOS}\label{results5}

In the previous subsections, it is assumed that the shear rate is uniform in space.  For constitutive models which have non-monotonic stress-shear rate relationships (such as Giesekus \citep{Giesekus1982} or Rolie-Poly \citep{Likhtman2003}), the material in a shear flow can shear band such that two distinct regions of shear rate exist in the flow. Neither the FENE-P or sPTT models will shear band in SSSF since their underlying stress-shear rate curves are both monotonic. However, as we have highlighted already, during LAOS, the FENE-P model behaves in a much more non-linear manner than during steady-shear and aspects of the response of the model, such as the presence of strong stress overshoots and self-intersecting secondary loops, resemble those of models that do exhibit shear-banding. Moreover, recent studies have highlighted that shear banding can occur due to stress overshoots in transient flows, even in models which have monotonic underlying stress-shear rate constitutive curves \citep{Adams2009, Carter2016, Moorcroft2011, Moorcroft2013}.  It is therefore sensible to check whether the FENE-P model is capable of shear banding in LAOS. 

Using the 1D modelling approach, and enforcing true creeping flow, the stress response at the top boundary will match the results obtained in the previous subsections if there is no shear-banding. However, if shear-banding occurs, we will be able to observe the heterogeneous velocity gradient in the gap, and the stress response at the top wall will not match with the previous results. In order to avoid the known problems of stress selection \citep{Lu2000,Olmsted2008} and discontinuous strain rates in shear banding, we add a small diffusive term ($\kappa \nabla^2 \mathsfbi{A}$) to the right hand side of the constitutive models during the simulations. The value of $\kappa$ was fixed at $10^{-9}$. Note that $\kappa$ here is dimensionless and is given by $\kappa = \lambda D/H^2$ where $D$ is the diffusion coefficient in $m^2/s$. In the Supplementary Material, we show that this methodology is capable of capturing shear-banding in LAOS using the Rolie-Poly (ROuse LInear Entangled POLYmers) model.

\begin{figure}
\centering
\large 
$\tau_{p,12} \ $ \textit{vs} $\ \dot{\gamma}$

\includegraphics[width=0.9\textwidth]{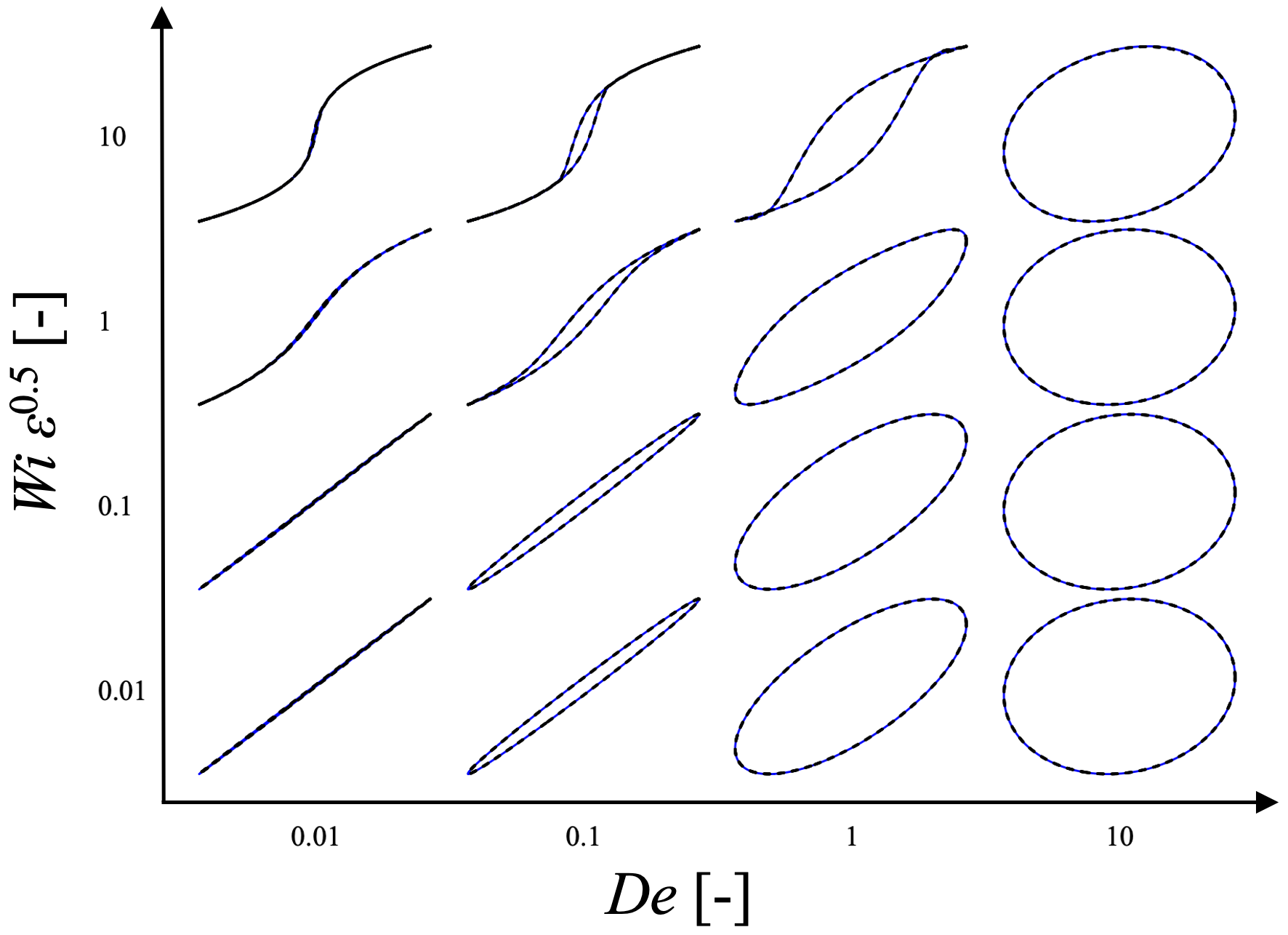}
\caption{Lissajous-Bowditch plots (viscous projection) for the sPTT model. Blue solid lines show the 0-D approximation solution from the previous section, black dashed lines show the results from the 1D simulations where the stress is computed at the top (moving) boundary.}
\label{FullFlowCompPTT}
\end{figure}

\begin{figure}
\centering
\large 
$\tau_{p,12} \ $ \textit{vs} $\ \dot{\gamma}$

\includegraphics[width=0.9\textwidth]{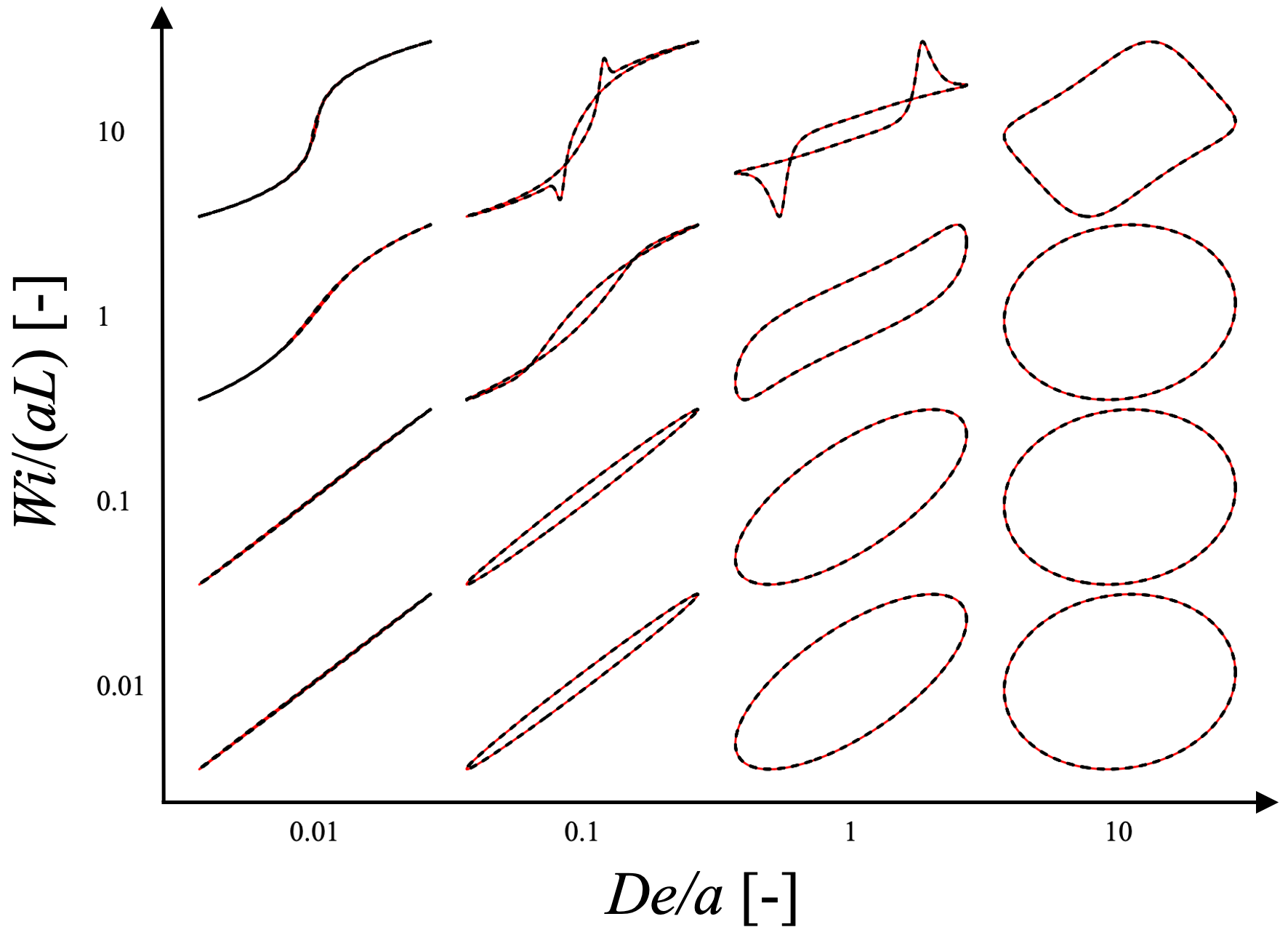}
\caption{Lissajous-Bowditch plots (viscous projection) for the FENE-P model ($L^2 = 100$). Red solid lines show the 0D approximation solution from the previous section, black dashed lines show the results from the 1D MOL simulations where the stress is computed at the top (moving) boundary.}
\label{FullFlowCompFENE}
\end{figure}

Figures \ref{FullFlowCompPTT} and \ref{FullFlowCompFENE} show the viscous Lissajous-Bowditch plots for the sPTT and FENE-P with $L^2 1/\epsilon = 100$ models, respectively, where the stress response has been computed with the 0D approximation (black dashed lines) and 1D simulation (solid lines). For both models, both the 0D and 1D methods give practically identical responses, indicating that the 0D approximation is adequate for properly describing the model responses and that no shear-banding is occurring in the 1D simulations. Since the methodology used for the 1D simulations is capable of predicting shear-banding, it is reasonable to assume that the FENE-P model does not shear band in LAOS despite the fact that the model response is significantly more non-linear in LAOS than it is in SSSF at least in the range of $De/a$ and $Wi/(aL)$ investigated. 

\section{Conclusions}

We have compared the response of the sPTT and FENE-P constitutive models in LAOS when the parameters in the models are chosen such that the models are mathematically identical for steady and homogeneous flows. The results show that the FENE-P model behaves in a significantly more non-linear manner than the sPTT model in LAOS, as it does in other transient flows such as start-up flow. The FENE-P model exhibits strong stress overshoots and large self-intersecting secondary loops in the viscous Lissajous curves, whereas the sPTT model does not. The stress overshoots and self intersecting secondary loops arise from the FENE-P model primarily due to the extensibility function being multiplied with $\mathsfbi{A}$ instead of $(\mathsfbi{A}-\mathsfbi{I})$ in the conformation tensor form of the constitutive model, which can be thought of in terms of network theory as the system exhibiting faster rates of micro-structural destruction than creation. This drives a change in $A_{22}$ and $A_{33}$ during the oscillation which, due to the use of the upper-convected derivative, causes the elastic recoil of $A_{12}$ to exceed the growth of $A_{12}$ for increasing rates of deformation, which ultimately leads to the pronounced stress overshoots. It is important to note that the differences between these two models arise for both Eulerian unsteady flows (such as LAOS) and Lagrangian unsteady (or inhomogenous) flows, and so the differences between the model responses such as the stress overshoots will also occur in Eulerian steady flows which are Lagrangian unsteady due to, perhaps, expansions and contractions in a complex geometrical domain. Such complex geometries are often encountered in many industrial flows and processes. In fluid regions with strong accelerations, one can expect much sharper changes in the polymeric stress with the FENE-P model than the sPTT model. 

Although it has previously been shown analytically for the FENE-P and sPTT models that the stress-strain rate curves scale with $Wi/(aL)$ and $Wi\sqrt{\epsilon}$, respectively, for simple steady-shear, we have been able to show with our numerical results that the sPTT LAOS response also scales with $Wi\sqrt{\epsilon}$, but the FENE-P response in LAOS only scales with $Wi/(aL)$ at large enough values of $L^2$.  We have also been able to explain this analytically. This is shown to be caused by the specific functional form of the extensibility function as well as its position in the FENE-P constitutive model (inside the brackets in the recoil term rather than on the outside).

Using the Sequence of Physical Processes framework, we highlight the differences in the model responses in terms of the transient moduli $G^{'}_t$ and $G^{''}_t$. The Cole-Cole plots show that the FENE-P model exhibits significantly more complex rheological behaviour during the oscillation. A key result obtained from the SPP analysis is that the FENE models (both FENE-P and FENE-CR) exhibit backflow (i.e. negative $G^{''}_t$) both before and after the stress overshoot. The region of backflow before the stress overshoot is shown to be linked to the presence of the extensibility function in the stress-conformation tensor relationship. This highlights that the SPP framework can be particularly useful for identifying the correct form of constitutive models for viscoelastic materials from LAOS data.

Although the FENE-P model in LAOS behaves more like models that exhibit shear banding, such as Giesekus and Rolie-Poly, than it does in steady-shear, the FENE-P model does not appear to be capable of shear banding in LAOS. This was investigated by solving both the momentum equation and the constitutive model in a 1D gap of fluid using the method of lines technique. For all cases with the sPTT and FENE-P models, the velocity gradient remained constant across the gap, and the Lissajous-Bowditch plots were almost identical when the 1D and 0D solutions were compared. 

\vspace{0.3cm} 

\noindent \textbf{Acknowledgements}. The authors acknowledge the developers of the SPP software for kindly providing us with a copy of the software. We would also like to thank all of the reviewers for their feedback and comments. We thank Reviewer 3 specifically for showing us how to prove the universal scaling of the sPTT model LAOS response in Section \ref{results1}.

\vspace{0.3cm}

\noindent \textbf{Funding}. The authors acknowledge the financial support of the Center in Advanced Fluid Engineering for Digital Manufacturing, UK (CAFE4DM) project (Grant No. EP/R00482X/1)

\vspace{0.3cm}

\noindent \textbf{Declaration of interests}. The authors report no conflict of interest.

\appendix

\section{Comparing the FENE-CR and sPTT models}\label{appA}

\begin{figure}
\centering
\includegraphics[width=0.6\textwidth]{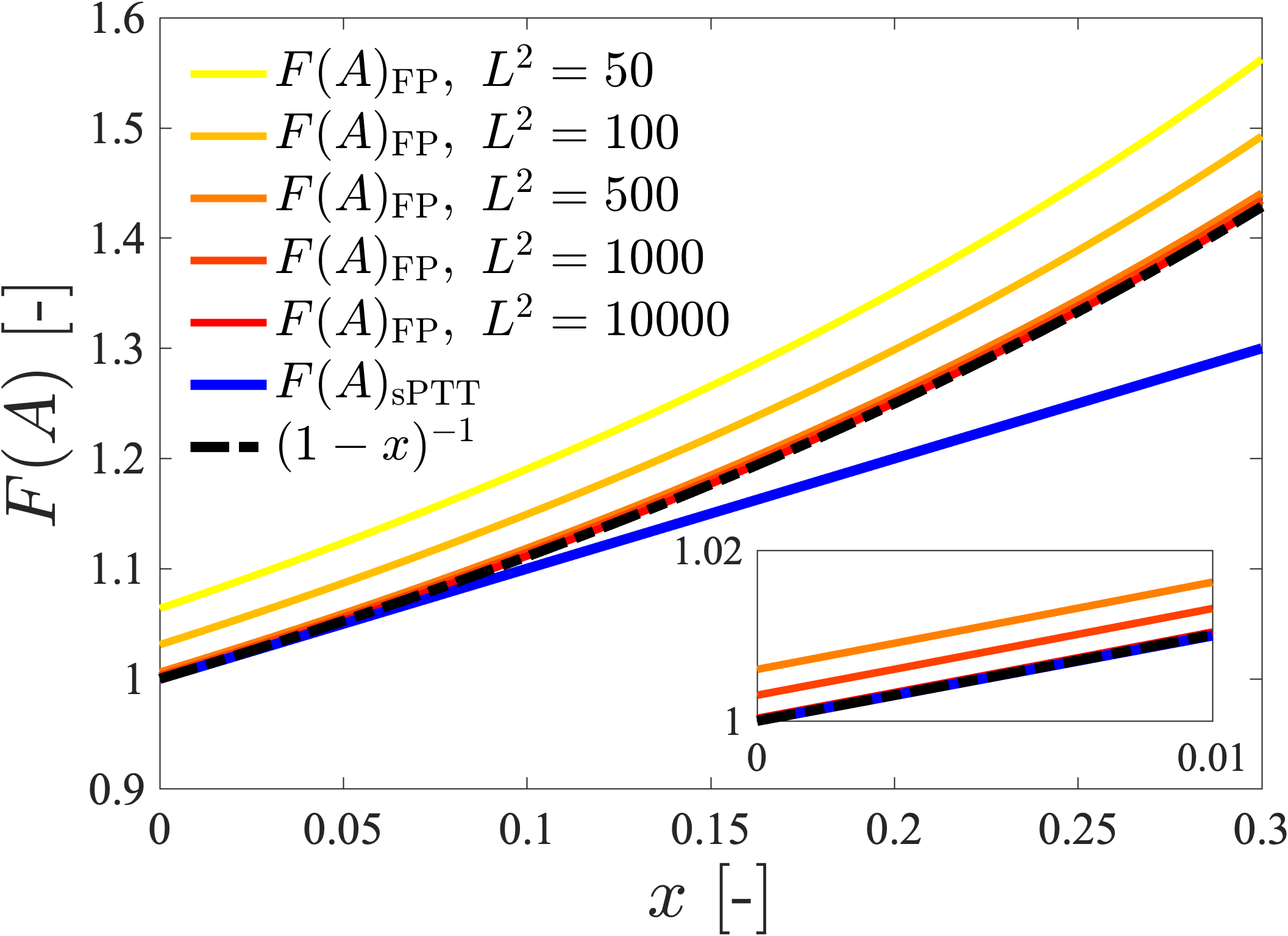}
\caption{Extensibility function $F(A)$ \textit{versus} $x$ where $x = \epsilon (A_{11}-1)$ for the sPTT model and $x = (A_{11}-1)/L^2$ for the FENE-CR model. Black dashed line shows the expression for $F(A)_{\mathrm{FP}}$ in the case that $L^2 \gg 3$. Yellow to red lines show $F(A)$ for the FENE-CR model where $A_{22} = A_{33} = 1$ and so $F(A) = (1-x-3/L^2)^{-1}$.}
\label{feneCRfuncfig}
\end{figure}

\begin{figure}
\centering
\subfloat[$~$]{
\includegraphics[width=0.47\textwidth]{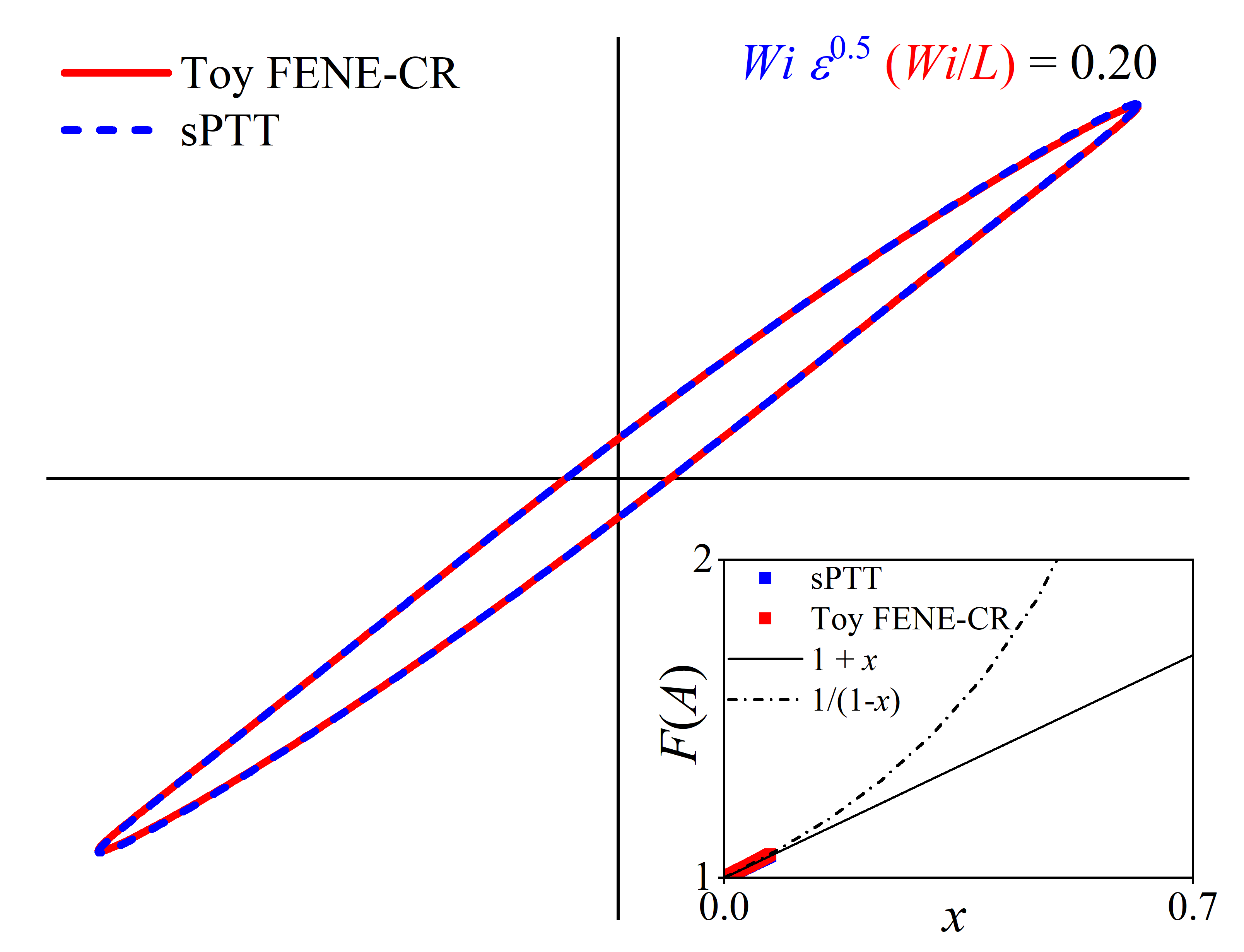}}
\subfloat[$~$]{
\includegraphics[width=0.47\textwidth]{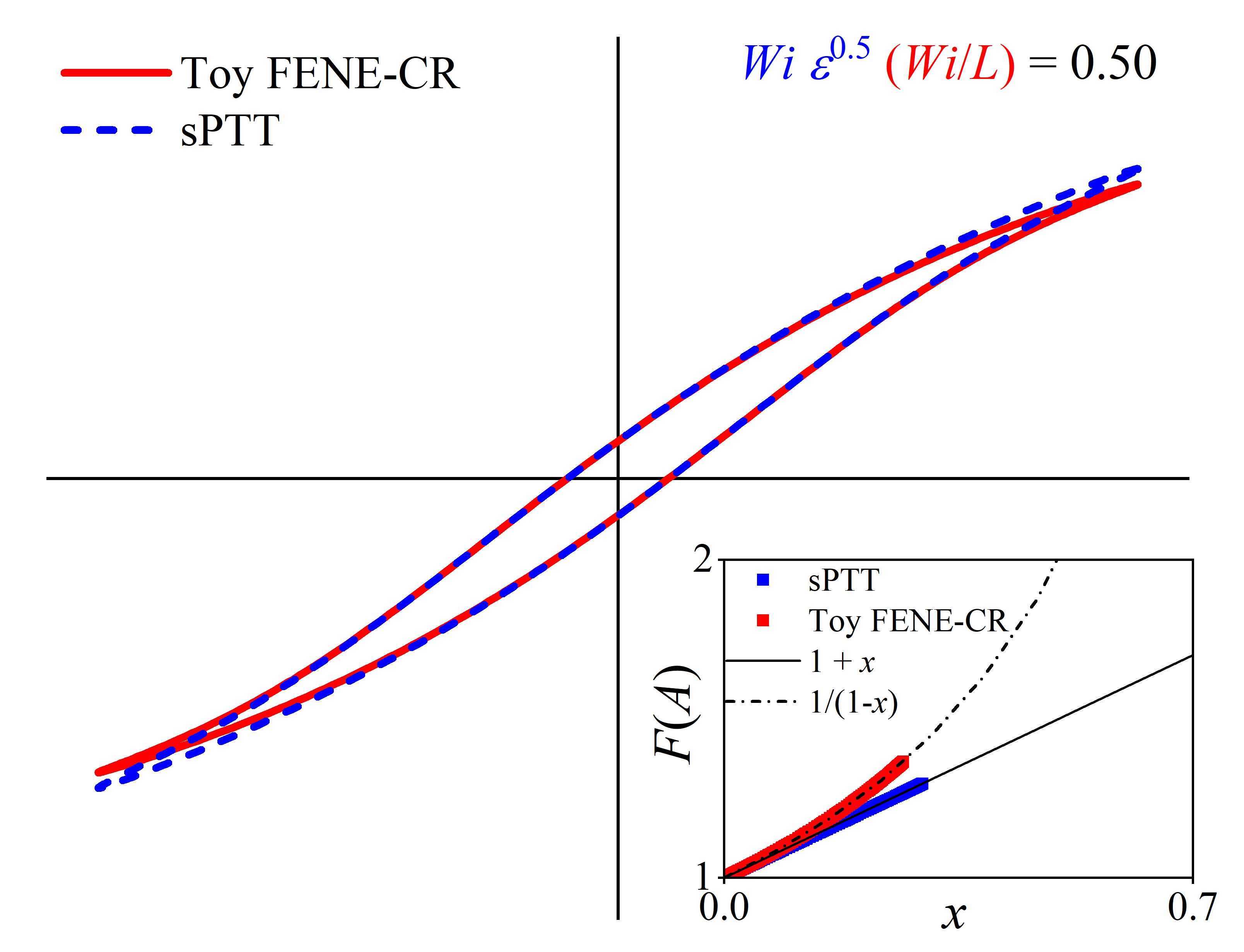}}

\subfloat[$~$]{
\includegraphics[width=0.47\textwidth]{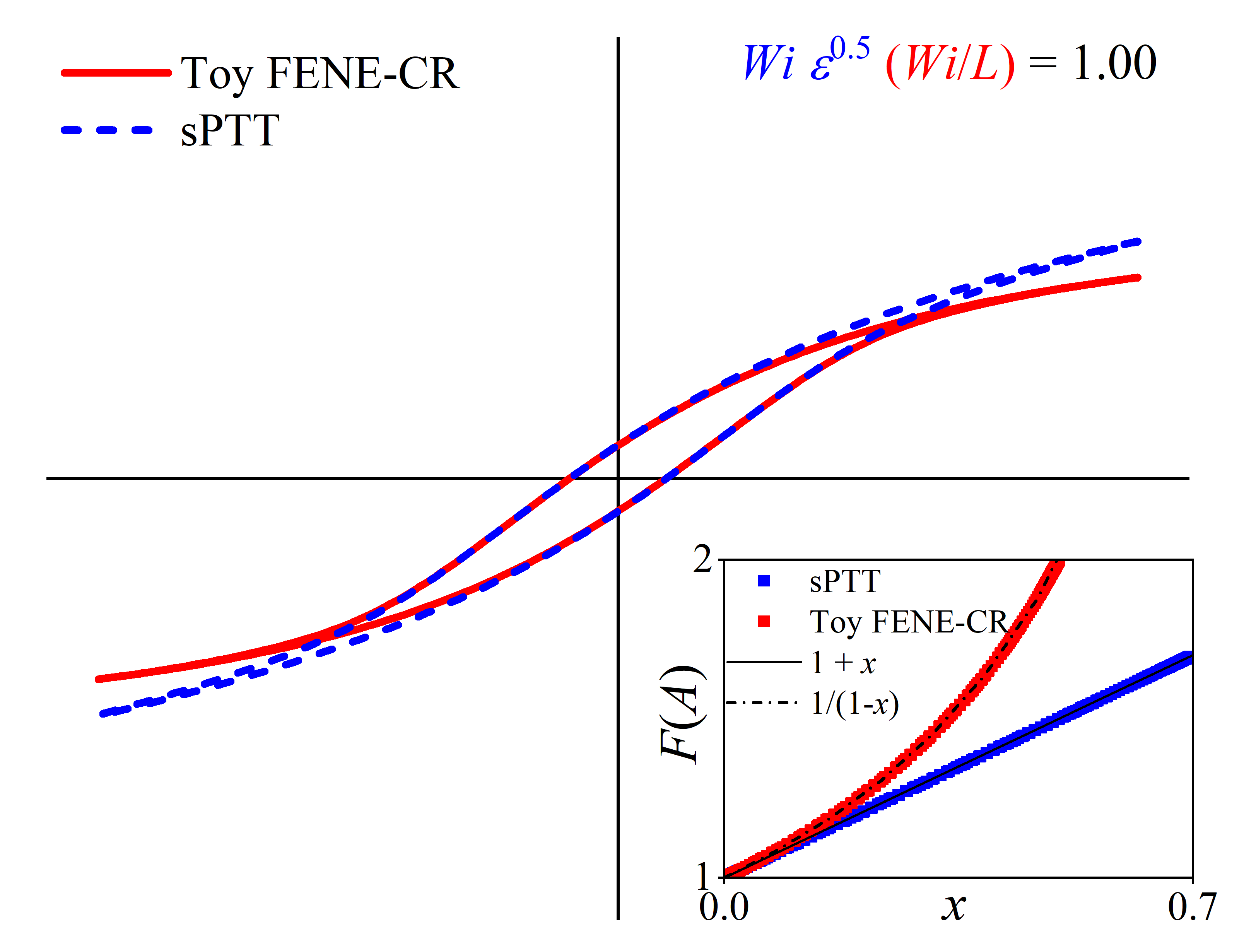}}
\caption{Viscous Lissajous curves ($\tau_{p,12}$ \textit{versus} $\dot{\gamma}$) for sPTT model (toy FENE-CR) model for $De = 0.2$ and varying $Wi\sqrt{\epsilon} \ (Wi/L)$. Inset shows the range of $F(A)$ for each model during the oscillation. $b = 0$ and $L^2 = 10^5$ for the toy FENE-CR model.}
\label{FENE_CRvssPTT}
\end{figure}

In Section \ref{results1} we highlight that the FENE-CR model only scales universally with $Wi/L$ for $L^2 \gg 3$.  We also point out that the evolution of $\mathsfbi{A}$ for the sPTT and FENE-CR models become identical for $L^2 \gg 3$ and in the case that terms of $\mathcal{O}(x^2)$ can be neglected in the expansion of $F(A)_{\mathrm{FP}}$, where $x = (A_{11}-1)/L^2$, which corresponds to the MAOS regime.  This is also highlighted in Figure \ref{feneCRfuncfig}. We introduce here a toy FENE-CR model which is defined as

\begin{subequations}
\begin{gather}
De \frac{\partial}{\partial t} \mathsfbi{A} - Wi (\mathsfbi{A} \cdot \nabla \boldsymbol{u} + \nabla \boldsymbol{u}^{\mathrm{T}} \cdot \mathsfbi{A} - \boldsymbol{u} \bcdot \nabla \mathsfbi{A}) = - F(A)_{\mathrm{FP}} (  \mathsfbi{A} -  \mathsfbi{I} )  \\ 
\boldsymbol{\tau}_p = \frac{(1-\beta)}{Wi} (F(A)_{\mathrm{FP}})^b(\mathsfbi{A} - \mathsfbi{I})
\end{gather}
\end{subequations}

\noindent where $0 \leq b \leq 1$.  We remind the reader that $F(A)_{\mathrm{FP}} = L^2/(L^2 - \mathrm{tr}(\mathsfbi{A}))$. When $b = 0$, the only difference between the toy FENE-CR and sPTT models is the specific form of $F(A)$ used (i.e. $1+\epsilon(\mathrm{tr}(\mathsfbi{A})-3)$ \textit{versus} $L^2/(L^2 - \mathrm{tr}(\mathsfbi{A}))$).  In this case, the stress response of the toy FENE-CR model will now also become equivalent to that of the sPTT model in the MAOS regime. We show this in Figure \ref{FENE_CRvssPTT}. For $Wi\sqrt{\epsilon} \ (Wi/L) = 0.2$, the responses of both models are non-linear, due to the fact that $F(A)$ becomes transient and larger than unity during the oscillation. However, since $x$ is small during the oscillation and $L^2 \gg 3$, $F(A)_{\mathrm{FP}}$ is essentially linear and equivalent to $F(A)_{\mathrm{sPTT}}$, and so the stress responses of both models are practically identical. For larger values of $Wi\sqrt{\epsilon} \ (Wi/L)$, $F(A)_{\mathrm{FP}}$ starts to deviate from $F(A)_{\mathrm{sPTT}}$ at points in the oscillation due to the increased values of $A_{11}$, and so the stress responses begin to differ. As is likely expected, the toy FENE-CR model response appears more non-linear than the sPTT response, owing to the increased non-linearity in $F(A)$. The responses are, however, at least qualitatively similar even for moderate values of $Wi \sqrt{\epsilon} \ (Wi/L)$. 

\begin{figure}
\centering
\subfloat[$~$]{
\includegraphics[trim={1.2cm 0cm 3.4cm 0cm},clip,width=0.33\textwidth]{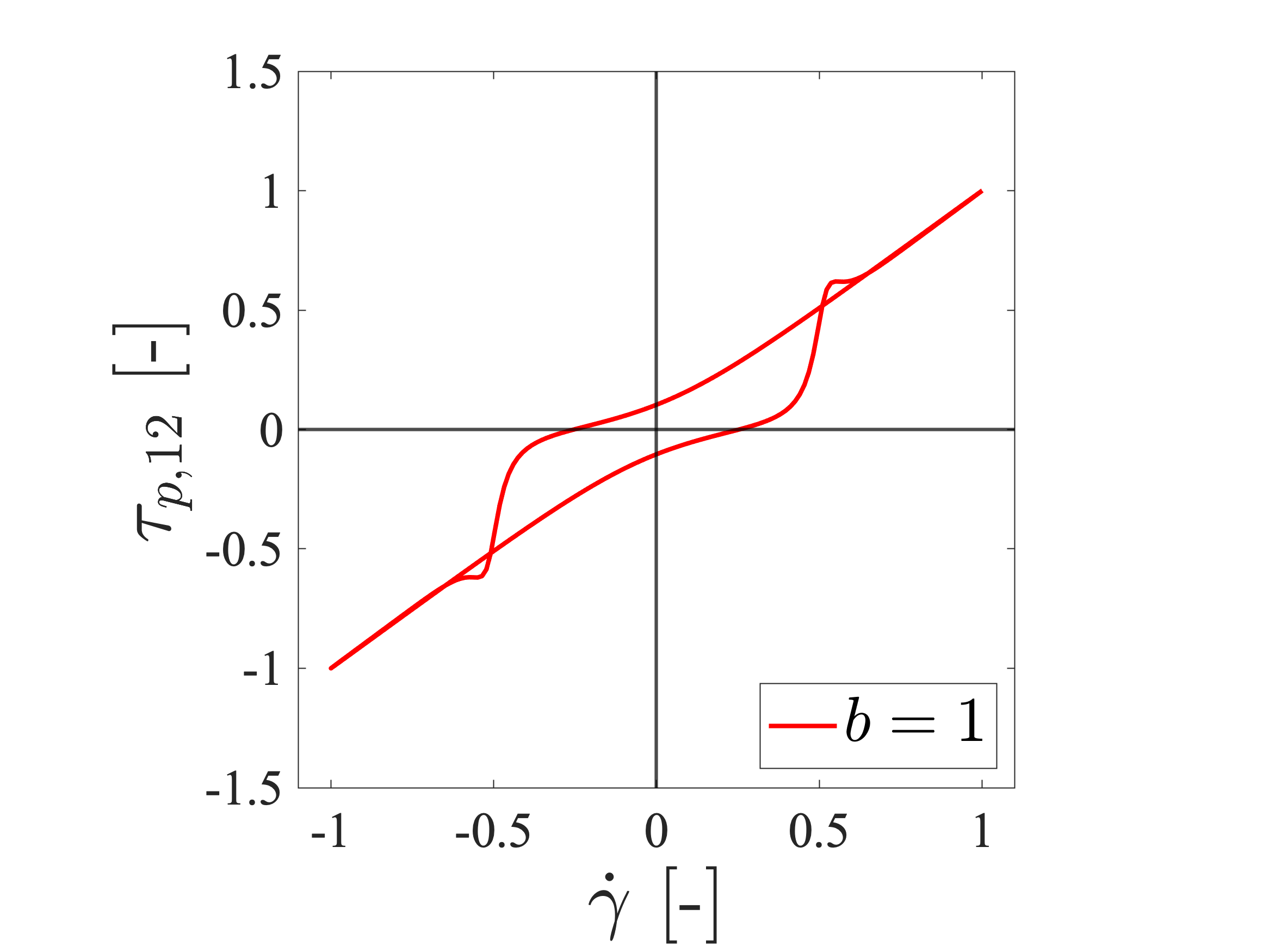}}
\subfloat[$~$]{
\includegraphics[trim={1.2cm 0cm 3.4cm 0cm},clip,width=0.33\textwidth]{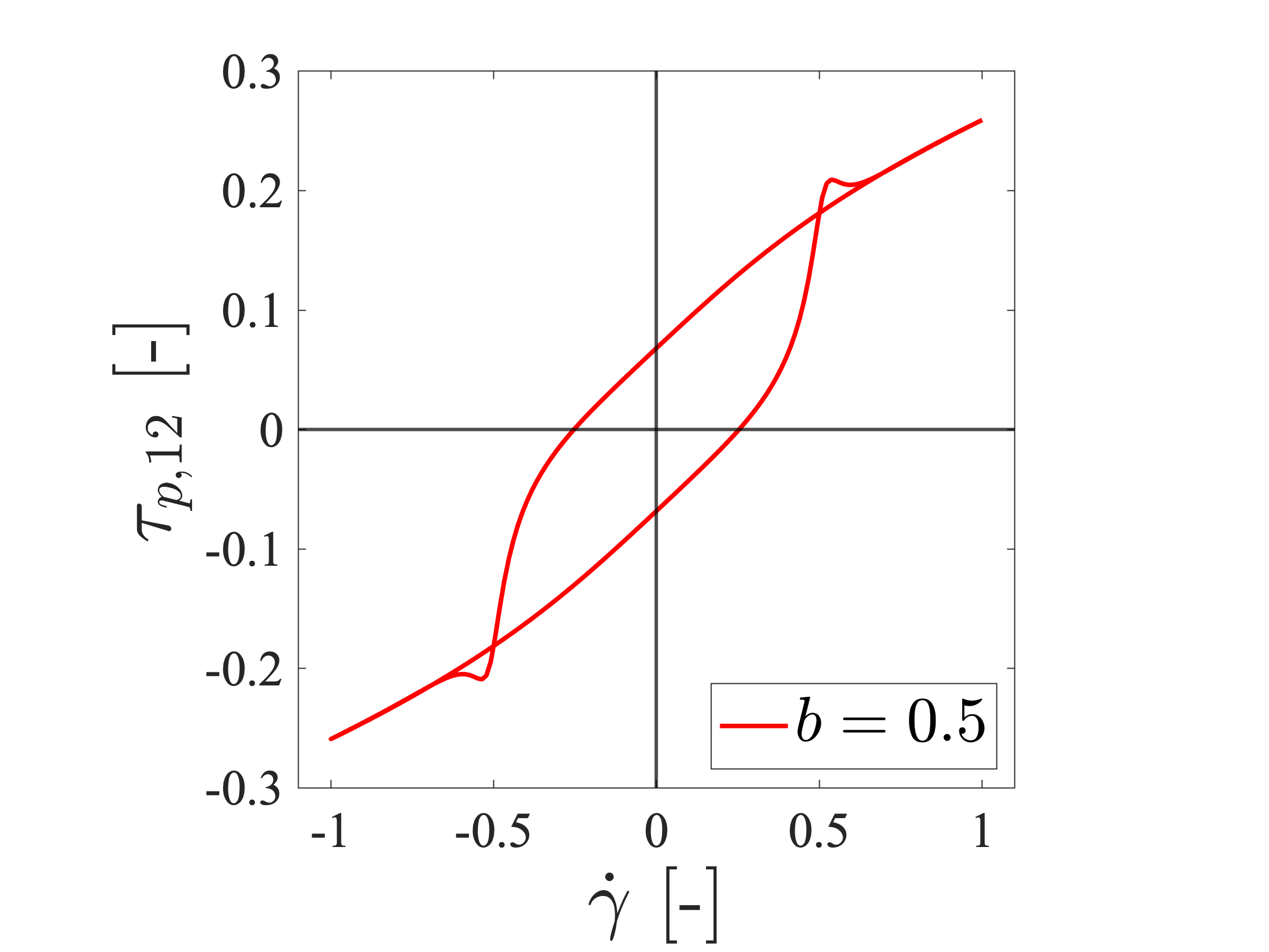}}
\subfloat[$~$]{
\includegraphics[trim={1.2cm 0cm 3.4cm 0cm},clip,width=0.33\textwidth]{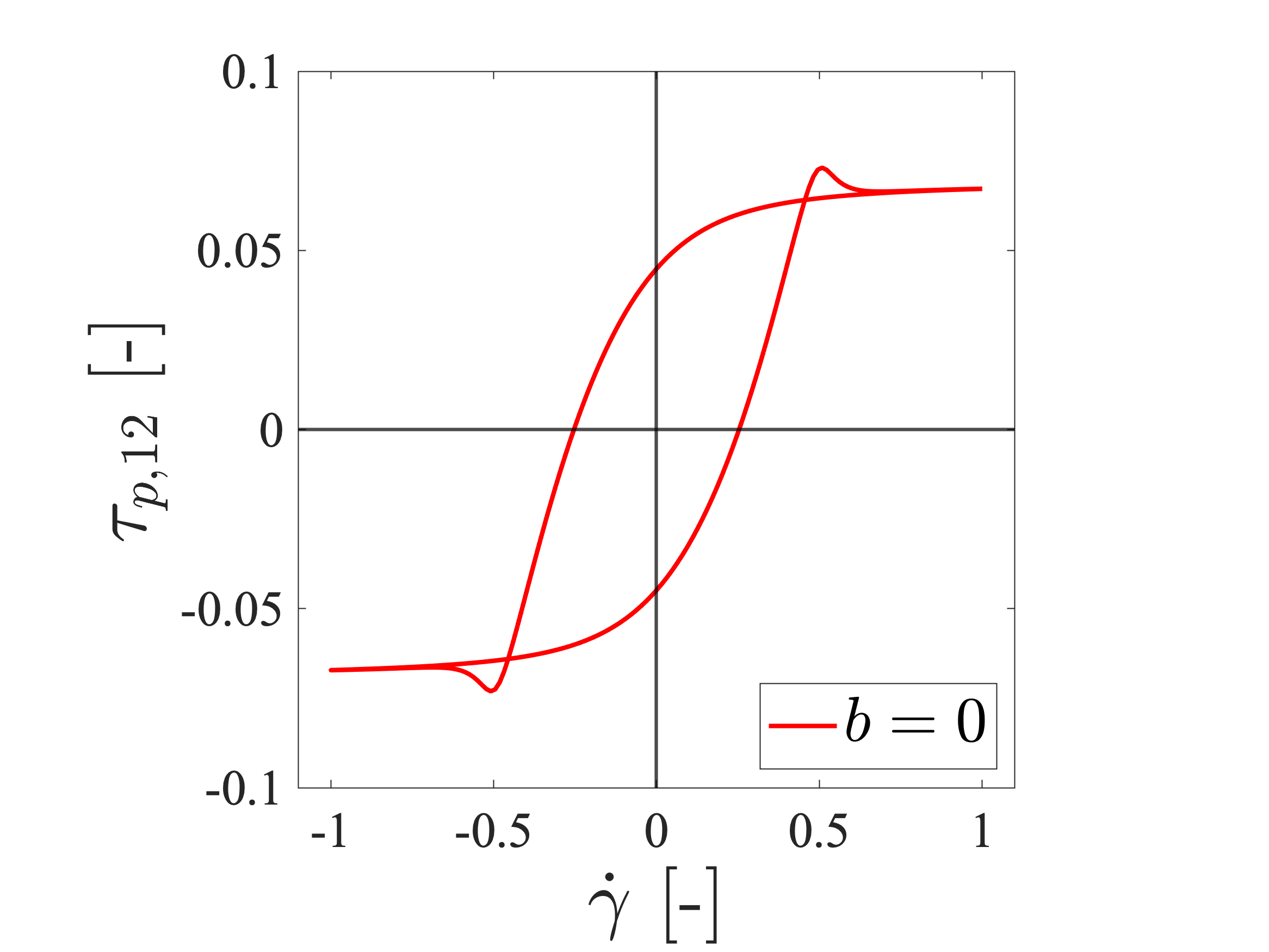}}
\caption{Viscous Lissajous curves ($\tau_{p,12}$ \textit{versus} $\dot{\gamma}$) for the toy FENE-CR model with (a) $b=1$, (b) $b=0.5$, and (c) $b=0$. Note that small stress overshoots are observed in all cases.}
\label{FENE_CR_Lissajous}
\end{figure}

Figure \ref{FENE_CR_Lissajous} shows the viscous Lissajous curves for the toy FENE-CR model with varying values of $b$ at $De = 1$, $Wi = 100$ and $L^2 = 100$. For all values of $b$ in the toy FENE-CR model, despite the fact that $Q_g$ is linear, a small stress overshoot is observed due to the non-linearity of $F(A)_{\mathrm{FP}}$, and hence non-linearity of $Q_r$.  This highlights again that the pronounced stress overshoots, which are observed in the FENE-P, toy FENE-P, and toy sPTT model responses, are closely linked with the non-linearity of $Q_g$ and thus the transient nature of $A_{22}$.  This becomes particularly clear when Figures \ref{FENE_CR_Lissajous}(c) and \ref{ToysFENE}(f) are compared. Note when comparing these two figures that $F(A)$ is moving from the inside to the outside of the brackets in the recoil term, and $F(A)$ does not appear in the $\boldsymbol{\tau}_p$-$\mathsfbi{A}$ relationship in either case. 

\begin{figure}
\centering
\subfloat[$~$]{
\includegraphics[trim={1.2cm 0cm 3.4cm 0cm},clip,width=0.33\textwidth]{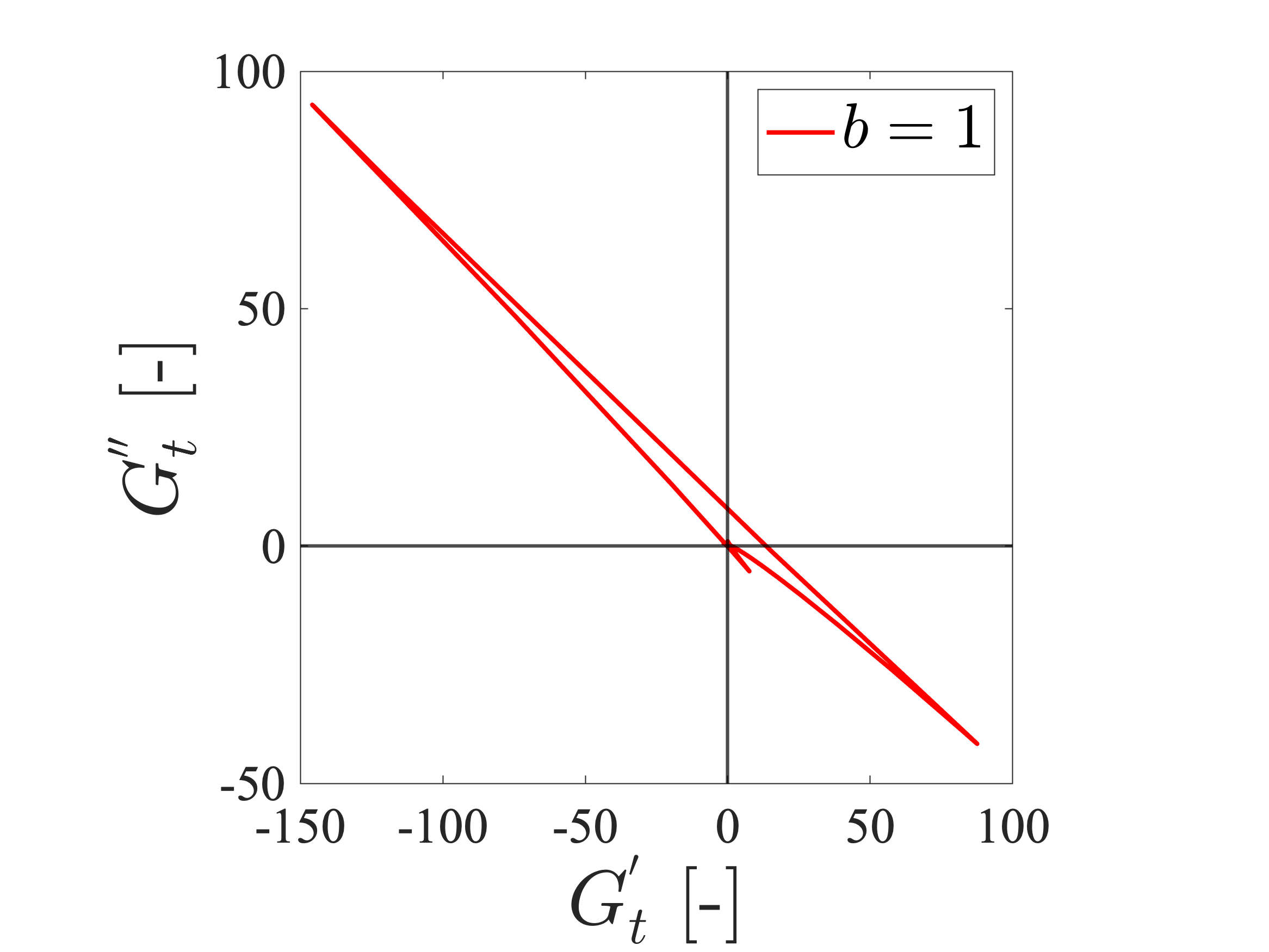}}
\subfloat[$~$]{
\includegraphics[trim={1.2cm 0cm 3.4cm 0cm},clip,width=0.33\textwidth]{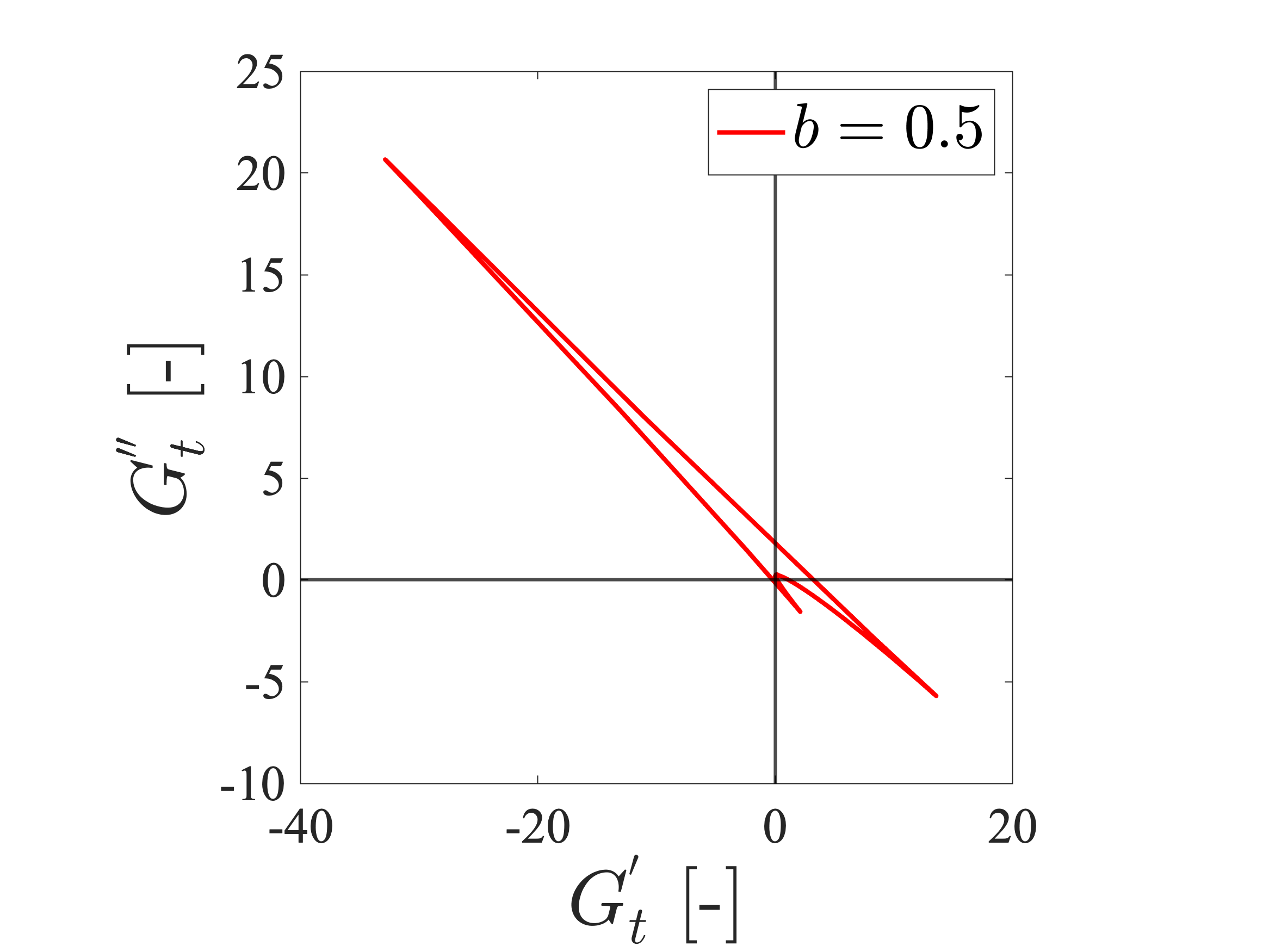}}
\subfloat[$~$]{
\includegraphics[trim={1.2cm 0cm 3.4cm 0cm},clip,width=0.33\textwidth]{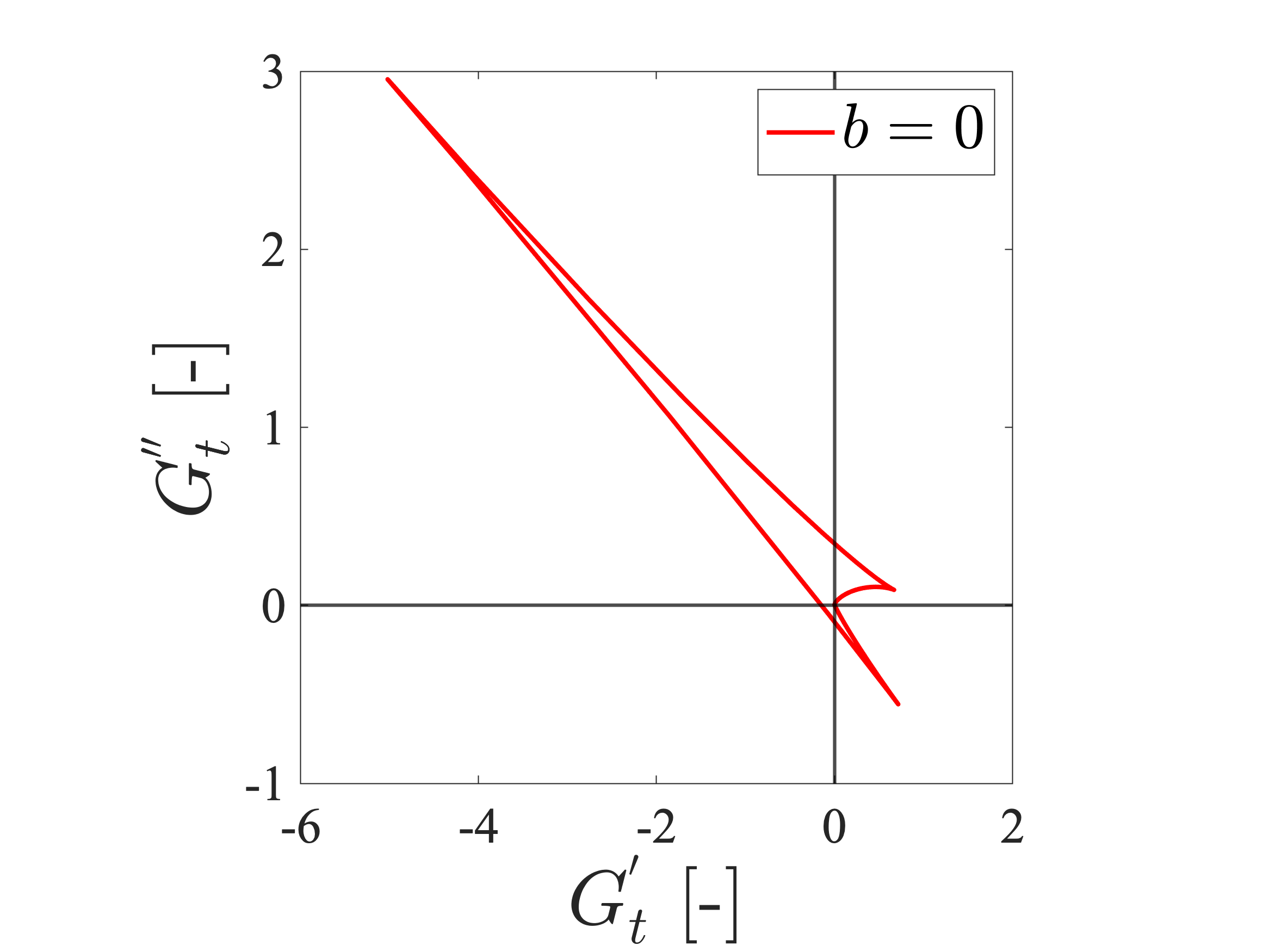}}
\caption{Cole-Cole plots ($G^{''}_t$ \textit{versus} $G^{'}_t$) for the toy FENE-CR model with (a) $b=1$, (b) $b=0.5$, and (c) $b=0$. Note that for $b=0$, only one region of viscous backflow (i.e. $G^{''}_t < 0$) is observed, whereas two regions of backflow are observed for $b=0.5$ and $b=1$.}
\label{FENE_CR_cc}
\end{figure}

Figure \ref{FENE_CR_cc} shows the Cole-Cole plots for the toy FENE-CR model responses displayed in Figure \ref{FENE_CR_Lissajous}.  For $b > 0$, a region of backflow is observed before the small stress overshoot, which is also observed in the toy FENE-P model response for $b > 0$.  This further highlights that the SPP framework can clearly identify the presence of $F(A)$ in the $\boldsymbol{\tau}_p - \mathsfbi{A}$ relationship, and thus help to identify a suitable form of constitutive model from LAOS data.

\bibliographystyle{jfm}
\bibliography{bib,bib2}

\end{document}